%% file: debyelayer.tex
\documentclass[11pt, a4paper]{article}

\usepackage{fullpage}
\usepackage{authblk}
\usepackage{amsmath, amsfonts, amssymb}
\usepackage{graphicx}
\usepackage{subfigure}
\usepackage{color}
\usepackage[small,bf]{caption}
\usepackage[sort&compress,square,comma,numbers]{natbib}

\usepackage{our_commands}

\title{The electric double layer at the interface between a polyelectrolyte gel and salt bath}

\author[1,2]{Matthew G.~Hennessy\thanks{matthew.hennessy@bristol.ac.uk}}
\author[2]{Giulia~L.~Celora}
\author[2]{Andreas~M\"unch}
\author[3]{Barbara~Wagner}
\author[2]{Sarah~L.~Waters}

\affil[1]{Department of Engineering Mathematics, University of Bristol, University Walk, Bristol, BS8 1TW, United Kingdom}
\affil[2]{Mathematical Institute, Woodstock Road, University of Oxford, Oxford, OX2 6GG, United Kingdom}
\affil[3]{Weierstrass Institute, Mohrenstrasse 39, 10117 Berlin, Germany}

\numberwithin{equation}{section}

\begin{document}

\maketitle
\allowdisplaybreaks

\begin{abstract}
\input{abstract.tex}

\end{abstract}

\input{introduction.tex}

\input{formulation.tex}

\input{asymptotics3d.tex}


\input{cylindrical.tex}

\input{conclusions.tex}

\input{appendix.tex}

\bibliographystyle{abbrv}
\bibliography{refs}

\end{document}

%% file: abstract.tex
The electric double layer (EDL) that forms at the interface between a polyelectrolyte gel and a salt bath is studied using asymptotic and numerical methods. Specifically, matched asymptotic expansions, based on the smallness of the Debye length relative to the typical gel dimensions, are used to construct solutions of the governing equations and derive electroneutral models with consistent jump conditions across the gel-bath interface. A general approach for solving the equations of incompressible nonlinear elasticity in a curved boundary layer is developed and used to resolve the gel mechanics in the EDL. A critical feature of the model is that it accounts for phase separation within the gel, which gives rise to diffuse interfaces with a characteristic thickness described by the Kuhn length.  We show that the solutions of the electroneutral model
can only be asymptotically matched to the solutions in the EDL,
in general, when the
Kuhn length greatly exceeds the Debye length. Conversely, if the Debye length
is similar to or larger than the Kuhn length, then the entire gel
can self-organise into periodic, electrically charged domains via phase
separation. The breakdown of electroneutrality demonstrates that the commonly
invoked electroneutral assumption must be used with caution, as it generally
only applies when the Debye length is much smaller than the Kuhn length.

%% file: introduction.tex
\section{Introduction}

Polyelectrolyte gels are soft, electro-active materials that are used
in a wealth of applications including
smart materials~\cite{Dong2006, Sidorenko2007},
fuel cells \cite{Komoto2019}, gel diodes \cite{Yamamoto2014},
regenerative medicine~\cite{Kwon2006}, and drug-delivery
systems~\cite{hydroex4}. A polyelectrolyte gel
consists of a network of deformable polymers that is swollen with fluid.
The polymers carry a fixed electric charge
and can therefore electrostatically interact with ions that are
dissolved in the imbibing fluid. Typically, polyelectrolyte gels are
surrounded by a bath consisting of a salt solution,
which allows for solvent and
ion exchange across the gel-bath interface until an equilibrium is established.
This equilibrium sets the degree of swelling that occurs in the gel and can
be controlled through a number of factors such as temperature and
electric fields, as well as the pH and salt content in the surrounding
bath~\cite{Ahn2008}.
Slight alterations in the environmental parameters can trigger
enormous changes in the gel volume.
In some cases, the volume of the gel
will undergo a discontinuous change, a phenomenon that is called a
volume phase transition~\cite{Dimitriyev_2019, Mussel2019}.
Environmental stimuli can also induce
phase separation, whereby a homogeneous gel spontaneously separates into
co-existing phases with different
compositions~\cite{Kramarenko1998, Style2018}.
Phase separation 
been proposed as a facile means of self-assembling nanostructures in
polyelectrolyte gels~\cite{WuJhaOlveraDeLaCruz2010,Wu2012}.


When a polyelectrolyte gel is surrounded by a salt solution, ions from the
solution will migrate to the free surface of the gel and form a diffuse layer
of electric charge known as the electric double layer (EDL). Generally, the
EDL has two components, the Stern layer and the diffuse layer, that
collectively act to screen the electric charges on the polymer chains.
The thickness
of the EDL is described by the Debye length and is often on the
order of tens of nanometers. An interesting feature of polyelectrolyte gels
is that the EDL is diffuse on both sides of the gel-bath interface due
to the mobile ions in the gel migrating to counter the accumulation of
charge in the surrounding bath. 


Despite the intricate structure of the EDL, it is generally believed to
play a passive role in the gel dynamics and is often neglected in
studies that aim to construct new models of polyelectrolyte
gels~\cite{drozdov_modeling_2015-3, Drozdov2016, Zhang2020, hua_theory_2012}
or employ existing models to interpret experimental
data~\cite{Horkay2001, Yu2017, TanakaOhmine1982}.
The few exceptions include the works by
Hong \etal\cite{Hong2010} and Wang and
Hong~\cite{Wang2010}, who compute solutions in the EDL for a limited range
of parameters by prescribing an
ad-hoc form of the deformation gradient tensor.
The motivation for neglecting the EDL stems from the
smallness of the Debye length (tens of nanometers) relative to the typical
dimensions of a polyelectrolyte gel (microns to centimeters); thus,
any impact of the EDL on the gel dynamics is assumed to be confined to an
extremely thin region near the free surface. 

To ease the computational burden of resolving the thin EDL,
it is common to simplify the governing equations by taking
the electroneutral limit, in which the ratio of the Debye length to the
characteristic gel size is asymptotically set to zero.
The name of the electroneutral limit derives from the fact that, to a very good
approximation, the gel and the bath are electrically neutral outside of the
EDL.  Thus, the electroneutral limit involves collapsing the EDL to a
region of zero thickness to produce equations that govern electrically
neutral materials. 
Using matched asymptotic
expansions, the bulk equations
in the electroneutral limit can be supplemented with jump
conditions across the EDL to produce the so-called electroneutral model.
Although the electroneutral limit is used extensively when modelling
polyelectrolyte gels,
very little attention is paid to
computing the solution in the EDL and checking that it can be asymptotically
matched to the solution of the electroneutral model.
Moreover, the jump conditions across the EDL are rarely derived
despite being highly non-trivial, as demonstrated by the celebrated
Helmholtz--Smoluchowski slip condition for ionic solutions in
contact with a rigid solid~\cite{Yariv2009}.
Mori \etal\cite{Mori2013} used matched asymptotics to derive an electroneutral
model for a polyelectrolyte gel but did not compute solutions to it nor
study the EDL in detail.

The aims of this paper are to use matched asymptotic
expansions to: (i) revisit the assumption that the EDL plays a  passive
role in the dynamics of polyelecrolyte gels
and (ii) ascertain the validity of the electroneutral limit. In particular,
we will compute the electroneutral model and explore when its solutions can be
asymptotically matched to the solutions in the EDL. The main result
of our work is that asymptotic matching of solutions cannot always be
carried out because the EDL can trigger a mode of phase separation
that leads to a breakdown of electroneutrality across the entire gel.

Our asymptotic analysis of the EDL builds on that of
Yariv \cite{Yariv2009} by accounting for the electro-chemo-mechanics
of the gel,
which requires reformulating and solving the equations
of three-dimensional nonlinear elasticity in a curved and evolving boundary
layer. By using a general form of the deformation gradient tensor in the
analysis, we are
able to elucidate how the simplified form proposed by Hong
\etal\cite{Hong2010} and Wang and Hong~\cite{Wang2010} arises.
Another crucial feature of our analysis is that it is based
on a phase-field model of a polyelectrolyte gel that can capture phase separation. The use of a phase-field model introduces a new length
scale into the problem, the Kuhn length, which charactersises the thickness
of diffuse interfaces that arise from phase separation.
Most models in the literature do not account for phase separation
and thus take the Kuhn length to be zero. However, we find that
the electroneutral limit is only asymptotically consistent, in general,
when the Kuhn length greatly exceeds the Debye length, which prevents
the emergence of electrically charged domains in the gel due
to phase separation.  Thus, we argue that particular care
must be taken when applying electroneutral models to experimental data.

The paper is organised as follows. 
In Sec.~\ref{sec:model} the governing  equations for a polyelectrolyte gel
are presented along with those of the surrounding bath. 
In Sec.~\ref{sec:asymptotics} we carry out the asymptotic analysis of the EDL
for a general three-dimensional configuration assuming the Kuhn length
is much larger than the Debye length. 
In Sec.~\ref{sec:omega}, we discuss how the analysis differs if the Kuhn
length is zero, which is more typical across the literature. 
The asymptotic framework is then applied to cylindrical polyelectrolyte gels
in Sec.~\ref{sec:cylinder}. The paper concludes in Sec.~\ref{sec:conclusions}.

%% file: formulation.tex
\section{Mathematical model}
\label{sec:model}

\begin{figure}
  \centering
  \includegraphics[width=0.7\textwidth]{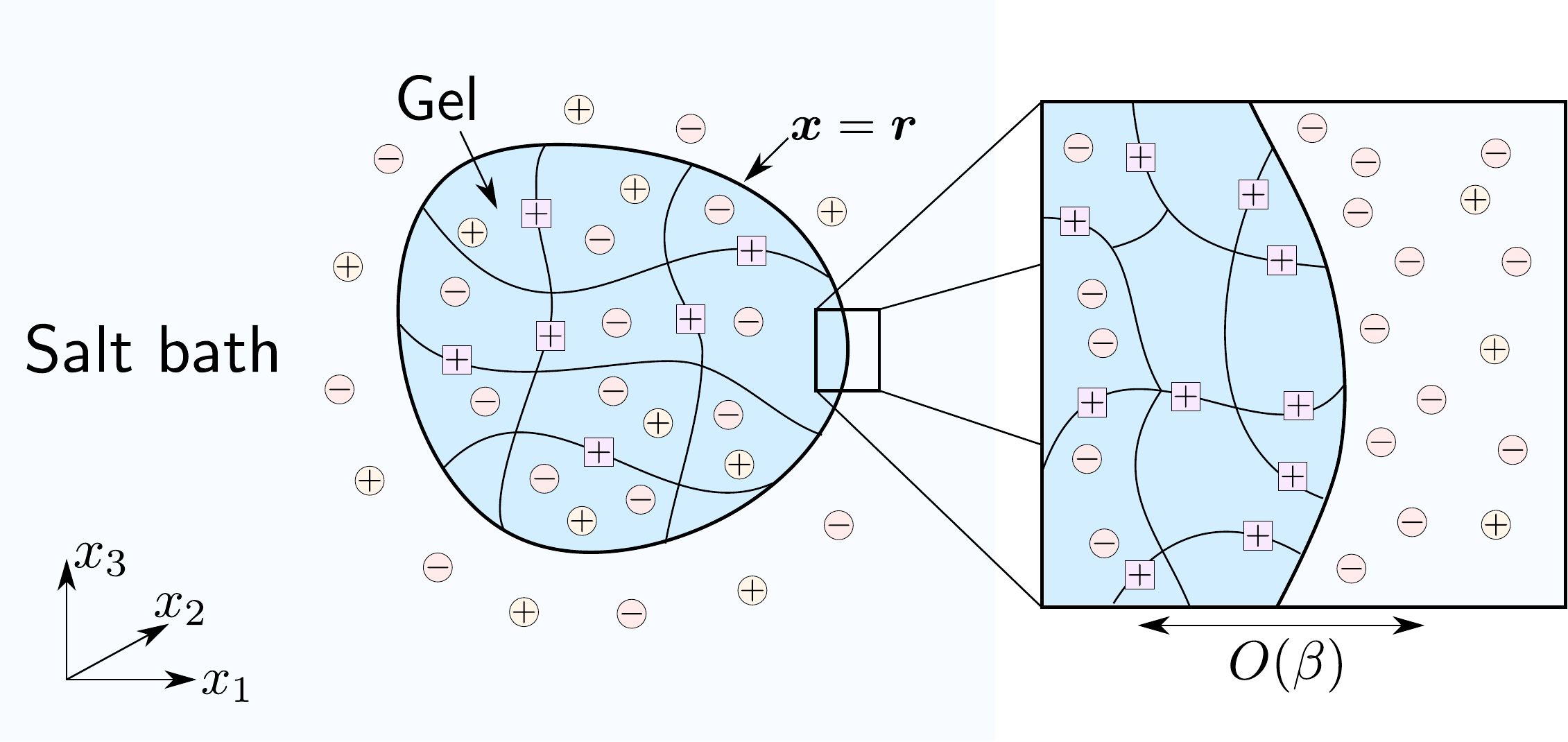}
  \caption{A swollen polyelectrolyte gel surrounded by a bath.
    The bath consists of a solvent and a dissolved binary salt.
    The polymers of the gel
    carry an electric charge, which is assumed to be positive.
    An electric double layer of thickness $O(\beta)$
    forms near the gel-bath interface, located at $\vec{x} = \vec{r}$,
    where charge neutrality is violated. 
    The non-dimensional Debye length $\beta$ is defined
    in \eqref{nd:params}.}
  \label{fig:schematic}
\end{figure}

We consider a polyelectrolyte gel that is surrounded by a bath, as shown in
Fig.~\ref{fig:schematic}. The bath consists of a solvent and a dissolved binary
salt such as NaCl or CaCl$_2$. The gel is composed of a crosslinked network of
deformable polymers that carry electric charges of the
same sign. 

Our analysis is based on the thermodynamically consistent model of a
polyelectrolyte gel that is surrounded by a viscous bath that has been
derived by Celora \etal\cite{Celora_modelling_2021}. 
For brevity, we only present the non-dimensional form of the governing
equations in the main text; however, the dimensional model is provided in Appendix \ref{app:dimensional}.
In the equations below, the subscript $m$ is used to represent
quantities associated with the solvent ($s$), cation ($+$), or the anion ($-$).
The subscript $n$ refers to the polymer network. 
The set $\mathbb{M} = \{s, +, -\}$ contains all of the mobile species that move
relative to the polymers. We let $\mathbb{I} = \{+,-\}$ denote the ionic
species. 

In non-dimensionalising the model, spatial variables are scaled with a
characteristic length scale $L$, which, for example, might
represent the size of the gel in its dry or as-prepared states.
We choose a time scale associated with solvent diffusion in
the gel, $t \sim L^2 / D_s^0$ where $D_s^0$ is a reference value of the
diffusivity. This time scale imparts a velocity scale for each species:
$\vec{v}_k \sim D_s^0 / L$. 
The chemical potentials of the mobile species are written as
$\mu_m = \mu_m^0 + k_B T \mu_m'$, where $\mu_m^0$ is a reference chemical
potential, $k_B$ is Boltzmann's constant, and $T$ is the absolute temperature.
The diffusives fluxes in the gel and the bath scale like
$\vec{j}_m \sim D_s^0 / (\nu L)$ and $\vec{q}_m \sim D_s^0 / (\nu  L)$,
respectively, where $\nu$ is a typical molecular volume (assumed to the be
same for each mobile species).
The electric potential in the bath and the gel is scaled with the thermal
voltage, $\Phi \sim k_B T / e$, where $e$ is the elementary charge.
Pressure gradients in the gel are assumed to balance the elastic
stress, $p \sim G$, where $G$ is the shear modulus of the polymer network.
In the bath, pressure gradients are balanced with the Maxwell stress, leading
to $p \sim \epsb (k_B T / e)^2 / L^2$, with $\epsb$ denoting the
electrical permitivity of the bath, which is assumed to be constant.

This scaling introduces four key dimensionless parameters given by
\begin{align}
  \G = \frac{\nu G}{k_B T},
  \quad
  \omega = \frac{L_K}{L},
  \quad
  \beta = \frac{L_D}{L},
  \quad
  \N = \frac{\eta D_s^0}{\epsb (k_B T / e)^2},
  \label{nd:params}
\end{align}
where $L_K$ is the Kuhn length, $L_D = (\nu \epsg k_B T)^{1/2}/e$ is
the Debye length, with $\epsg$ denoting the electric permittivity of the
gel and $\eta$ the kinematic viscosity of the bath, both of which are
assumed to be independent of composition. The parameter $\G$ characterises
the energetic cost of elastically deforming the gel relative
to the energy that is
released upon insertion of a solvent molecule into the polymer network. 
The parameters $\omega$ and
$\beta$ describe the thickness of diffuse internal interfaces and EDL
relative to $L$, respectively. Alternatively, $\omega$ can be related to the
energetic cost of gradients in the solvent concentration; see
Celora \etal\cite{Celora_modelling_2021} for details. 
Finally, $\N$ represents the ratio of the viscous stress to the Maxwell stress
in the bath. The magnitudes of these numbers will be estimated in
Sec.~\ref{sec:parameters}.

\subsection{Governing equations for the gel}

The governing equations for the gel are formulated in terms of Eulerian 
coordinates $\vec{x} = x_i \vec{e}_i$ associated with the current state of
the system, where $\vec{e}_i$ are Cartesian basis vectors.
An Eulerian coordinate system enables the equations to be
written in a physically intuitive way and it facilitates coupling the gel
and bath models via boundary conditions.
A detailed account of Eulerian-based
hydrogel modelling is provided by Bertrand \etal\cite{Bertrand2016}. In Eulerian coordinates, the deformation gradient tensor $\tens{F}$,
which describes the distortion of material elements relative to the dry state of
the gel, is more
readily expressed through its inverse,
\begin{align}
  \tens{F}^{-1} = \nabla \vec{X},
  \label{nd:gel:F_X}
\end{align}
where $\vec{X}(\vec{x},t) = X_I \vec{E}_I$ are Lagrangian coordinates associated
with the reference (dry) state of the gel,
$\vec{E}_I$ are Cartesian basis vectors in the reference state,
and $\nabla = \vec{e}_i\,\pdf{}{x_i}$. The adopted conventions for computing
derivatives of vectors and tensors are given in Appendix~\ref{app:conventions}.
The quantity $\vec{X}(\vec{x},t)$ provides the Lagrangian coordinates
of the material element that is located at the point $\vec{x}$ in the current state
at time $t$.
The determinant $J = \det \tens{F}$ characterises volumetric changes in
material elements.
Both the polymers and the imbibed salt solution are assumed to be
incompressible. As a result, any volumetric change in a solid element must be due
to a variation in the amount of fluid contained within that element.
This leads to
the so-called molecular incompressibility condition
\begin{align}
  J = \left(1 - \sum_{\allm} \phi_m\right)^{-1} = \phi_n^{-1},
  \label{nd:gel:J}
\end{align}
where $\phi_k$ represent the volume fraction of species $k$. The volume of fixed
charges on the polymers is accounted for in the network fraction $\phi_n$.
Since $J$ describes the volume of swollen material elements
relative to their dry volume, we also refer to it as the swelling ratio.
The Lagrangian coordinates $\vec{X}$ are convected with material elements
and thus satisfy the equation
\begin{align}
  \pd{\vec{X}}{t} + \vec{v}_n \cdot \nabla \vec{X} = 0,
  \label{nd:gel:convected_X}
\end{align}
where $\vec{v}_n$ is the velocity of the polymer network. 
Equation \eqref{nd:gel:convected_X} can be rearranged to obtain an expression for
the velocity $\vec{v}_n$ given by
\begin{align}
  \vec{v}_n = -\tens{F}\,\pd{\vec{X}}{t},
  \label{nd:gel:v_n}
\end{align}
where \eqref{nd:gel:F_X} has been used to write $\nabla \vec{X}$ in terms of
$\tens{F}$.

Conservation of polymer, solvent, and ions leads to
\subeq{
  \begin{align}
    \pd{\phi_n}{t} + \nabla \cdot (\phi_n \vec{v}_n) = 0, \\
    \pd{\phi_m}{t} + \nabla \cdot (\phi_m \vec{v}_n + \vec{j}_m) = 0,
    \label{nd:gel:phi_m}
\end{align}}
where $\vec{j}_m = \phi_m(\vec{v}_m - \vec{v}_n)$ is the diffusive flux
and $\allm$.
The volume-averaged mixture velocity in the gel, $\vec{v}$, is defined as,
and satisfies,
\begin{align}
  \vec{v} \equiv \phi_n \vec{v}_n + \sum_{\allm} \phi_m \vec{v}_m = \vec{v}_n +
  \sum_{\allm} \vec{j}_m.
  \label{nd:gel:v}
\end{align}
Diffusive transport of solvent and ions is described by a 
Stefan--Maxwell model. The fluxes are thus given by
\subeq{
  \label{nd:gel:j}
\begin{align}
  \vec{j}_s &= -\D_s(J) \sum_{\allm} \phi_m \nabla \mu_m, \\
  \vec{j}_\pm &= -\D_{\pm} \phi_{\pm} \nabla \mu_{\pm} + \frac{\phi_{\pm}}{\phi_s}\vec{j}_s,
\end{align}
}
where $\D_s(J) = D_s(J) / D_s^0$ and $\D_{\pm} = D_{\pm} / D_s^0$. The
dimensional parameters $D_s$ and $D_\pm$ denote the solvent diffusivity relative
to the polymer network and the ionic diffusivity relative to a pure solvent
bath, respectively.
The dependence of $\D_s$ on $J$ reflects the change in diffusivity
(or permeability) that occurs as the polymer network is
deformed~\cite{Bertrand2016}. The chemical potentials can be written as
\subeq{
\begin{align}
  \mu_s &= \Pi_s + \G p - \omega^2 \nabla^2 \phi_s, \label{nd:gel:mu_s}
  \\
  \mu_{\pm} &= \Pi_{\pm} + \G p + z_{\pm} \Phi,
\end{align}
}
where $z_{\pm}$ is the valence of the ions and $\Pi_m$ are osmotic pressures
defined as
\subeq{
  \label{nd:gel:Pi}
\begin{align}
  \Pi_s &= \log \phi_s + \chi\,J^{-1}(1 - \phi_s) + J^{-1}, \\
  \Pi_\pm &= \log \phi_\pm + J^{-1}(1 - \chi \phi_s).
            \label{nd:gel:Pi_pm}
\end{align}
}
Here, $\chi$ is the Flory interaction parameter, which describes (unfavourable)
enthalpic interactions between the solvent molecules and the polymers.
The electric potential satisfies
\begin{align}
  -\beta^2 \nabla^2 \Phi = z_+\phi_+ + z_{-} \phi_- + z_f \phi_f,
  \label{nd:gel:Phi}
\end{align}
where $\phi_f$ represents the volume fraction of fixed charges on the polymer network and $z_f$ denotes the valence of these charges. The nominal volume
fraction of fixed charges is
$\alphaf = \phi_f J$. We will focus on cationic gels with
positive fixed charges, $z_f > 0$.

The conservation of linear momentum in the gel leads to
\begin{align}
  \nabla \cdot \tens{T} = \vec{0},
  \label{nd:gel:div_T}
\end{align}
where $\tens{T}$ is the Cauchy stress tensor, which can be decomposed according to
\subeq{
  \label{nd:gel:T}
\begin{align}
  \tens{T} = \tens{T}_e + \tens{T}_K + \tens{T}_M - p \tens{I}.
\end{align}
The first contribution, $\tens{T}_e$, represents the elastic stress tensor
and is calculated by assuming the polymer network behaves
as a neo-Hookean material. This leads to
\begin{align}
  \tens{T}_e &= J^{-1}(\tens{B} - \tens{I}),
               \label{nd:gel:T_e}
\end{align}
where $\tens{B} = \tens{F} \tens{F}^T$ is the left Cauchy--Green deformation tensor.
The second and third contributions, $\tens{T}_K$ and $\tens{T}_M$,
correspond to the Korteweg and Maxwell stress tensors, respectively,
which capture the force generated within the bulk of the gel due to
internal interfaces and electric fields. These tensors can be written as
  \begin{align}
    \tens{T}_K &= \G^{-1} \omega^2 \left[\left(\frac{1}{2}|\nabla \phi_s|^2 + \phi_s \nabla^2 \phi_s\right)\tens{I} - \nabla \phi_s \otimes \nabla \phi_s\right],
                 \label{nd:gel:T_K} \\
    \tens{T}_M &= \G^{-1} \beta^2 \left(\nabla \Phi \otimes \nabla \Phi - \frac{1}{2}|\nabla \Phi|^2 \tens{I}\right).
                 \label{nd:gel:T_M}
  \end{align}}
The final contribution to the Cauchy stress tensor represents an isotropic stress induced by the fluid pressure.

\subsection{Governing equations for the bath}
Conservation of solvent and ions in the bath is given by
\begin{align}
  \pd{\phi_m}{t} + \nabla \cdot (\phi_m \vec{v} +\jb_m) = 0,
  \label{nd:bath:phi_m}
\end{align}
where $\allm$, $\vec{v}$ is the mixture velocity
\begin{align}
  \vec{v} = \sum_{\allm} \phi_m \vec{v}_m,
\end{align}
and $\vec{q}_m = \phi_m(\vec{v}_m - \vec{v})$ are the diffusive fluxes. Unlike the
gel, the diffusive fluxes in the bath are defined relative to the mixture velocity.
The bath is assumed to be free of voids and incompressible, which leads to
the following conditions:
\begin{align}
  \sum_{\allm} \phi_m = 1, \quad \quad \nabla \cdot \vec{v} = 0.
  \label{nd:bath:no_void}
\end{align}
The diffusive fluxes in the bath are also described using a Stefan--Maxwell
model and given by
\subeq{
  \label{nd:bath:fluxes}
\begin{align}
  \jb_{\pm} &= -\D_{\pm} \phi_{\pm}\left(\nabla \mu_{\pm} - \sum_{\allm} \phi_m
  \nabla \mu_m\right) + \frac{\phi_{\pm}}{\phi_s}\jb_s, 
  \label{nd:bath:fluxes_pm}
	\\
  \jb_s &= -\jb_{+} - \jb_-.
\end{align}
}
The chemical potentials of the solvent and ions are
\subeq{
\begin{align}
  \mu_s &= \log \phi_s + \epsilon_r \beta^2 p, \\
  \mu_{\pm} &= \log \phi_{\pm} + \epsilon_r \beta^2 p  + z_{\pm} \Phi,
\end{align}
}
where $\epsilon_r = \epsb / \epsg$. The electric potential satisfies
\begin{align}
  -\epsilon_r \beta^2  \nabla^2 \Phi = z_+\phi_+ + z_{-} \phi_-.
  \label{nd:bath:Phi}
\end{align}
Conservation of linear momentum in the bath implies that
\begin{align}
  \nabla \cdot \tens{T} = \vec{0},
  \label{nd:bath:div_T}
\end{align}
where the Cauchy stress tensor is
\subeq{\label{nd:bath:T}
\begin{align}
  \tens{T} &= \tens{T}_v + \tens{T}_M - p \tens{I}.
\end{align}
The first component captures the viscous stresses in the bath, which is assumed
to be a Newtonian incompressible fluid; thus,
\begin{align}
  \tens{T}_v &= \N(\nabla \vec{v} + \nabla \vec{v}^T).
\end{align}
The Maxwell stress tensor for the bath reads
\begin{align}
  \tens{T}_M &=  \nabla \Phi \otimes \nabla \Phi - \frac{1}{2}|\nabla \Phi|^2 \tens{I}. \label{nd:T_m_bat}
\end{align}}
By combining
\eqref{nd:bath:div_T}--\eqref{nd:bath:T}, we can write the stress
balance in non-conservative form,
\begin{align}
  \nabla \cdot \tens{T}_v + \nabla^2 \Phi \nabla \Phi = \nabla p,
  \label{nd:bath:div_T_nc}
\end{align}
which will be advantageous for the asymptotic analysis of the double layer.

\subsection{Boundary conditions at the gel-bath interface}

In the current configuration, the gel-bath interface is defined by the surface
$\vec{x} = \vec{r}(s_1, s_2,t)$, which is parametrised by $s_1$ and $s_2$.
The tangent vectors to the interface are defined as
$\vec{t}_\alpha = \pdf{\vec{r}}{s_\alpha}$, $\alpha = 1,2$. 
The normal vector to the interface is denoted by
$\vec{n} = (\vec{t}_1 \times \vec{t}_2) / |\vec{t}_1 \times \vec{t}_2|$
and assumed to point from the gel into the bath.
The normal velocity of the interface is written as $V_n$. We use
the notation $\vec{x} \to \vec{r}^{\pm}$ to denote approaching the interface
from the interior of the bath ($+$) and gel ($-$). 

The kinematic boundary condition is imposed on the polymer network
\begin{align}
  \left[\vec{v}_n \cdot \vec{n} - V_n\right]_{\vec{x}=\vec{r}^{-}} = 0.
  \label{nd:bc:kinematic}
\end{align}
Conservation of solvent and ions across the moving boundary of the gel
implies that
\begin{align}
  \left[\jg_m \cdot \vec{n}\right]_{\vec{x}=\vec{r}^{-}} = A_m = \left[\jb_m\cdot \vec{n} + \phi_m (\vec{v}_m \cdot  \vec{n} - V_n)\right]_{\vec{x}=\vec{r}^{+}},
  \label{nd:bc:mass_m}
\end{align}
where the $A_m$ are introduced to facilitate the asymptotic matching in
Sec.~\ref{sec:asymptotics}. 
By summing \eqref{nd:bc:mass_m} over $\allm$ and using \eqref{nd:gel:v} and \eqref{nd:bc:kinematic}, we find that the normal component of the mixture velocity is continuous at the interface,
\begin{align}
  \left[\vec{v}\cdot\vec{n}\right]_{\vec{x} = \vec{r}^{-}} =
  \left[\vec{v}\cdot\vec{n}\right]_{\vec{x} = \vec{r}^{+}},
\end{align}
which is a reflection of the conservation of total mass.

Continuity of the chemical potential across the interface leads to
\begin{align}
  \left.\mu_m\right|_{\vec{x}=\vec{r}^{-}} = M_m = \left.\mu_m\right|_{\vec{x}=\vec{r}^{+}}.
  \label{nd:bc:mu_m}
\end{align}
Due to the non-local term in the solvent chemical potential
\eqref{nd:gel:mu_s}, an additional boundary condition on the solvent
fraction in the gel is required. We impose the variational condition
\begin{align}
  \left[\nabla \phi_s \cdot \vec{n}\right]_{\vec{x} = \vec{r}^{-}} = 0.
  \label{nd:bc:grad_phi}
\end{align}
From a physical point of view, this condition implies
that the solvent does not preferentially wet or dewet the interface, both of
which would lead to a localised gradient in the solvent composition.

After non-dimensionalisation, momentum conservation at the interface leads to
\begin{align}
  \left[\G \tens{T} \cdot \vec{n}\right]_{\vec{x} = \vec{r}^{-}} =  \left[\epsilon_r \beta^2 \tens{T} \cdot \vec{n}\right]_{\vec{x} = \vec{r}^{+}}.
  \label{nd:bc:stress}
\end{align}
The asymptotic analysis will reveal that the stresses in the bath are
$O(\beta^{-1})$ in size. As discussed in Sec.~\ref{sec:parameters}, typically 
$\beta \ll \G$ and $\epsilon_r \simeq 1$,
meaning that
\eqref{nd:bc:stress} can be reduced to a stress-free condition for the gel:
\begin{align}
  \left[\tens{T} \cdot \vec{n}\right]_{\vec{x} = \vec{r}^{-}} = \vec{0}.
  \label{nd:bc:stress_simp}
\end{align}
The final boundary condition that must be imposed on the mechanical problem
is a form of slip condition. Here we simply impose continuity of the
tangential components of the mixture velocity:
\begin{align}
  \left[\vec{v}\cdot\vec{t}_\alpha\right]_{\vec{x} = \vec{r}^{-}} = U_\alpha = 
  \left[\vec{v}\cdot\vec{t}_\alpha\right]_{\vec{x} = \vec{r}^{+}}.
  \label{nd:bc:slip}
\end{align}
However, this is just one option of several possible consistent conditions.
For instance,
Mori \etal\cite{Mori2013} opted for a Navier slip condition on the solvent velocity in their kinetic model of a polyelectrolyte gel, whereas
Feng and Young~\cite{Feng2020} used thermodynamics to derive two different
slip conditions for non-ionic gels. 
The choice of slip condition will not have a significant impact on the
asymptotic analysis.


We assume there are no surface charges on the interface and
therefore impose continuity of the electric potential and electric
displacement:
\subeq{
\begin{align}
  \left.\Phi\right|_{\vec{x} = \vec{r}^{-}} &= \left.\Phi\right|_{\vec{x} = \vec{r}^{+}}, \\
  \left[\nabla \Phi \cdot \vec{n}\right]_{\vec{x} = \vec{r}^{-}} &= \left[\epsilon_r\nabla \Phi\cdot \vec{n}\right]_{\vec{x} = \vec{r}^{+}}.
\end{align}
}

\subsection{Parameter estimation}
\label{sec:parameters}

We assume that the molecular volume of solvent and ions is
$\nu \sim 10^{-28}$~m\unit{3} \cite{Yu2017}, the system is held at a
temperature of $T = 300$~K, and the gels have a length scale of
$L\sim 1$~cm. Horkay \etal\cite{Horkay2001} measured the shear moduli
of polyelectrolyte gels to be around $G \sim 10$~kPa, which
leads to $\G \sim 10^{-4}$. Yu \etal\cite{Yu2017} reported values of
$\G \sim 10^{-3}$.

We assume that the electrical permittivity of the gel and the bath are
approximately the same as water due to the ions being dilute. Thus, we set
$\epsg \simeq \epsb \simeq 80\,\epsilon_0$, where
$\epsilon_0$ is the permittivity of free space. Hence,
$\epsilon_r = \epsg / \epsb \simeq 1$. The non-dimensional width 
of the EDL is then $\beta \sim 10^{-8}$, corresponding
to a dimensional value of $0.1$~nm. However, we will show in
Sec.~\ref{sec:cylinder} that this value underestimates the width of
the EDL computed from the model.

The dimensionless parameter $\omega$ is difficult to estimate due to
uncertanties in the values of the Kuhn length. Hua \etal\cite{hua_theory_2012}
set $L_K = 0.9$~nm in their modelling study. Similarly, Wu \etal\cite{Wu2012}
take $L_K = 1$~nm. Both values lead to an estimate of
$\omega \sim 10^{-7}$. The estimated values of $\beta$ and
$\omega$ suggest that the Debye and Kuhn lengths will be comparable. 

Drozdov \etal\cite{Drozdov2016b} report
solvent diffusion coefficients ranging from
$D_s^0 \sim 10^{-11}$~m\unit{2}$\cdot$s\unit{-1} to
$D_s \sim 10^{-9}$~m\unit{2}$\cdot$s\unit{-1}. In a dilute solution, the
ionic diffusivities are on the order of
$D_{\pm} \sim 10^{-9}$~m\unit{2}$\cdot$s\unit{-1} \cite{SherwoodThomasK1975Mt}.
Thus, we expect $\D_{\pm}$ to range from $1$ to $100$. Assuming
the solvent is water, which has a viscosity $\eta_w \sim 10^{-3}$~Pa$\cdot$s, and that the concentration of ions in the bath is small compared to the concentration of solvent molecules, i.e. the bath is a dilute solution, then we can approximate the mixture viscosity $\eta$ with $\eta_w$.
Hence, we find that $\N$ ranges from $10^{-2}$ to $1$. 
The (nominal) volume fraction of fixed charges is reported to range from $\alphaf \sim 10^{-3}$ to $\alphaf \sim 10^{-1}$ \cite{Hong2010, Yu2017}.
The Flory interaction parameter $\chi$ is generally a function of the
gel composition and temperature. However, we treat $\chi$ as a constant,
which is a common simplification in the literature. 
Yu \etal\cite{Yu2017} use constant values of $\chi$ that range from 0.1 to 1.6.

%% file: asymptotics3d.tex
\section{Asymptotic analysis for large Kuhn lengths}
\label{sec:asymptotics}

Matched asymptotic expansions in the limit $\beta \to 0$ will now be used to
formulate and, in some cases, solve the governing equations away from and
within the EDL at the gel-bath interface. The analysis in this section will
focus on the case when the Kuhn length is much
larger than the Debye length; thus, we will consider the limit
$\beta \to 0$ with $\beta \ll \omega$. Although our estimates
suggests that $\omega$ and $\beta$ are similar in magnitude and hence the
limit $\beta \to 0$ with $\omega = O(\beta)$ may be more physically
accurate, we will show that the asymptotic solutions cannot generally
be matched in this case. Analysing the case when $\beta \ll \omega$ provides
mathematical and physical insights into why the matching fails.

The asymptotic analysis is split into three parts.
In Sec.~\ref{sec:outer}, we reduce the
model in the outer region away from the gel-bath interface and in doing
so formulate the bulk equations for the electroneutral model.
In Sec.~\ref{sec:inner}, we formulate the problem in the inner region near
the gel-bath interface to resolve the EDL. Finally, in Sec.~\ref{sec:matching},
we derive asymptotically consistent jump conditions across the EDL for the
electroneutral model.

\subsection{The outer problem}
\label{sec:outer}

\subsubsection{Electroneutral equations for the bath}
Taking $\beta \to 0$ in \eqref{nd:bath:Phi} leads to the
electroneutrality condition
\begin{align}
  z_+ \phi_+ + z_- \phi_- = 0.
  \label{outer:bath:en}
\end{align}
When \eqref{outer:bath:en} is combined with the no-void condition
\eqref{nd:bath:no_void}, the volume fractions of solvent $\phi_s$ and
anions $\phi_{-}$ can be eliminated from the problem. 
By manipulating the ion balances in \eqref{nd:bath:phi_m},
we can arrive at
  \begin{align}
    \nabla \cdot \left(z_+ \jb_+ + z_{-} \jb_- \right) = 0,
    \label{outer:bath:Phi}
  \end{align}
  which we interpret as an elliptic equation for
  the electric potential $\Phi$ in the bath. The volume fraction of cation
  evolves
  according to
  \begin{align}
    \pd{\phi_{+}}{t} + \vec{v} \cdot \nabla \phi_{+}
    + \nabla \cdot \jb_+ = 0.
    \label{outer:bath:phi}
  \end{align}
  The fluxes in the bath are given by
  \eqref{nd:bath:fluxes} and the chemical potentials reduce to
\subeq{
  \begin{align}
    \mu_s &= \log \phi_s, \\
    \mu_{\pm} &= \log \phi_{\pm} + z_{\pm} \Phi,
  \end{align}
}
which show that the contribution from the pressure can be neglected.
Finally, the mixture velocity $\vec{v}$ satisfies 
\subeq{
  \label{outer:bath:v}
\begin{align}
  \N \nabla^2 \vec{v} + \nabla^2 \Phi \nabla \Phi &= \nabla p,
  \label{outer:bath:mom}                                                    
  \\
  \nabla \cdot \vec{v} &= 0,
\end{align}
}
where $\N$ has been assumed to be independent of composition. The form
  of \eqref{outer:bath:mom} shows that Maxwell stresses
  enter the leading-order
  momentum balance despite the bath being electrically neutral.

\subsubsection{Electroneutral equations for the gel}
Taking $\beta \to 0$ in \eqref{nd:gel:Phi} leads to the electroneutrality
condition in the gel,
\begin{align}
  z_+ \phi_+ + z_- \phi_- = - z_f \phi_f.
  \label{outer:gel:ne}
\end{align}
Using $\phi_f = \alphaf / J$ along with \eqref{nd:gel:J} in \eqref{outer:gel:ne},
the anion fraction $\phi_-$ can be eliminated from the outer problem. 
By multiplying the conservation
equation for each ion by their respective valence number $z_i$ and adding, we
find that 
\begin{align}
  \nabla \cdot (z_+ \jg_{+} + z_{-} \jg_{-}) = z_f \left(\pd{\phi_f}{t} + \nabla \cdot (\phi_f \vec{v}_n)\right) = 0,   \label{outer:gel:Phi}
\end{align}
which determines the electric potential $\Phi$ in the gel.
The second equality is obtained by writing $\phi_f = \alphaf / J$, assuming that
$\alphaf$ is uniform in the reference state, and then using the identity~\cite{Gonzalez2008}
\begin{align}
  \pd{J}{t} + \vec{v}_n \cdot \nabla J = J \nabla \cdot \vec{v}_n.
\end{align}
The solvent and cation fractions satisfy the equations
\subeq{
  \begin{align}
    \pd{\phi_s}{t} + \nabla \cdot \left( \phi_s \vec{v}_n + \jg_s\right)
    &= 0, \\
    \pd{\phi_+}{t} + \nabla \cdot \left( \phi_+ \vec{v}_n + \jg_+\right) &= 0,
  \end{align}
}
where the fluxes and chemical potentials are given by \eqref{nd:gel:j}--\eqref{nd:gel:Pi}. The network velocity $\vec{v}_n$ is obtained by solving the
mechanical problem,
which consists of the kinematic relations in \eqref{nd:gel:F_X} and
\eqref{nd:gel:v_n} and the stress balance
\begin{align}
  \nabla \cdot \tens{T}_e
  + \omega^2 \G^{-1} \phi_s \nabla \nabla^2 \phi_s = \nabla p,
  \label{outer:gel:div_T}
\end{align}
where the elastic stress tensor is given by \eqref{nd:gel:T_e}.
  Contrary to the bath problem, the form of
  \eqref{outer:gel:div_T} shows that the Maxwell stresses do not contribute
  to the leading-order stress balance in the gel.

\subsection{The inner problem}
\label{sec:inner}

\begin{figure}
  \centering
  \includegraphics[width=\textwidth]{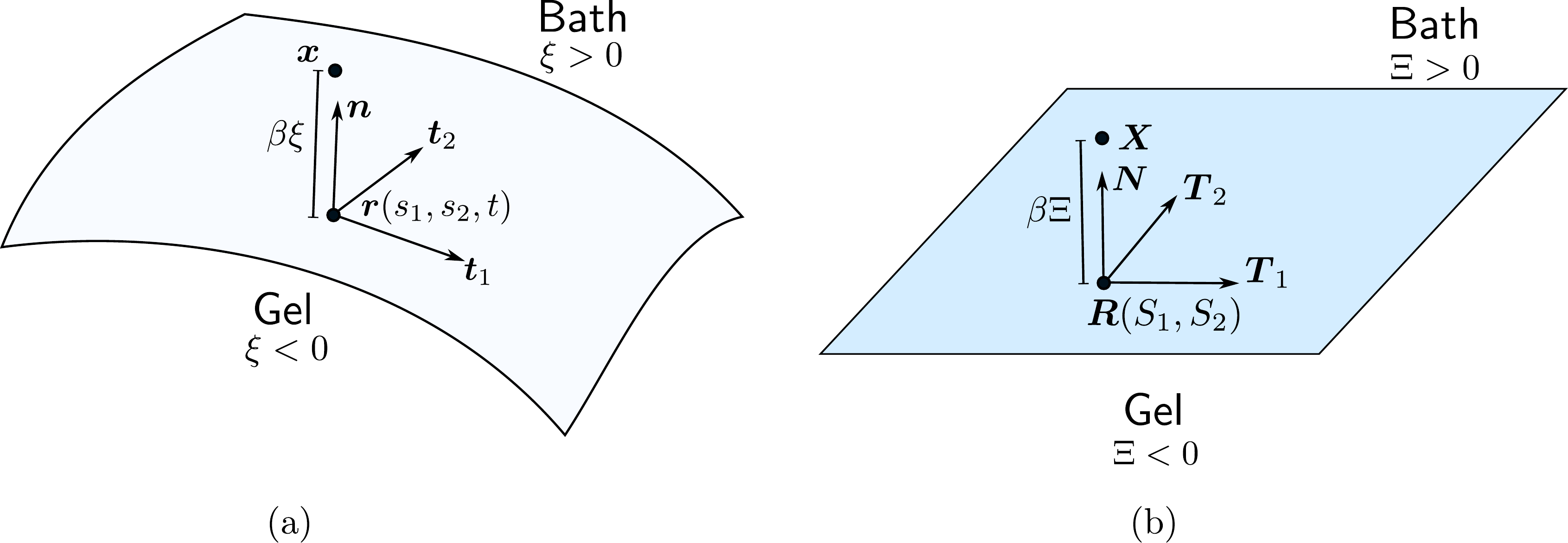}
  \caption{A schematic diagram of the (a) Eulerian and (b) Lagrangian
    coordinate systems used to formulate the inner problem. The vectors
    $\vec{r}$ and $\vec{R}$, $\vec{t}_\alpha$ and $\vec{T}_\alpha$,
    and $\vec{n}$ and $\vec{N}$ represent the gel-bath interface,
    tangent vectors, and unit normal vectors.
    The interface is parametrised by $s_\alpha$ and $S_\alpha$,
    and $\xi$ and $\Xi$ represent
    coordinates in the normal direction. The gel and bath domains are defined
    by $\xi,\Xi < 0$ and $\xi, \Xi > 0$, respectively.}
  \label{fig:curve_fitted}
\end{figure}

The inner problem is formulated using a surface-fitted coordinate system.
This allows a point $\vec{x}$ to be represented in
terms of its normal distance from the interface and its position along the
interface. We thus make the change of variable
\begin{align}
  \vec{x} &= \vec{r}(s_1, s_2, t') + \beta \xi \vec{n}(s_1, s_2, t'), \quad t = t',
            \label{inner:x}
\end{align}
where
$\xi$ is a coordinate in the normal direction. By convention, the normal
vector $\vec{n}$ points from the gel to the bath; therefore,
$\xi > 0$ corresponds
to the regions in the bath whereas $\xi < 0$ corresponds to regions in the gel. An illustration of this coordinate system is provided
in Fig.~\ref{fig:curve_fitted}~(a).
Under this change of variable, the spatial and time derivatives become
(see Appendix \ref{app:inner} for details)
\subeq{
  \label{eqn:inner_diff}
\begin{align}
  \nabla &= \beta^{-1} \vec{n} \pd{}{\xi} + \nabla_s + O(\beta), \\
  \nabla^2 &= \beta^{-2} \pdd{}{\xi} + 2 \beta^{-1} \kappa \pd{}{\xi} + \nabla_s^2 - \xi (\kappa_\alpha \kappa_\alpha)\pd{}{\xi} + O(\beta), \\
  \pd{}{t} &= -\beta^{-1} V_n \pd{}{\xi} + \pd{}{t'} - \pd{\vec{r}}{t'}
             \cdot \nabla_s + O(\beta),
\end{align}
}
where $\nabla_s$ and $\nabla_s^2$ are the surface gradient and surface
Laplacian, defined in \eqref{app:inner:n0} and \eqref{app:inner:lap_s};
$\kappa_1$ and $\kappa_2$ are the principal curvatures of
the interface;
and $\kappa = (\kappa_1 + \kappa_2)/2$ is the mean curvature.
In deriving \eqref{eqn:inner_diff}, we have assumed that the non-dimensional
curvatures satisfy $\kappa_\alpha = O(1)$ as $\beta \to 0$,
i.e.\ the dimensional
curvature is $O(L^{-1})$ where $L$ is the typical length scale of the gel.
In the calculations that follow, the prime on $t'$ will be dropped. 

Tildes are used to denote dependent variables in the inner region, which 
are generally expanded as $\t{f} = \t{f}^{(0)} + \beta \t{f}^{(1)} + O(\beta^2)$,
where $f$ is an arbitrary quantity (scalar, vector, tensor). 
However, additional rescaling is required in some cases; this will be made
explicit in the proceeding discussion. 
Near the interface, the
outer solutions for the bath and gel expanded as
\subeq{
\begin{align}
  \lim_{\xi \to 0^{+}}f(\vec{r} + \beta \xi \vec{n},t) &= \bath{f}(\vec{r},t) + O(\beta), \\
  \lim_{\xi \to 0^{-}} f(\vec{r} + \beta \xi \vec{n},t) &= \gel{f}(\vec{r},t) + O(\beta),
\end{align}
}
which will be used for asymptotic matching.


\subsubsection{Inner problem for the bath}

\paragraph{Mass conservation}
The $O(\beta^{-1})$ contributions to \eqref{nd:bath:phi_m} in inner coordinates
must satisfy
\begin{align}
  -V_n \pd{\t \phi_m^{(0)}}{\xi} + \pd{}{\xi}\left(\t{\phi}_m^{(0)} \t{\vec{v}}^{(0)}\cdot \vec{n} + \t{\jb}_m^{(0)} \cdot \vec{n}\right) = 0,
\end{align}
where we have used the fact that $\vec{n}$ is independent of $\xi$. 
Integrating these equations gives
\begin{align}
  \t{\jb}_m^{(0)} \cdot \vec{n} + \t{\phi}_m^{(0)} \left( \t{\vec{v}}^{(0)}\cdot \vec{n} - V_n\right) = A_m(s_1, s_2, t)
  \label{inner:bath:mass_m}
\end{align}
where the integration constant $A_{m}$ is determined by matching to the
outer solution:
\begin{align}
  A_m(s_1, s_2 ,t) = \bath{\jb}_m \cdot \vec{n} + \bath{\phi}_s (\bath{\vec{v}}_m \cdot \vec{n} - V_n).
  \label{inner:bath:A_m}
\end{align}
The leading-order part of the incompressibility condition for the bath
\eqref{nd:bath:no_void} is given by
\begin{align}
  \pd{}{\xi}\left(\t{\vec{v}}^{(0)}\cdot \vec{n}\right) = 0.
  \label{inner:bath_mass_ic}
\end{align}
Integrating and matching to the outer solution as $\xi \to \infty$ gives
\begin{align}
  \t{\vec{v}}^{(0)}\cdot\vec{n} = \bath{\vec{v}}\cdot \vec{n}.
  \label{inner:bath:mass_mixture}
\end{align}

\paragraph{Momentum conservation}

After transforming the Maxwell and viscous stress tensors using
\eqref{eqn:inner_diff}, we anticipate that $\tens{T}_M = O(\beta^{-2})$ and
$\tens{T}_v = O(\beta^{-1})$ as the electric potential
$\t{\Phi}$ and mixture velocity $\t{\vec{v}}$ should
remain $O(1)$ in size across the EDL. Moreover, we expect that the
pressure will scale like the Maxwell stress so that
$p = O(\beta^{-2})$. The pressure scaling can be motivated
by considering a situation in which the fluid is motionless; in this case,
mechanical equilibrium demands that the fluid pressure balances the
Maxwell stress, as these are the only two forces at play.
Therefore, we write
$\tens{T}_M = \beta^{-2} \t{\tens{T}}_M$,
$\tens{T}_v = \beta^{-1} \t{\tens{T}}_v$,
and $p = \beta^{-2} \t{p}$. Consequently, the Cauchy stress tensor
must also be scaled as $\tens{T} = \beta^{-2} \t{\tens{T}}$.
Expanding $\t{\tens{T}}$ in powers of $\beta$ and matching to the
far field leads to the stress-free conditions
$\t{\tens{T}}^{(0)}\cdot \vec{n} \to \vec{0}$ and
$\t{\tens{T}}^{(1)}\cdot\vec{n} \to \vec{0}$
as $\xi \to \infty$. Taking the normal component of the former and the
tangential component of the latter leads to
\subeq{
  \begin{align}
    \tilde{p}^{(0)} \to 0, \quad \xi \to \infty,
    \label{inner:bath:stress_match_1}\\
    \vec{t}_\alpha \cdot \t{\tens{T}}_v^{(0)} \cdot \vec{n} \to 0,
    \quad \xi \to \infty,
    \label{inner:bath:stress_match_2}
  \end{align}}
where we have exploited the fact that $\p_{\xi} \t{\Phi}^{(0)} \to \infty$ as
$\xi \to 0$ to simplify the contributions arising from the Maxwell stresses.

The local form of the stress balance \eqref{nd:bath:div_T_nc} is given by
\begin{align}
  &\beta \pd{}{\xi}\left(\t{\tens{T}}^{(0)}_v \cdot \vec{n}\right)
  + \pd{\t{\Phi}^{(0)}}{\xi}\pdd{\t{\Phi}^{(0)}}{\xi}\,\vec{n}
   - 2 \beta \kappa \left(\pd{\t{\Phi}^{(0)}}{\xi}\right)^2\,\vec{n}
    + \beta \pd{\t{\Phi}^{(0)}}{\xi}\pdd{\t{\Phi}^{(1)}}{\xi} \vec{n}
  \notag \\ 
  &+ \beta \pdd{\t{\Phi}^{(0)}}{\xi} \left(\nabla_s \t{\Phi}^{(0)} +
    \pd{\t{\Phi}^{(1)}}{\xi}\vec{n}\right)
    =
    \pd{\t{p}^{(0)}}{\xi}\vec{n} + \beta \left(\nabla_s \t{p}^{(0)} + \pd{\t{p}^{(1)}}{\xi}\vec{n}\right) + O(\beta^2).
\end{align}
The $O(1)$ contribution can be integrated to obtain a solution for the pressure,
\begin{align}
  \t{p}^{(0)} = \vec{n}\cdot \t{\tens{T}}_M^{(0)} \cdot \vec{n} = \frac{1}{2}\left(\pd{\t{\Phi}^{(0)}}{\xi}\right)^2, \label{inner:bath:pressure_0}
\end{align}
where the constant of integration has been set to zero using
\eqref{inner:bath:stress_match_1}.
Thus, the pressure in the bath balances the normal
component of the Maxwell stresses, as expected. Using the solution for
the pressure \eqref{inner:bath:pressure_0} to evaluate the leading-order
component of the Cauchy stress tensor reveals that
$\t{\tens{T}}^{(0)}\cdot \vec{n} \equiv \vec{0}$,
implying that the normal stresses in the bath are $O(\beta^{-1})$ in size.
This validates reducing the stress-continuity condition \eqref{nd:bc:stress}
to the stress-free condition on the gel \eqref{nd:bc:stress_simp}.

The $O(\beta)$  problem involves the leading-order contribution
to the viscous stress tensor, which is given by
\begin{align}
  \t{\tens{T}}_v^{(0)} = \N\left(\pd{\t{\vec{v}}^{(0)}}{\xi} \otimes \vec{n} + \vec{n} \otimes \pd{\t{\vec{v}}^{(0)}}{\xi}\right).
\end{align}
Using the incompressibility condition \eqref{inner:bath_mass_ic}, we find that
\begin{align}
  \t{\tens{T}}_v^{(0)}\cdot \vec{n} &= \N \pd{\t{\vec{v}}^{(0)}}{\xi}.
                                      \label{inner:gel:Tv}
\end{align}
By using \eqref{inner:bath:pressure_0} and \eqref{inner:gel:Tv}, the tangential
components of the stress balance can be written as
\subeq{\label{inner:bath:slip}
\begin{align}
  \N \pdd{}{\xi}\left(\t{\vec{v}}^{(0)}\cdot \vec{t}_\alpha\right)
  + \pdd{\t{\Phi}^{(0)}}{\xi}\nabla_s\t{\Phi}^{(0)} \cdot \vec{t}_\alpha
  -\pd{\t{\Phi}^{(0)}}{\xi}\pd{}{\xi}\left(\nabla_s \t{\Phi}^{(0)}\cdot \vec{t}_\alpha\right)
  = 0.
\end{align}
This equation has also been derived by Yariv~\cite{Yariv2009}. It
can be solved with the boundary conditions
\begin{align}
  \left.\t{\vec{v}}^{(0)} \cdot \vec{t}_\alpha\right|_{\xi \to 0^{+}} = U_\alpha,
  \qquad
  \left.\pd{}{\xi}\left(\t{\vec{v}}^{(0)} \cdot \vec{t}_\alpha\right)\right|_{\xi \to \infty}
  = 0, \label{inner:bath:slip_bcs}
\end{align}}
where $U_\alpha$ can be computed from the mechanical problem for the gel.
The conditions in \eqref{inner:bath:slip_bcs} arise from imposing
the slip condition at the gel-bath interface \eqref{nd:bc:slip} and the matching
condition \eqref{inner:bath:stress_match_2}.

\paragraph{Chemical potentials and fluxes}

Expanding the chemical potentials gives, to leading order,
\subeq{\label{inner:bath:mu_0}
\begin{align}
  \t{\mu}_s^{(0)} &= \log \t{\phi}_s^{(0)} + \epsilon_r \t{p}^{(0)},
                    \label{inner:bath:mu_s_0}\\
  \t{\mu}^{(0)}_{\pm} &= \log \t{\phi}_{\pm}^{(0)} + \epsilon_r \t{p}^{(0)} + z_{\pm} \t{\Phi}^{(0)}. \label{inner:bath:mu_pm_0}
\end{align}}
The $O(\beta^{-1})$ contributions to the flux relation \eqref{nd:bath:fluxes_pm}
gives
\begin{align}
  \pd{\t{\mu}_{\pm}^{(0)}}{\xi} - \sum_{\allm} \t{\phi}_m^{(0)}\pd{\t{\mu}_m^{(0)}}{\xi} = 0.
  \label{inner:bath:j_m1}
\end{align}
The summation in this equation represents a local form of the Gibbs--Duhem relation
and is equal to zero. To show this, we first calculate through substitution
of \eqref{inner:bath:mu_0} that
\begin{align}
  \sum_{\allm} \t{\phi}_m^{(0)}\pd{\t{\mu}_m^{(0)}}{\xi} = \epsilon_r \pd{\t{p}^{(0)}}{\xi}
  + \left(z_+ \t{\phi}_+^{(0)} + z_- \t{\phi}_-^{(0)}\right)\pd{\t{\Phi}^{(0)}}{\xi}.
  \label{inner:bath:gd_0}
\end{align}
Inserting the solution for the pressure \eqref{inner:bath:pressure_0} and making use
of the leading-order part of the Poisson problem for the voltage,
\begin{align}
  -\epsilon_r \pdd{\t{\Phi}^{(0)}}{\xi} = z_+ \t{\phi}_+^{(0)} + z_- \t{\phi}_{-}^{(0)},
  \label{inner:bath:ep_0}
\end{align}
results in the terms on the right-hand side of \eqref{inner:bath:gd_0} cancelling
out. Therefore, we obtain
\begin{align}
  \sum_{\allm} \t{\phi}_m^{(0)}\pd{\t{\mu}_m^{(0)}}{\xi} = 0.
  \label{inner:bath:gd}
\end{align}
From \eqref{inner:bath:j_m1} and \eqref{inner:bath:gd}, we can deduce that the
leading-order chemical potentials are uniform across the EDL,
giving
\subeq{\label{inner:bath:mu}
\begin{align}
  \log \t{\phi}_{\pm}^{(0)} + \epsilon_r \t{p}^{(0)}  + z_{\pm} \t{\Phi}^{(0)}
  = M_{\pm}(s_1, s_2, t) \label{inner:bath:mu_pm}
  \\
  \log \t{\phi}_{s}^{(0)} + \epsilon_r \t{p}^{(0)} = M_s(s_1, s_2, t).
                      \label{inner:bath:mu_s}
\end{align}
}
By imposing the matching conditions $\t{\mu}^{(0)}_m \to \bath{\mu}_m$, $\t{\phi}_{m}^{(0)} \to \bath{\phi}_m$, $\t{p}^{(0)} \to 0$, and $\t{\Phi}^{(0)} \to \bath{\Phi}$ as $\xi \to \infty$,
we obtain
\subeq{\label{inner:bath:M_m}
\begin{align}
  M_\pm(s_1, s_2, t) &= \bath{\mu}_{\pm} = \log \bath{\phi}_{\pm} + z_{\pm} \bath{\Phi},
                        \label{inner:bath:M_pm} 
  \\
  M_s(s_1, s_2, t) &=  \bath{\mu}_s = \log \bath{\phi}_s. \label{inner:bath:M_s}
\end{align}
}
Equating \eqref{inner:bath:mu_pm} with \eqref{inner:bath:M_pm} provides
an expression for the ion fractions in the EDL,
\begin{align}
  \t{\phi}_{\pm}^{(0)} = \bath{\phi}_{\pm} \exp\left[z_{\pm}(\bath{\Phi} - \Phi^{(0)}) - \epsilon_r \t{p}^{(0)}\right]. \label{inner:bath:phi_i}
\end{align}

\paragraph{The electrical problem in the bath}

The leading-order electrical problem is obtained by combining
\eqref{inner:bath:ep_0} with the ionic volume fractions \eqref{inner:bath:phi_i}
to obtain a modified Poisson--Boltzmann equation given by
\begin{align}
     -\epsilon_r \pdd{\t{\Phi}^{(0)}}{\xi} = \exp\left[-(\epsilon_r/2) (\pdf{\t{\Phi}^{(0)}}{\xi})^2\right]\sum_{\alli}z_i \bath{\phi}_i\exp\left(z_{i}(\bath{\Phi} - \t{\Phi}^{(0)})\right),
     \label{inner:bath:Phi_ode}
\end{align}
where we have used \eqref{inner:bath:pressure_0} to eliminate the pressure.
The exponential prefactor on the right-hand side of
\eqref{inner:bath:Phi_ode} is non-standard and results from the
ionic chemical potentials depending on the pressure.
Equation \eqref{inner:bath:Phi_ode} can be integrated once and the conditions $\pdf{\t{\Phi}^{(0)}}{\xi} \to 0$ and $\t{\Phi}^{(0)} \to \bath{\Phi}$ as $\xi \to \infty$ used to obtain
\begin{align}
  \pd{\t{\Phi}^{(0)}}{\xi} = \mp \sqrt{\frac{2}{\epsilon_r}\log\left\{
  1 + \sum_{\alli}\bath{\phi}_i\left[\exp\left(z_i(\bath{\Phi} - \t{\Phi}^{(0)})\right)
  -1
  \right]\right\}}.
  \label{inner:bath:Phi_first_int}
\end{align}
The minus sign is taken if $\gel{\Phi} - \bath{\Phi} > 0$, which will generally
be the case if the fixed charges on the polymer chains are positive,
as assumed here.


\subsubsection{Inner problem for the gel}

\paragraph{Mass conservation}

Following the same approach as in the bath, the leading-order mass balance for
the polymer network leads to
\begin{align}
  -\beta^{-1} V_n \pd{\t{\phi}_n^{(0)}}{\xi} + \beta^{-1}\pd{}{\xi}\left(\t{\phi}_n^{(0)} \t{\vec{v}}_n^{(0)} \cdot \vec{n}\right) = 0.
\end{align}
Integrating and imposing the kinematic boundary condition
\eqref{nd:bc:kinematic} at the gel-bath interface ($\xi = 0$) gives
\begin{align}
  \t{\vec{v}}_n^{(0)} \cdot \vec{n} = V_n.
  \label{inner:gel:mass_n}
\end{align}
Similarly, by expressing \eqref{nd:gel:phi_m} in inner coordinates, integrating
the $O(\beta^{-1})$ contribution, and using \eqref{inner:gel:mass_n}, we
find that the diffusive fluxes are uniform and given by
\begin{align}
  \t{\jg}_m^{(0)}\cdot \vec{n} = A_m(s_1,s_2, t) = \gel{\jg}_m \cdot \vec{n}.
  \label{inner:gel:A_m}
\end{align}
where the $A_m$ are the same as in \eqref{inner:bath:mass_m}
and \eqref{inner:bath:A_m} due to the boundary conditions \eqref{nd:bc:mass_m}.
The second equality in \eqref{inner:gel:A_m} comes from matching to the
outer solution.


\paragraph{Chemical potentials and fluxes}

The $O(\beta^{-1})$ contributions to the constitutive relations for the flux
\eqref{nd:gel:j} give
  \begin{align}
    \pd{\t{\mu}_s^{(0)}}{\xi} = 0, \quad
    \pd{\t{\mu}_{\pm}^{(0)}}{\xi} = 0,
  \end{align}
implying the chemical potentials in the gel are also constant across
the EDL. Thus, we have that
\begin{align}
  \t{\mu}^{(0)}_m(\xi,s_1, s_2,t) = M_m(s_1, s_2 ,t) = \gel{\mu}_m,
  \label{inner:gel:M_m}
\end{align}
where $M_m$ are the same as in~(\ref{inner:bath:mu}). The $O(1)$ contributions to \eqref{nd:gel:j} provide expressions for the
tangential components of the diffusive fluxes,
\subeq{\label{inner:gel:jt}
  \begin{align}
    \t{\vec{j}}_s^{(0)}\cdot \vec{t}_\alpha &= -\D_s(\t{J}^{(0)})\sum_{\allm}
    \t{\phi}^{(0)}_m\, \nabla_s \gel{\mu}_m \cdot \vec{t}_\alpha, \\
    \t{\vec{j}}_{\pm}^{(0)}\cdot \vec{t}_\alpha &= -\D_\pm \t{\phi}^{(0)}_\pm \nabla_s \gel{\mu}_\pm \cdot \vec{t}_\alpha + \frac{\t{\phi}_{\pm}}{\t{\phi}_s^{(0)}}\, \t{\vec{j}}^{(0)}_s \cdot \vec{t}_\alpha,
  \end{align}}
which will be used in calculating the tangential mixture velocity;
see \eqref{inner:gel:vt}.

The chemical potential of solvent can be expanded as
\begin{align}
  \t{\mu}_s^{(0)} = \t{\Pi}_s^{(0)} + \G \t{p}^{(0)}
  - \beta^{-2}\omega^2\Bigg[&\pdd{}{\xi}\left(
    \t{\phi}_s^{(0)} + \beta \t{\phi}_s^{(1)} + \beta^2 \t{\phi}_s^{(2)}\right)
    + 2 \beta \kappa \pd{}{\xi}\left(\t{\phi}_s^{(0)} + \beta \t{\phi}_s^{(1)}\right) \notag \\ 
  &+ \beta^2\nabla_s^2\t{\phi}_s^{(0)} - \beta^2 (\kappa_\alpha \kappa_\alpha) \xi \pd{\t{\phi}_s^{(0)}}{\xi}\Bigg] + O(\beta).
  \label{inner:gel:mu_0}
\end{align}
Similarly, the boundary condition at the gel-bath interface \eqref{nd:bc:grad_phi}
can be expanded to give $\pdf{\t{\phi}_s^{(n)}}{\xi} = 0$ at $\xi = 0$
for $n = 0, 1, 2$.
The $O(\beta^{-2})$ and $O(\beta^{-1})$ contributions to \eqref{inner:gel:mu_0} along with the boundary and matching conditions show that
the solvent concentration is uniform to leading and next order,
\begin{align}
  \t{\phi}_s^{(0)}(\xi,s_1, s_2, t) = \gel{\phi_s}(s_1, s_2,t), \quad \t{\phi}_s^{(1)}(\xi, s_1, s_2, t) = \phi_s^{(1)}(s_1, s_2,t),
  \label{inner:gel:phi_s}
\end{align}
which is a distinguishing feature of the asymptotic limit in which $\beta \to 0$ with $\omega \gg \beta$.
Physically, this result is a consequence of gradients in the solvent
concentration having a high energy cost when the Kuhn length is large.
Using \eqref{inner:gel:phi_s} in \eqref{inner:gel:mu_0},
we find that the solvent chemical potential simplifies to
\begin{align}
  \t{\mu}_s^{(0)}(\xi,s_1, s_2,t) = \t{\Pi}_s^{(0)} + \G \t{p}^{(0)} - \omega^2\left(\pdd{\t{\phi}_s^{(2)}}{\xi}
  + \nabla_s^2 \gel{\phi_s}\right) = M_s(s_1, s_2,t).
  \label{inner:gel:mu_s}
\end{align}
By matching to the outer solution we find that
\begin{align}
  M_s(s_1, s_2,t) = \gel{\Pi}_s + \G \gel{p} - \omega^2 \nabla^2 \gel{\phi}_s.
  \label{inner:gel:M_s}
\end{align}
The chemical potentials of the ions can be expanded as
\begin{align}
  \t{\mu}_{\pm}^{(0)}(s_1, s_2,t) = \log \t{\phi}_{\pm}^{(0)} + \frac{1}{\t{J}^{(0)}}(1 - \chi \gel{\phi}_s) + \G \t{p}^{(0)} + z_{\pm} \t{\Phi}^{(0)} = M_{\pm}(s_1, s_2,t),  
  \label{inner:gel:mu_pm}
\end{align}
where matching gives
\begin{align}
  M_\pm(s_1, s_2,t) = \log \gel{\phi}_{\pm} + \frac{1}{\gel{J}}(1 - \chi \gel{\phi}_s) + \G \gel{p} + z_{\pm} \gel{\Phi}.
  \label{inner:gel:M_pm}
\end{align}
By combining \eqref{inner:gel:mu_pm} and \eqref{inner:gel:M_pm} and using
\eqref{inner:gel:phi_s} we find that 
\begin{align}
  \t{\phi}_{\pm}^{(0)} = \gel{\phi}_{\pm}\exp\left[z_{\pm}(\gel{\Phi}-\t{\Phi}^{(0)})
    + \G(\gel{p}-\t{p}^{(0)}) + \left(\frac{1}{\gel{J}}-\frac{1}{\t{J}^{(0)}}\right)(1-\chi \gel{\phi}_s)\right].
  \label{inner:gel:phi_pm}
\end{align}
Although this appears to be a closed-form expression for the volume fraction of ions,
it is important to recall that the Jacobian determinant $J$ also depends on the
these quantities; see \eqref{nd:gel:J}.


\paragraph{Kinematics}
Before proceeding with the stress balance in the gel, we derive local forms of
the deformation gradient tensor, displacement, and velocity of the polymer network that are valid for arbitrary deformations. In
Appendix~\ref{sec:ps}, the results are specialised to plane-strain
problems.

We first consider the Lagrangian representation
of the free surface, which is written as $\vec{X} = \vec{R}(S_1, S_2)$, where
$S_1$ and $S_2$ are parameters. The Lagrangian tangent and unit normal
vectors are
denoted by $\vec{T}_\alpha = \pdf{\vec{R}}{S_\alpha}$ and $\vec{N}$, respectively,
where we adopt the convention that Greek indices are equal to 1 or 2 and the Einstein summation convention for repeated indices.
We now write the Lagrangian coordinates $\vec{X}$ using an analogous
representation as in \eqref{inner:x} for Eulerian coordinates,
\begin{align}
  \vec{X} = \vec{R}(S_1, S_2) + \beta \Xi \vec{N}(S_1, S_2),
  \label{inner:X}
\end{align}
where $\Xi$ is the Lagrangian counterpart to $\xi$; see Fig.~\ref{fig:curve_fitted} for an illustration. 
Due to our formulation of the
governing equations in terms of Eulerian coordinates, we have, in the notation of inner variables, $S_\alpha = \t{S}_\alpha(s_1, s_2, \xi, t)$ and $\Xi = \t{\Xi}(s_1, s_2, \xi, t)$. We further impose that $\Xi = 0$ when $\xi = 0$. The
deformation gradient tensor can be written as
\begin{align}
  \t{\tens{F}}^{-1} = \beta^{-1} \pd{\t{S}_\alpha}{\xi} \vec{T}_\alpha \otimes \vec{n}
  + \pd{}{\xi}\left(\t{\Xi} \vec{N}\right)\otimes \vec{n} +
  \nabla_s \vec{R} + O(\beta).
  \label{inner:gel:F_full}
\end{align}
As before, we now expand $\t{S}_\alpha$, $\t{\Xi}$, $\t{\vec{X}}$, and
$\t{\tens{F}}$ in powers of $\beta$. The $O(\beta^{-1})$ components of
\eqref{inner:gel:F_full} imply that $\t{S}^{(0)}_\alpha$ are independent of
$\xi$ and thus $\t{S}^{(0)}_{\alpha} = \gel{S}_\alpha(s_1,s_2,t)$.
Using the expression for the surface gradient in \eqref{app:inner:n0}, we 
can define the (inverse) surface deformation gradient tensor as
\begin{align}
  \t{\tens{F}}_s^{-1} \equiv \nabla_s \vec{R}(\gel{S}_1, \gel{S}_2)
  = g^{\delta \gamma} \pd{\gel{S}_\alpha}{s_\gamma} \vec{T}_{\alpha} \otimes \vec{t}_\delta,
\end{align}
where $g^{\delta \gamma}$ are components of the inverse metric tensor.
The tensor $\t{\tens{F}}_s$ contains information about the stretching of material
elements in  the tangential directions.
The $O(1)$ component of \eqref{inner:gel:F_full} can be written as
\begin{align}
  \left(\t{\tens{F}}^{(0)}\right)^{-1} = \pd{\t{\Xi}^{(0)}}{\xi} \vec{N} \otimes \vec{n} + \pd{\t{S}^{(1)}_{\alpha}}{\xi} \vec{T}_\alpha \otimes \vec{n}
  + \t{\tens{F}}_s^{-1}.
  \label{inner:gel:F_inv}
\end{align}
We will show below that the tangential stress balances in the gel lead to
$\pdf{\t{S}^{(1)}_\alpha}{\xi} = 0$, which allows the deformation gradient
tensor to be expressed as
\begin{align}
  \t{\tens{F}}^{(0)} = \left(\pd{\t{\Xi}^{(0)}}{\xi}\right)^{-1} \vec{n} \otimes \vec{N} +
  \t{\tens{F}}_s. \label{inner:gel:F_simp}
\end{align}
The deformation gradient tensor used by Hong \etal\cite{Hong2010, Wang2010}
can be obtained
from \eqref{inner:gel:F_simp} by setting
$\t{\tens{F}}_s = \lambda_s (\vec{t}_1 \otimes \vec{T}_1 + \vec{t}_2 \otimes \vec{T}_2)$ where $\lambda_s$ is an
imposed stretch along the tangential directions.
Taking the determinant of \eqref{inner:gel:F_simp} leads to
\begin{align}
  \t{J}^{(0)} = \left(\pd{\t{\Xi}^{(0)}}{\xi}\right)^{-1} \det \t{\tens{F}}_s,
  \label{inner:gel:J}
\end{align}
which can be equated to \eqref{nd:gel:J} to eliminate $\t{\Xi}^{(0)}$ from the
problem. By matching \eqref{inner:gel:F_simp} to the outer solution for the
deformation gradient tensor, we find that
\begin{align}
  \vec{t}_\delta \cdot \t{\tens{F}}_s \cdot \vec{T}_\gamma = \vec{t}_\delta \cdot \gel{\tens{F}} \cdot \vec{T}_\gamma,
  \label{inner:gel:S_0}
\end{align}
which provides a system of differential equations that can be used to determine
$\gel{S}_\alpha$. 

The inner expansion of the velocity of the polymer network can be
calculated from \eqref{nd:gel:v_n} using the representations for $\vec{X}$
and $\tens{F}$ given by \eqref{inner:X} and \eqref{inner:gel:F_simp}.
The leading-order contribution can be expressed as
\begin{align}
  \t{\vec{v}}_n^{(0)} = V_n \vec{n} + \tens{I}_s \cdot \pd{\vec{r}}{t} -  \tilde{\tens{F}}_s \cdot \left(\vec{T}_{\alpha} \pd{\gel{S}_{\alpha}}{t}\right),
\end{align}
where $\tens{I}_s = \t{\tens{F}}_s \t{\tens{F}}_s^{-1} = g^{\alpha \nu} \vec{t}_\alpha \otimes \vec{t}_\nu$ is the surface identity tensor.
The tangential components of the mixture velocity can then
be evaluated using \eqref{nd:gel:v} as
\begin{align}
  \t{\vec{v}}^{(0)}\cdot \vec{t}_\nu =
  \pd{\vec{r}}{t} \cdot \vec{t}_\nu
  -\vec{t}_\nu \cdot \t{\tens{F}}_s\cdot \left(\vec{T}_{\alpha} \pd{\gel{S}_{\alpha}}{t}\right)
  +
  \sum_{\allm} \t{\vec{j}}_m^{(0)}\cdot \vec{t}_\nu,
  \label{inner:gel:vt}
\end{align}
where the tangential components of the flux are given by \eqref{inner:gel:jt}.
Taking the limit of \eqref{inner:gel:vt} as $\xi \to 0^{-}$ enables
the quantity $U_\alpha$ in \eqref{inner:bath:slip_bcs} to be determined.

\paragraph{Momentum conservation}
The leading-order part of the stress balance
in the gel \eqref{nd:gel:div_T}, expressed in inner coordinates, is
\begin{align}
  \pd{}{\xi}\left(\t{\tens{T}}^{(0)}\cdot \vec{n}\right)
  = 0.
  \label{inner:gel:div_T}
\end{align}
Thus, by integrating \eqref{inner:gel:div_T} and imposing the simplified boundary
condition \eqref{nd:bc:stress_simp}, we find that
\begin{align}
  \t{\tens{T}}^{(0)}\cdot \vec{n} =   \t{\tens{T}}_e^{(0)}\cdot \vec{n}  + \t{\tens{T}}_M^{(0)}\cdot \vec{n} +   \t{\tens{T}}_K^{(0)}\cdot \vec{n} - \t{p}^{(0)}\vec{n}= 0
  \label{inner:gel:T}
\end{align}
across the EDL. The leading-order elastic, Maxwell, and
Korteweg stress tensors are
\subeq{
  \begin{align}
	  \t{\tens{T}}_e^{(0)} &= \frac{1}{\t{J}^{(0)}}\left(\t{\tens{B}}^{(0)} - \tens{I}\right), \label{inner:gel:Te} \\
    \t{\tens{T}}_M^{(0)} &= \frac{1}{\G}\left(\pd{\t{\Phi}^{(0)}}{\xi}\right)^2
                           \left(\vec{n} \otimes \vec{n} - \frac{1}{2}\,\tens{I}\right), \\
    \t{\tens{T}}_K^{(0)} &= \frac{\omega^2}{\G}\left\{\left[\frac{1}{2}|\nabla_s \gel{\phi}_s|^2 + \gel{\phi}_s\left(\pdd{\t{\phi}_s^{(2)}}{\xi} + \nabla_s^2 \gel{\phi}_s\right)\right]\tens{I} - \nabla_s \gel{\phi}_s\otimes \nabla_s\gel{\phi}_s\right\}.
                           \label{inner:gel:Tk}
  \end{align}
}
Since 
$\vec{t}_\gamma \cdot \t{\tens{T}}_M^{(0)} \cdot \vec{n} = 0$ and
$\vec{t}_\gamma \cdot \t{\tens{T}}_K^{(0)} \cdot \vec{n} = 0$,
the tangential component
of \eqref{inner:gel:T} implies that
\begin{align}
  \vec{t}_\gamma \cdot \t{\tens{B}}^{(0)} \cdot \vec{n} = 0.
  \label{inner:gel:tan_stress}
\end{align}
By inverting
\eqref{inner:gel:F_inv} and calculating $\t{\tens{B}}^{(0)}$, we find that
the tangential stress balances \eqref{inner:gel:tan_stress} imply that
$\pdf{\t{S}^{(1)}_\alpha}{\xi} = 0$, as previously claimed.
The normal component of \eqref{inner:gel:T} implies that
\begin{align}
  \t{p}^{(0)} &= \vec{n}\cdot \left(\t{\tens{T}}_e^{(0)} + \t{\tens{T}}_K^{(0)} + \t{\tens{T}}_M^{(0)}\right)\cdot \vec{n}.
                \label{inner:gel:p_0}
\end{align}
In order to evaluate the Korteweg stresses without explicitly solving
for $\t{\phi}_s^{(2)}$, the expression for the solvent
chemical potential \eqref{inner:gel:mu_s} can be used in \eqref{inner:gel:Tk}
to obtain
\begin{align}
  \vec{n}\cdot\t{\tens{T}}_K^{(0)}\cdot\vec{n} =  \frac{1}{\G}\left[\frac{\omega^2}{2}|\nabla_s \gel{\phi}_s|^2 + \gel{\phi}_s\left(\t{\Pi}_s^{(0)} - M_s\right)\right] +
  \gel{\phi}_s\t{p}^{(0)}.
  \label{inner:gel:nTkn}
\end{align}
where $M_s$ is given by \eqref{inner:gel:M_s}. Note that setting $\omega = 0$ in \eqref{inner:gel:nTkn} results in
$\vec{n}\cdot\t{\tens{T}}_K^{(0)}\cdot\vec{n} = 0$, as expected. This is because $\omega = 0$ leads to $\t{\Pi}_s^{(0)} + \G \t{p}^{(0)} - M_s = 0$ from
\eqref{inner:gel:mu_s}.
Substitution of
\eqref{inner:gel:nTkn} into \eqref{inner:gel:p_0} gives an algebraic
relation for the pressure $\t{p}^{(0)}$.


\paragraph{The electrical problem in the gel}

The leading-order electrical problem in the gel is given by
  \begin{align}
    -\pdd{\t{\Phi}^{(0)}}{\xi} = z_+ \t{\phi}_{+}^{(0)} + z_-\t{\phi}_-^{(0)} + z_f \t{\phi}_f^{(0)},
    \label{inner:gel:Phi_ode}
  \end{align}
  which is coupled to the algebraic equations for the volume fractions of ions
  \eqref{inner:gel:phi_pm} and the mechanical pressure
  \eqref{inner:gel:p_0}.   
  The electrical problems for the bath and gel
  can be decoupled by combining the first integral for the electric potential
  in the bath \eqref{inner:bath:Phi_first_int}
  with the electrostatic boundary conditions
  $\t{\Phi}^{(0)}(0^{-},s_1,s_2,t) = \t{\Phi}^{(0)}(0^{+},s_1,s_2,t)$ and
  $\p_{\xi} \t{\Phi}^{(0)}(0^{-},s_1,s_2,t) = \epsilon_r \p_{\xi} \t{\Phi}^{(0)}(0^{+},s_1,s_2,t)$ to obtain
  \subeq{    \label{inner:gel:Phi_bcs}
  \begin{align}
    \left.\pd{\t{\Phi}^{(0)}}{\xi}\right|_{\xi = 0^{-}} = 
    \left.\mp \sqrt{2\epsilon_r \log\left\{
  1 + \sum_{\alli}\bath{\phi}_i\left[\exp\left(z_i(\bath{\Phi} - \t{\Phi}^{(0)})\right)
  -1
    \right]\right\}}\right|_{\xi = 0^{-}},
  \end{align}
  which acts as a boundary condition for \eqref{inner:gel:Phi_ode}. The
  electrical problem in the gel is closed by imposing the matching
  condition
  \begin{align}
    \t{\Phi}^{(0)} \to \gel{\Phi}, \quad \xi \to -\infty.
  \end{align}}

\subsection{Jump conditions across the gel-bath interface}
  \label{sec:matching}

  Asymptotically consistent jump conditions across the EDL
  for the electroneutral model are derived by 
  connecting the inner and outer solutions in the bath and gel
  via the boundary conditions the gel-bath interface.

\paragraph{Kinematic conditions}
By imposing mass conservation at the gel-bath interface \eqref{nd:bc:mass_m},
we can equate \eqref{inner:bath:A_m} with \eqref{inner:gel:A_m} to obtain
\begin{align}
  \gel{\jg}_{m} \cdot \vec{n} = \bath{\jb}_m \cdot \vec{n} + \bath{\phi}_m (\bath{\vec{v}}\cdot \vec{n} - V_n). \label{jump:flux}
\end{align}
Moreover, by matching \eqref{inner:gel:mass_n}
to the solution as $\xi \to -\infty$, the outer
problem obeys the usual kinematic boundary condition \eqref{nd:bc:kinematic} on the network
\begin{align}
\gel{\vec{v}}_{n}\cdot \vec{n} - V_n = 0.
\end{align}
By summing \eqref{jump:flux} over $\allm$, the mixture velocities are found
to satisfy
\begin{align}
  \gel{\vec{v}}\cdot \vec{n} = \bath{\vec{v}}\cdot \vec{n}.
\end{align}

Assuming the anion fraction is eliminated from the outer problems, then
only the jump conditions for the solvent and cation in \eqref{jump:flux} 
need to be imposed. However, the jump conditions for the ions can be
combined to obtain 
  \begin{align}
    (z_+ \gel{\jg}_+ + z_- \gel{\jg}_-)\cdot \vec{n}
    &=
    (z_+ \bath{\jb}_+ + z_- \bath{\jb}_-)\cdot \vec{n},
  \end{align}
  which provides a condition on the normal derivatives of the
  electric potential.

  \paragraph{Continuity of chemical potentials}

  By combining \eqref{nd:bc:mu_m} along with \eqref{inner:bath:M_m}
  and \eqref{inner:gel:M_m}, continuity of chemical potential across
  the interface is recovered: $\gel{\mu}_m = \bath{\mu}_m$.
  Continuity of the solvent chemical potential means that
  \eqref{inner:gel:M_s} can be equated with 
  \eqref{inner:bath:M_s} to produce
\begin{align}
  \gel{\Pi}_s + \G \gel{p} - \omega^2 \nabla^2 \gel{\phi}_s = \log \bath{\phi}_s.
  \label{match:mu_s}
\end{align}
Equating the chemical potentials of the ions, i.e.\ \eqref{inner:bath:M_pm}
with \eqref{inner:gel:M_pm}, provides a jump condition
for the ionic volume fractions
\begin{align}
  \gel{\phi}_{\pm} = \bath{\phi}_\pm \exp\left[z_\pm(\bath{\Phi} - \gel{\Phi}) - \G \gel{p}
  - \frac{1}{\gel{J}}(1  - \chi \gel{\phi}_s)\right].
  \label{jump:phi_pm}
\end{align}
Using \eqref{jump:phi_pm} in the electroneutrality condition for the gel
\eqref{outer:gel:ne} produces a jump condition for the electrical potentials
\begin{align}
  \sum_{\alli} z_i \bath{\phi}_i
  \exp\left(z_i(\bath{\Phi} - \gel{\Phi})\right)
  = z_f \phi_f\exp\left[\G \gel{p} +
    \frac{1}{\gel{J}}(1 - \chi \gel{\phi}_s)\right].
\end{align}
These equations are also coupled to the molecular incompressibility
condition in the gel \eqref{nd:gel:J}, the no-void condition in the bath
\eqref{nd:bath:no_void}, and the electroneutrality condition in the bath
\eqref{outer:bath:en}.

\paragraph{Variational condition}

Matching the derivatives of the solvent fraction in the gel using
\eqref{inner:gel:phi_s} recovers the variational condition
\begin{align}
  \nabla \gel{\phi}_s \cdot \vec{n} = 0.
  \label{match:variational}
\end{align}

\paragraph{Continuity of stress}

By matching \eqref{inner:gel:T} with the outer solution, we obtain
stress-free conditions for the gel at the interface:
\begin{align}
  \gel{\tens{T}}\cdot \vec{n} = 0.
  \label{match:stress}
\end{align}

\paragraph{Slip condition}

The final boundary condition for the electroneutral model is a slip
condition on the mixture velocity of the bath. This is obtained by solving
\eqref{inner:bath:slip}, where the value of $U_\alpha$ can be derived from the gel problem
\eqref{inner:gel:vt} by taking the limit $\xi \to 0^{-}$,
and then imposing the matching conditions
$\t{\vec{v}}^{(0)}\cdot \vec{t}_\alpha \to \bath{\vec{v}}\cdot \vec{t}_\alpha$
as $\xi \to \infty$.
The final result is a 
Helmholtz--Smoluchowski slip condition for a deformable porous solid.


\section{Asymptotic analysis for Kuhn lengths of zero}
\label{sec:omega}

The asymptotic analysis of the inner region is slightly different for models
that neglect phase separation and set the Kuhn length to zero, $\omega = 0$.
From
\eqref{inner:gel:mu_0}, the leading-order contribution to the solvent chemical potential  $\mu_s$ in the EDL now becomes
\subeq{
  \label{m2:phi}
\begin{align}
  \t{\mu}_s^{(0)} = \t{\Pi}_s^{(0)} + \G \t{p}^{(0)} = \gel{\mu}_s,
  \label{omega:phi_s}
\end{align}
which can be interpreted as a nonlinear algebraic equation for $\t{\phi}^{(0)}_s = \phi_s^{(0)}(\xi,s_1,s_2,t)$. Importantly, the solvent fraction can now vary across the EDL as a result of the complex interplay between mechanics, electrostatics, and thermodynamics captured by \eqref{omega:phi_s}.
The corresponding ion fractions are given by
\begin{align}
  \label{omega:phi_pm}
  \t{\phi}_{\pm}^{(0)} = \gel{\phi}_{\pm}\exp\left[z_{\pm}(\gel{\Phi}-\t{\Phi}^{(0)})
  + \G(\gel{p}-\t{p}^{(0)}) + \frac{1-\chi \gel{\phi}_s}{\gel{J}}-\frac{1-\chi \t{\phi}^{(0)}_s}{\t{J}^{(0)}}\right].
\end{align}
}
The pressure in the gel across the EDL can be calculated directly from
\eqref{inner:gel:p_0} after neglecting the Korteweg stresses. The jump
conditions
across the EDL are the same as those in Sec.~\ref{sec:matching}, except
the $\nabla^2\gel{\phi}_s$ term in \eqref{match:mu_s}
and the variational condition in \eqref{match:variational} can be dropped.


%% file: cylindrical.tex
\section{Swelling of a constrained cylinder}
\label{sec:cylinder}

We now use our formulation to study the EDL forming in cylindrical
polyelectrolyte gels that are in equilibrium with an
external bath. Following the experimental setup considered by
Horkay \etal\cite{Horkay2001}, we assume the gel can freely swell
in the radial direction but is
confined in the axial direction. 
We consider axisymmetric equilibrium solutions and let
$r$ and $R$ denote the Eulerian and Lagrangian radial coordinates,
respectively.
The deformation gradient tensor can be written as
\begin{align}
\tens{F} = \lambda_r\, \vec{e}_r\otimes \vec{E}_R + \lambda_\theta\, \vec{e}_\theta\otimes
  \vec{E}_\Theta + \lambda_z \vec{e}_z \otimes \vec{E}_Z,
  \label{cyl:F}
\end{align}
where $\lambda_r = (\pdf{R}{r})^{-1}$, $\lambda_\theta = r/R$, and
$\lambda_z$ note the radial, orthoradial, and experimentally controlled
axial stretch, respectively. 
The normal and tangent vectors to the free surface are given by
$\vec{n} = \vec{e}_r$, $\vec{N} = \vec{E}_R$; $\vec{t}_1 = \vec{e}_\theta$,
$\vec{T}_1 = \vec{E}_\Theta$; $\vec{t}_2 = \vec{e}_z$ and
$\vec{T}_2 = \vec{E}_Z$.
We choose the
non-dimensionalisation such that the radius of the cylinder in the reference
configuration is scaled to unity. The radius in the current configuration
is denoted by $a$. We thus have that $R(r=0) = 0$ and $R(r=a) = 1$. 
We restrict our attention to
monovalent salts with $z_{\pm} = \pm 1$. At equilibrium, the solution to the
outer problem for the bath corresponds to a uniform composition
and electric potential.
Using the electroneutrality and no-void conditions for the bath, we obtain $\bath{\phi}_s = 1 - 2\bath{\phi}_{+}$. The cation fraction,
$\bath{\phi}_+$, is treated as a free parameter. The electric potential,
$\bath{\Phi}$, is treated as an arbitrary constant, which we assume
is non-zero for generality.

In Sec.~\ref{sec:c_outer}, the outer problem in the gel is formulated. This
consists of a system of nonlinear algebraic equations for
homogeneously swollen states that are in equilibrium with the bath.
In Sec.~\ref{sec:c_inner}, the corresponding inner problems are formulated
for models in the case $\omega = 0$ and $\omega \gg \beta$. 
The inner solution is validated against a full numerical solution in
Sec.~\ref{sec:validation} and used to explore the structure of the EDL
in Sec.~\ref{sec:structure}.

\subsection{Solution of the outer problem: homogeneous equilibria}
\label{sec:c_outer}
At equilibrium, the chemical potentials in the gel must be spatially
uniform, leading to $\mu_m = \gel{\mu}_m$ for $\allm$. 
We assume that the outer solution corresponds to a homogeneously swollen
cylindrical gel, in which case $\phi_m = \gel{\phi}_m$ for $\allm$.
The deformation gradient tensor is given by
\begin{align}
  \gel{\tens{F}} = (\gel{J}/\lambda_z)^{1/2} \left(\vec{n} \otimes \vec{N}
  + \vec{t}_1 \otimes \vec{T}_1\right) + \lambda_z
  \vec{t}_2 \otimes \vec{T}_2.
  \label{cyl:outer:F}
\end{align}
Consequently, the radial and orthoradial components of the elastic stress
tensor are
${\sf T}_{e,\theta\theta} = {\sf T}_{e,rr} = \lambda_z^{-1} - (\gel{J})^{-1}$.
The stress balance
in the hydrogel reduces to $\pdf{p}{r} = 0$. Imposing the
matching condition \eqref{match:stress} reveals that the
pressure balances the radial elastic stress, $\gel{p} = \gel{{\sf T}}_{e,rr}$.
The volume fraction of solvent and ions, as well as the electric potential,
are determined from the jump conditions 
\subeq{
  \label{cyl:eq}
\begin{align}
  \log \gel{\phi}_s + \frac{1}{\gel{J}} + \frac{\chi (1 - \gel{\phi}_s)}{\gel{J}} + \G\left(\frac{1}{\lambda_z} - \frac{1}{\gel{J}}\right) = \log(1 - 2 \bath{\phi}_+), \\
  \gel{\phi}_{\pm} = \bath{\phi}_+ \exp\left[\pm (\bath{\Phi} - \gel{\Phi})
  -\G\left(\frac{1}{\lambda_z}-\frac{1}{\gel{J}}\right) - \frac{1}{\gel{J}}\left(1 - \chi \gel{\phi}_s\right)\right], \\
  2 \bath{\phi}_+ \sinh(\bath{\Phi} - \gel{\Phi}) = -z_f \phi_f \exp\left[\G\left(\frac{1}{\lambda_z}-\frac{1}{\gel{J}}\right) + \frac{1}{\gel{J}}\left(1 - \chi \gel{\phi}_s\right)\right], \label{cyl:eq_en}
\end{align}
}
where $\phi_f = \alphaf / \gel{J}$ and $\gel{J}$ is given by \eqref{nd:gel:J}.
When the cation fraction in the bath is small, $\bath{\phi}_{+} \ll 1$, 
the nonlinear system \eqref{cyl:eq} can be reduced to a single equation,
as described in Appendix~\ref{app:dilute}.

\begin{figure}
  \centering
  \subfigure[]{\includegraphics[width=0.48\textwidth]{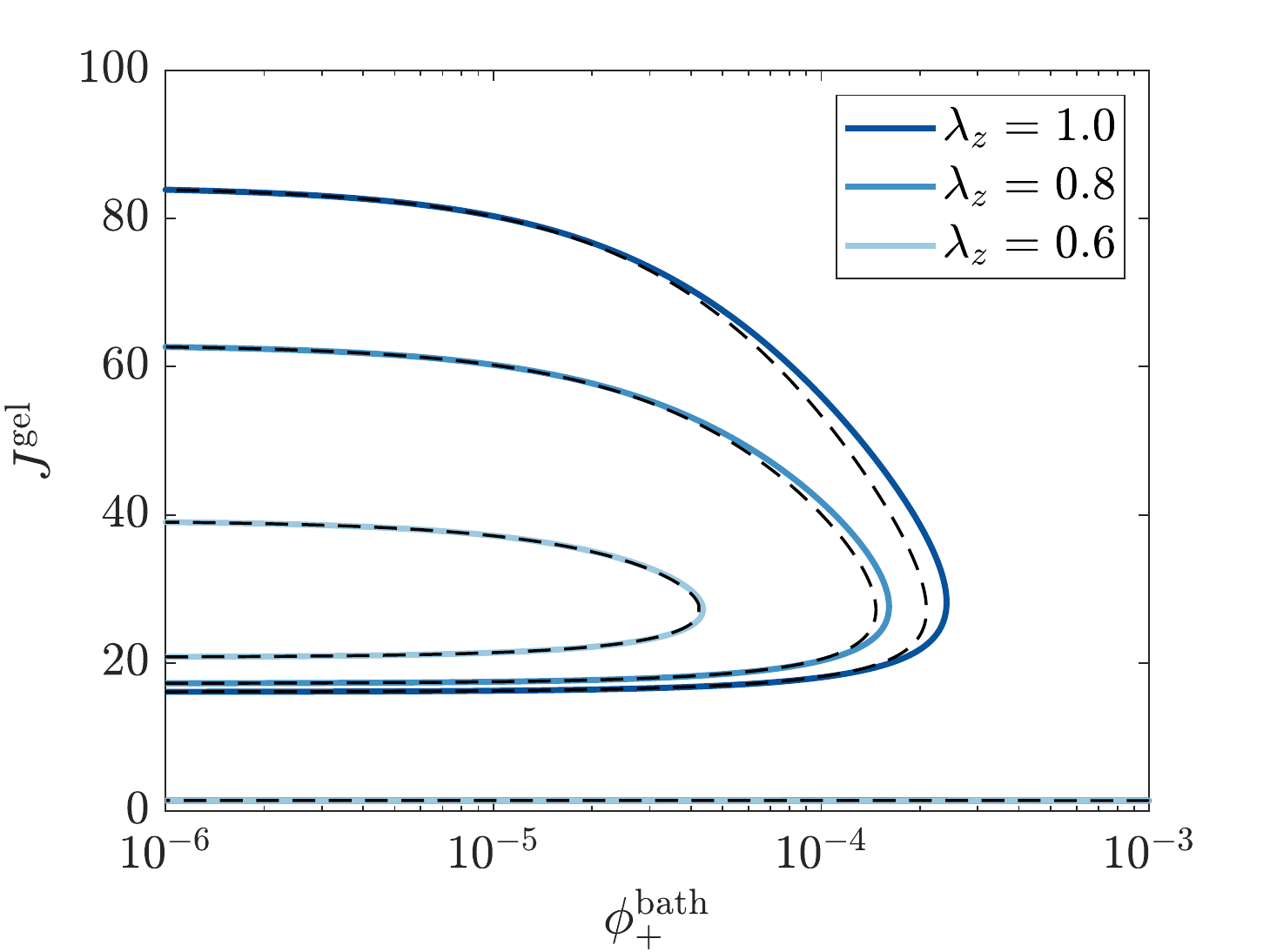}}
  \subfigure[]{\includegraphics[width=0.48\textwidth]{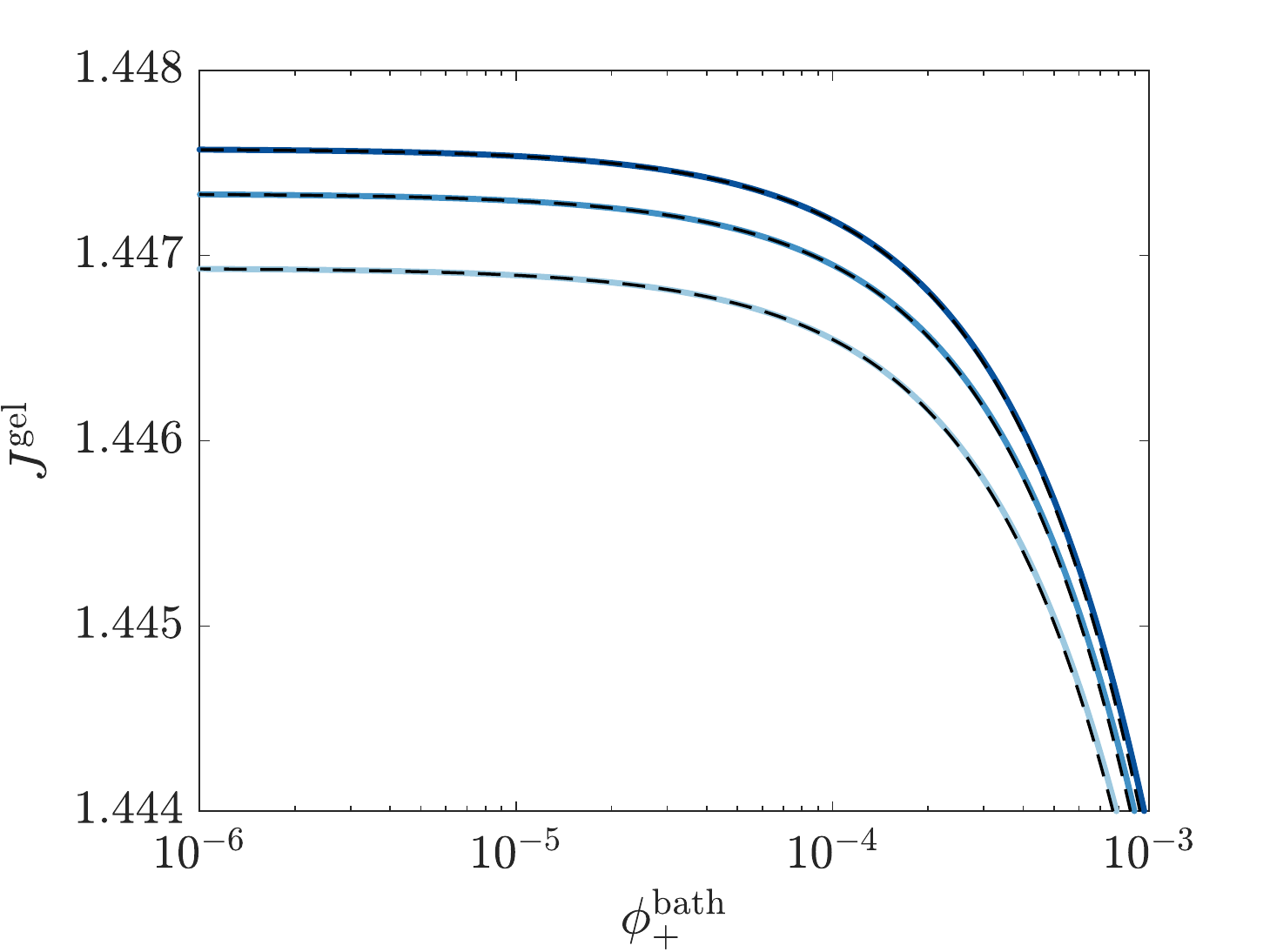}}
  \caption{(a) Equilibrium swelling ratio $\gel{J}$ as a function of cation
    fraction in the bath $\bath{\phi}_+$ showing swollen and collapsed branches.
    (b) The swelling ratio along the collapsed branch.
    Solid lines correspond to solutions of
    \eqref{cyl:eq}. Dashed lines represent solutions to the reduced
    equation \eqref{cyl:eq_red} for a dilute concentration of cations.
    The parameter values are $\G = 0.0005$, $\chi = 1.2$, $\alphaf = 0.05$
    $z_\pm = \pm 1$, $z_f = 1$.}
  \label{fig:cylindrical_eq}
\end{figure}

We numerically solve the nonlinear system of algebraic equations defining
the outer problem (i.e.\ the homogeneous equilibria)
given by \eqref{cyl:eq} using pseudo-arclength
continuation. The results are shown as solid curves
in Fig.~\ref{fig:cylindrical_eq}
for three different values of $\lambda_z \leq 1$, corresponding to gels in axial
compression.  The dashed black line represents numerical solutions to the
reduced model derived in Appendix~\ref{app:dilute}.  The figure shows 
there are two distinct solution branches, one of which
describes highly swollen gels ($\gel{J} > 10$), whereas the other corresponds
to weakly swollen gels ($\gel{J} \sim 1.4$).  
We refer to the former and latter as the swollen and collapsed branches,
respectively. The swollen branch folds back on itself at a critical salt
concentration, indicating that a volume phase transition can occur in
this system as the salt concentration increases in the bath,
which leads to a discontinuous decrease in the gel volume. 
Increasing the axial compression reduces the degree of swelling for a
given salt fraction as well as the critical salt fraction
at which the volume phase transition occurs, in agreement with
experimental observations~\cite{Horkay2001}. 
Due to the incompressibility of the gel, imposing an axial compression results in a radial stretch. The elastic energy cost of inserting a molecule
into a pre-stretched gel is greater than for a dry (or unstretched) gel. Hence,
the balance between the mixing and elastic energies is established
at smaller concentrations, resulting in the equilibrium swelling ratio $\gel{J}$ decreasing with the axial stretch $\lambda_z$.

\subsection{Formulation of the inner problem}
\label{sec:c_inner}
The self-contained inner problem for the gel is formulated by 
 accounting for
non-homogeneous deformations in the EDL due
to composition gradients. The
deformation gradient \eqref{cyl:F}
is expanded
as $\t{\tens{F}}^{(0)} = \t{\lambda}_r^{(0)} \vec{n} \otimes \vec{N} +
\t{\lambda}_\theta^{(0)} \vec{t}_1 \otimes \vec{T}_1 + 
\lambda_z \vec{t}_2 \otimes \vec{T}_2$. By matching to
\eqref{cyl:outer:F} as $\xi \to -\infty$ using the conditions
in \eqref{inner:gel:S_0}, we find that
$\t{\lambda}^{(0)}_\theta = (\gel{J}/\lambda_z)^{1/2}$. By calculating
$\t{J}^{(0)} = \det \t{\tens{F}}^{(0)}$ it is possible to eliminate
$\t{\lambda}_r^{(0)}$ and hence write the deformation gradient tensor as
\begin{align}
  \pl{\t{\tens{F}}}^{(0)} = \t{J}^{(0)}\left(\frac{1}{\lambda_z\gel{J}}\right)^{1/2}
  \vec{n}\otimes \vec{N} + \left(\frac{\gel{J}}{\lambda_z}\right)^{1/2} \vec{t}_1
  \otimes \vec{T}_1 + \lambda_z \vec{t}_2 \otimes \vec{T}_2.
\end{align}
The radial elastic stress can be calculated from \eqref{inner:gel:Te} as
\begin{align}
  \t{{\sf T}}_{e,rr}^{(0)} = \vec{n}\cdot \t{\tens{T}}^{(0)}_{e}\cdot\vec{n}
  = \frac{1}{\lambda_z} \frac{\t{J}^{(0)}}{\gel{J}} - \frac{1}{\t{J}^{(0)}},
\end{align}
which allows the pressure to be determined from \eqref{inner:gel:p_0}.

The inner problem for the gel can now be constructed using the results
from the previous sections. In particular, if $\omega = 0$, then the
governing equations for the gel can be condensed into 
\subeq{
  \label{cyl:inner_0}
\begin{align}
  \log \t{\phi}^{(0)}_s + \frac{1}{\t{J}^{(0)}} + \frac{\chi (1 - \t{\phi}_s^{(0)})}{\t{J}^{(0)}} + \G\t{p}^{(0)} = \log(1 - 2 \bath{\phi}_+),
  \label{cyl:inner0:phi_s}
  \\
  \t{\phi}^{(0)}_{\pm} = \bath{\phi}_+ \exp\left[\pm (\bath{\Phi} - \t{\Phi}^{(0)})
  -\G\t{p}^{(0)} - \frac{1}{\t{J}^{(0)}}\left(1 - \chi \t{\phi}^{(0)}_s\right)\right], \\
  -\pdd{\t{\Phi}^{(0)}}{\xi} = \t{\phi}_{+}^{(0)} - \t{\phi}_-^{(0)} + z_f \t{\phi}_f^{(0)}, \\
  \tilde{p}^{(0)} = \frac{1}{\lambda_z} \frac{\t{J}^{(0)}}{\gel{J}} - \frac{1}{\t{J}^{(0)}}  + \frac{1}{2\G}\left(\pd{\Phi^{(0)}}{\xi}\right)^2, \\
  \t{J}^{(0)} = (1 - \t{\phi}^{(0)}_s - \t{\phi}_{+}^{(0)} - \t{\phi}_{-}^{(0)})^{-1},
\end{align}
}
where $\t{\phi}_f^{(0)} = \alphaf / \t{J}^{(0)}$. In the case $\omega \gg \beta$,
Eqn~\eqref{cyl:inner0:phi_s} is replaced with $\t{\phi}_s^{(0)} = \gel{\phi}_s$,
resulting in the system 
\subeq{\label{cyl:inner_omega}
\begin{align}
  \t{\phi}^{(0)}_{\pm} = \bath{\phi}_+ \exp\left[\pm (\bath{\Phi} - \t{\Phi}^{(0)})
  -\G\t{p}^{(0)} - \frac{1}{\t{J}^{(0)}}\left(1 - \chi \gel{\phi}_s\right)\right], \\
  -\pdd{\t{\Phi}^{(0)}}{\xi} = \t{\phi}_{+}^{(0)} - \t{\phi}_-^{(0)} + z_f \t{\phi}_f^{(0)}, \\
  \G (1 - \gel{\phi}_s) \tilde{p}^{(0)} = \G\left(\frac{1}{\lambda_z} \frac{\t{J}^{(0)}}{\gel{J}} - \frac{1}{\t{J}^{(0)}}\right) + \gel{\phi}_s \left(\tilde{\Pi}_s^{(0)} - M_s\right) + \frac{1}{2}\left(\pd{\t{\Phi}^{(0)}}{\xi}\right)^2, \\
  \t{J}^{(0)} = (1 - \gel{\phi}_s - \t{\phi}_{+}^{(0)} - \t{\phi}_{-}^{(0)})^{-1}, \\
  \t{\Pi}_s^{(0)} = \log \gel{\phi}_s + \frac{\chi(1 - \gel{\phi}_s)}{\t{J}^{(0)}} - \frac{1}{\t{J}^{(0)}},
\end{align}}
where $M_s = \bath{\mu}_s = \log(1 - 2 \bath{\phi}_+)$. In both cases, the
boundary conditions for the electrical potential are given by \eqref{inner:gel:Phi_bcs}. Moreover, the expression for the hoop stress in the gel,
$\t{{\sf T}}_{\theta\theta}^{(0)} = \vec{t}_1 \cdot
\t{\tens{T}}^{(0)}\cdot \vec{t}_1$, is the same in both cases as well:
\begin{align}
  \t{{\sf T}}^{(0)}_{\theta\theta} = \frac{1}{\lambda_z}\left(\frac{\gel{J}}{\t{J}^{(0)}}
  - \frac{\t{J}^{(0)}}{\gel{J}}\right) - \G^{-1}\left(\pd{\t{\Phi}^{(0)}}{\xi}\right)^2.
\end{align}
The first term represents the elastic contribution to the total hoop stress,
which can be compressive or tensile. The second term captures the contribution
from the Maxwell stresses, which is always compressive.

\subsection{Validation of the asymptotic solution to the inner problem}
\label{sec:validation}

The systems \eqref{cyl:inner_0} and \eqref{cyl:inner_omega} are discretised
using finite differences and solved using Newton's method. 
Once the inner problem in the gel is
solved, the 
electric potential in the bath can be obtained by integrating
\eqref{inner:bath:Phi_first_int} and imposing continuity at the interface.
To validate the asymptotic approach, we also solve the full steady problem in
axisymmetric cylindrical coordinates, details of which are provided in
Appendix \ref{app:full_cylinder}.

We consider the case where the axial stretch and
salt content in the bath are set to
$\lambda_z = 1$ and $\bath{\phi}_{+} = 10^{-5}$, with
the remaining parameters being the same as those in
Fig.~\ref{fig:cylindrical_eq}. There
are three possible solutions to the outer problem. We are only concerned with
two of these, which correspond to the collapsed state ($\gel{J} \simeq 1.447$)
and the highly swollen state ($\gel{J} \simeq 82$). The other solution,
which has a swelling ratio $\gel{J} \simeq 60$, is expected to be
unstable~\cite{Celora_SIAP_2021}.
In this subsection, we focus on the inner solution when the outer solution
corresponds to the collapsed state. In Sec.~\ref{sec:structure},
we explore how the
solution to the inner problem is affected by the choice of outer solution.

\begin{figure}
  \centering
  \subfigure[]{\includegraphics[width=0.32\textwidth]{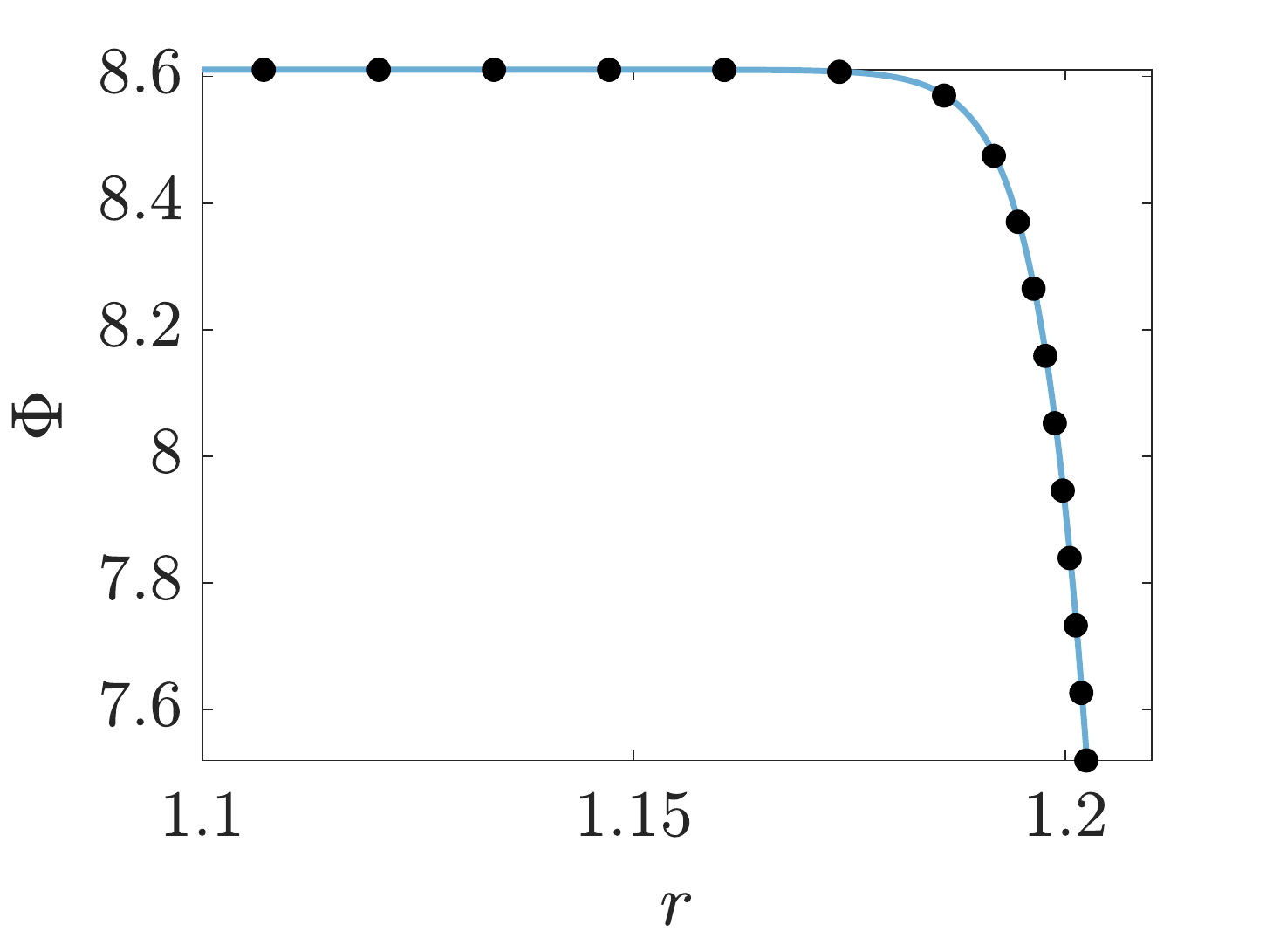}}
  \subfigure[]{\includegraphics[width=0.32\textwidth]{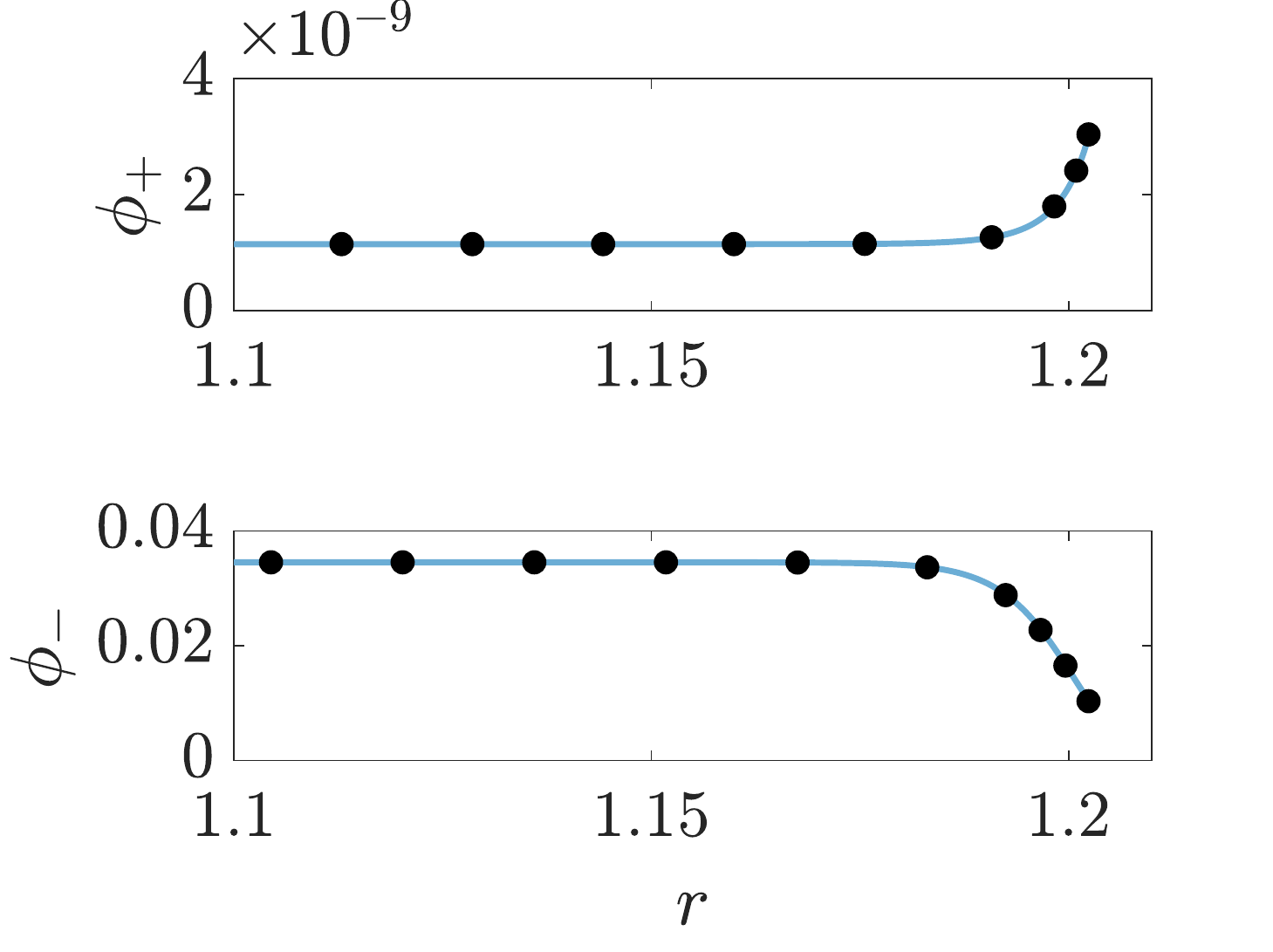}}
  \subfigure[]{\includegraphics[width=0.32\textwidth]{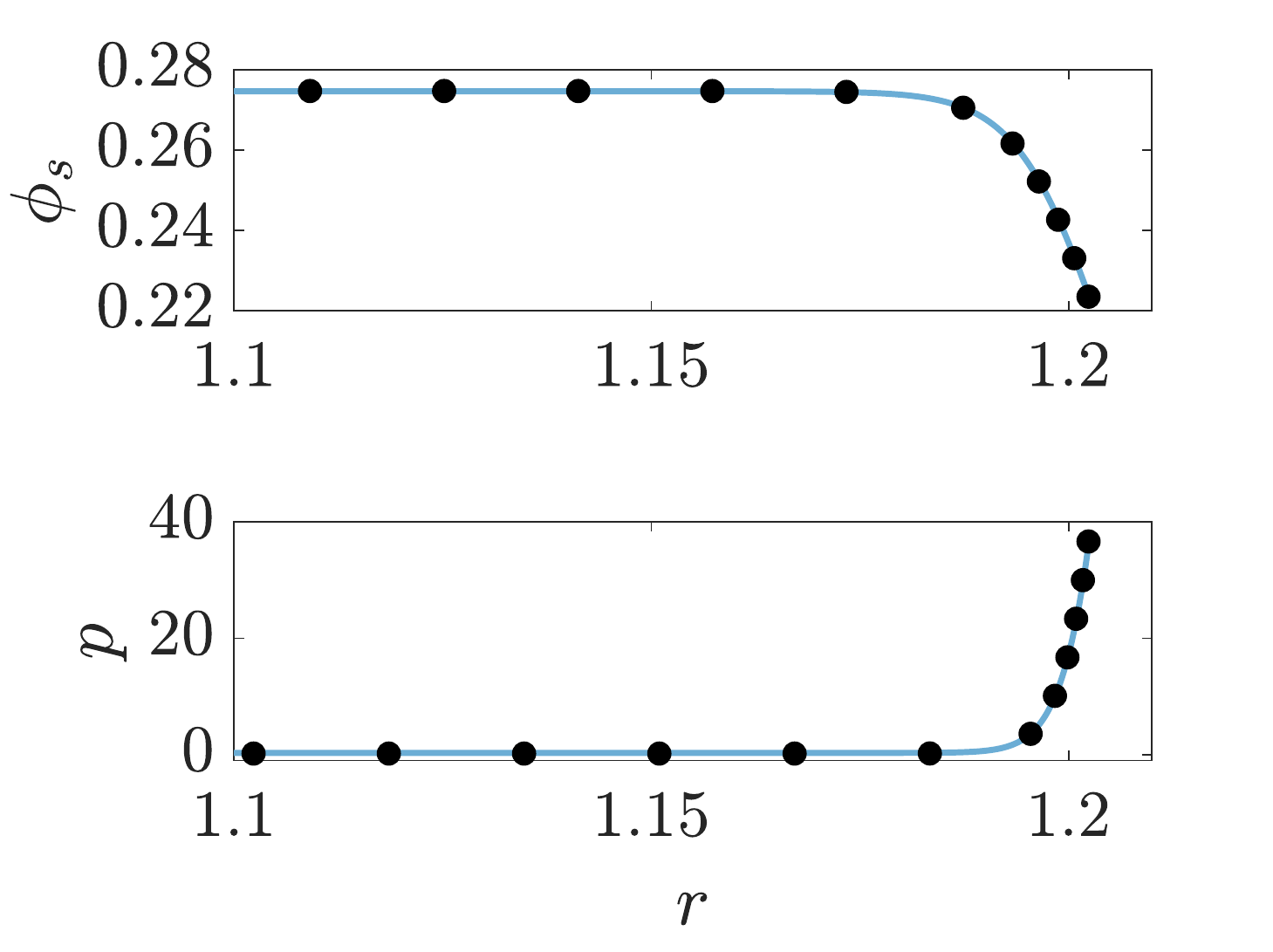}}
  \subfigure[]{\includegraphics[width=0.32\textwidth]{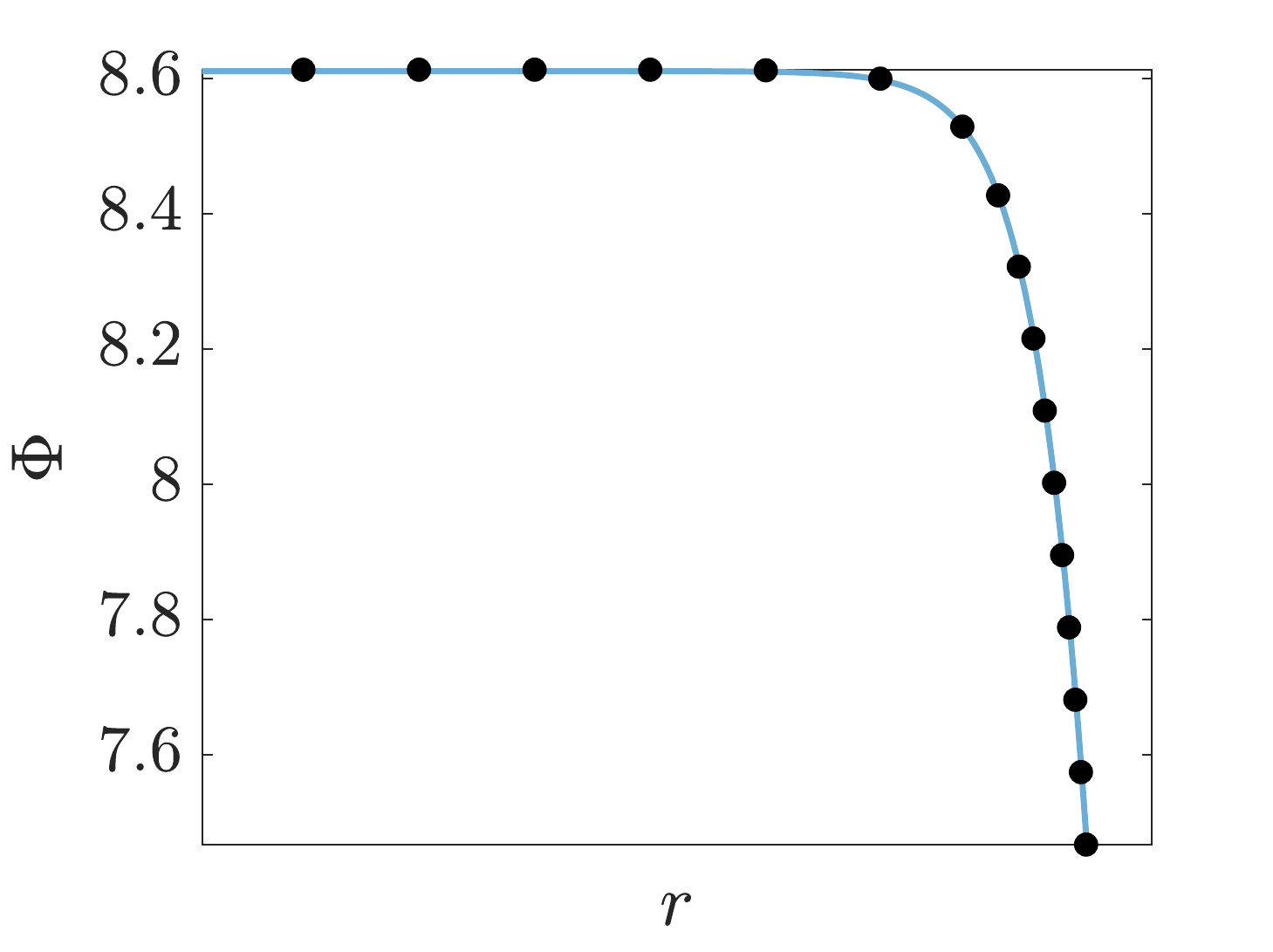}}
  \subfigure[]{\includegraphics[width=0.32\textwidth]{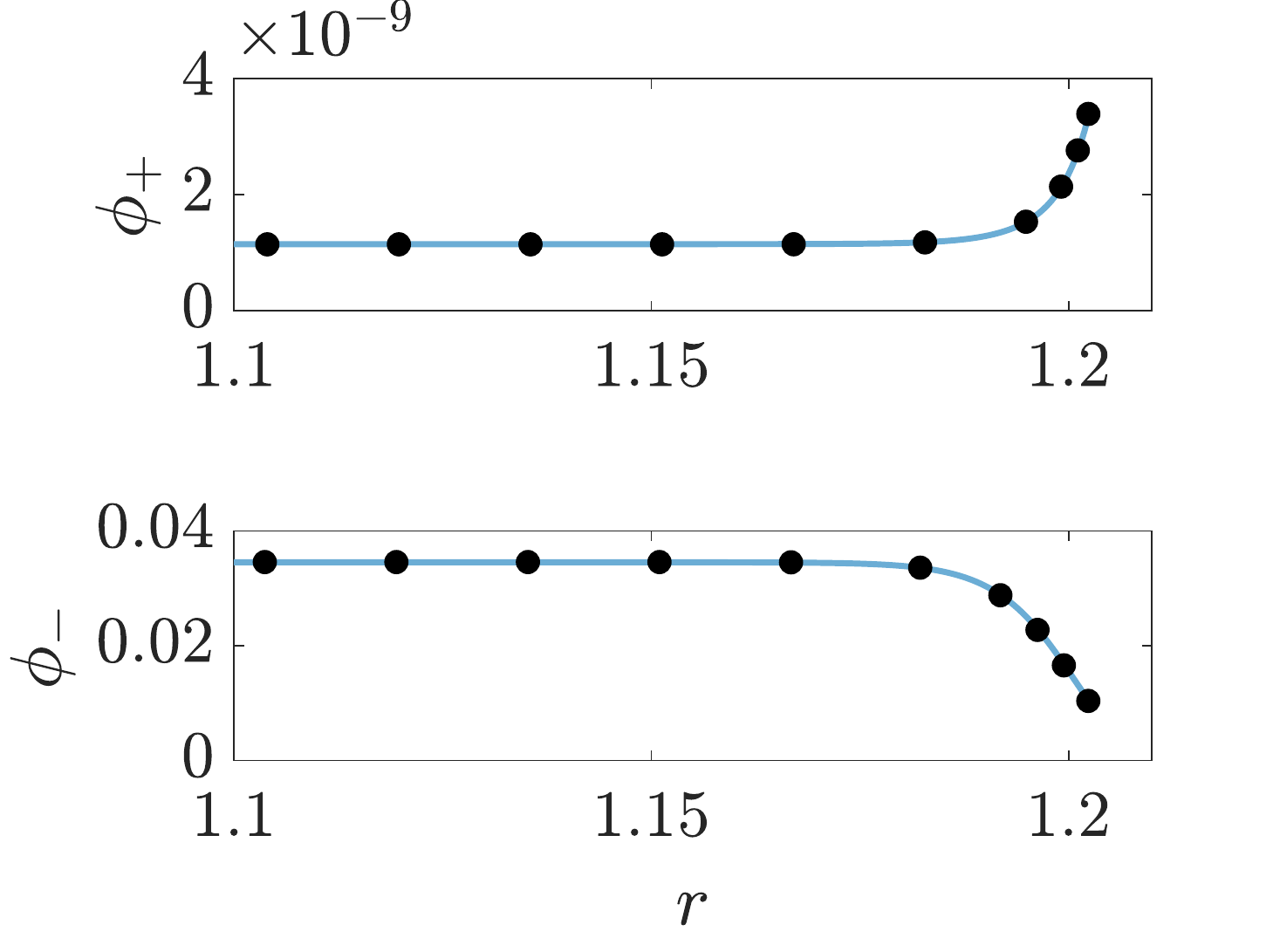}}
  \subfigure[]{\includegraphics[width=0.32\textwidth]{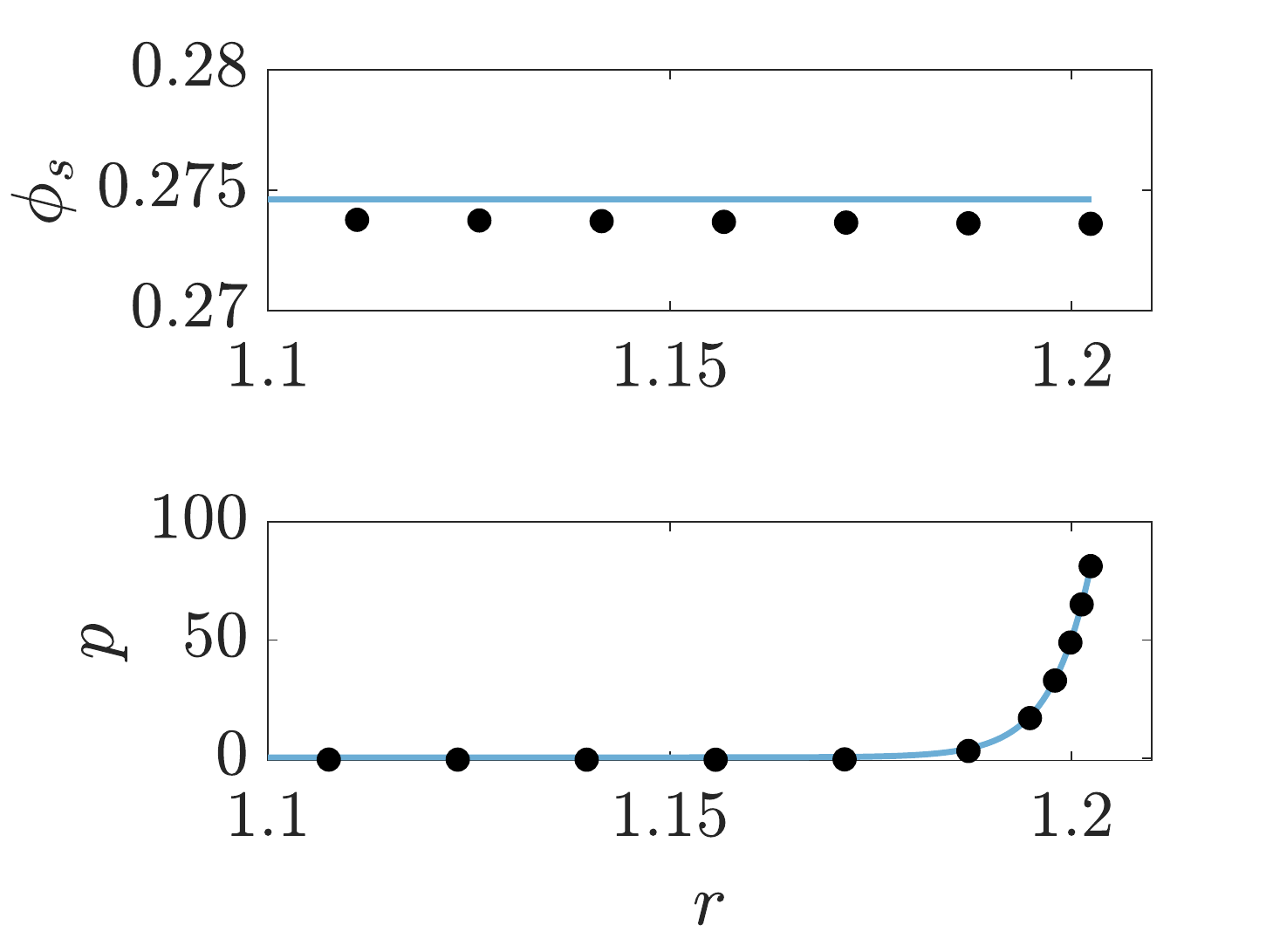}}
  \caption{Numerical solutions of the inner problem (lines) and the full steady
    problem (circles) showing the structure of the EDL.
    Only the solution to the gel problem is shown.
    The parameter values are
    $\chi = 1.2$, $\G = 5\cdot 10^{-4}$, $\alphaf = 0.05$, $\bath{\phi}_+ = 10^{-5}$, $\lambda_z = 1$, $\epsilon_r = 1$, $z_{\pm} = \pm 1$, $z_f = 1$, and $\beta = 10^{-3}$.
    Panels (a)--(c) correspond to the case when $\omega = 0$. Panels (d)--(f)
    correspond to the case when $\omega = 0.5 \gg \beta$.}
  \label{fig:full_valid}
\end{figure}

The inner solution that is computed by matching to the collapsed state is
compared with the solution of the full steady problem in
Fig.~\ref{fig:full_valid}. The non-dimensional Debye thickness has been
set to $\beta = 10^{-3}$. Although this is higher than the estimate given in
Sec.~\ref{sec:parameters}, it facilitates the numerical solution of the full
model. The solutions are plotted in terms of the radial coordinate $r$,
which acts as the outer variable for this geometry. The
outer and inner variables are related by $r = a + \beta \xi$.
Due to the formulation of the model in terms of Eulerian coordinates,
the gel radius $a$ is a free boundary. In the full steady problem, $a$
is calculated as part of the numerical solution;
in the asymptotic framework, it is determined from the outer solution
as $a = (\gel{J})^{1/2}$.

In Fig.~\ref{fig:full_valid} (a)--(c), we compare the solutions of the
full steady problem (circles) and the inner problem (lines) when
$\omega = 0$. The
electric potential shown in Fig.~\ref{fig:full_valid} (a) indicates that
the choice of non-dimensionalisation underestimates
the width of the EDL,
which is roughly $0.025$ or $25 \beta$. This underestimation is due to the
non-dimensionalisation not accounting for the
the small volume fractions of ions in the EDL;
see Fig.~\ref{fig:full_valid} (b).
Despite this, the solutions to the inner problem and the
full problem are in excellent agreement.

The comparison between the inner and full solutions in the case of
$\omega \gg \beta$ is shown in Fig.~\ref{fig:full_valid} (d)--(f).
To ensure a sufficient separation between the Debye length and
the width of diffuse interfaces, we have taken $\omega = 0.5 = 500 \beta$.
Overall, there is good agreement between the solutions, with the main
discrepancy occurring in the solvent fraction;
see Fig.~\ref{fig:full_valid} (f).

\subsection{Investigating the structure of the electric double layer}
\label{sec:structure}

The inner solution is now used to explore the
structure of the EDL and how this
depends on the outer solution,
i.e., the degree of swelling that occurs in the bulk of the gel.
We begin by fixing the parameter
values to be those in Fig.~\ref{fig:cylindrical_eq} with
$\lambda_z = 1$ and $\bath{\phi}_{+} = 10^{-5}$. We then solve the inner
problem by matching to the two outer solutions that represent the collapsed
and highly swollen states described in Sec.~\ref{sec:validation}.

\begin{figure}
  \centering
  \subfigure[]{\includegraphics[width=0.48\textwidth]{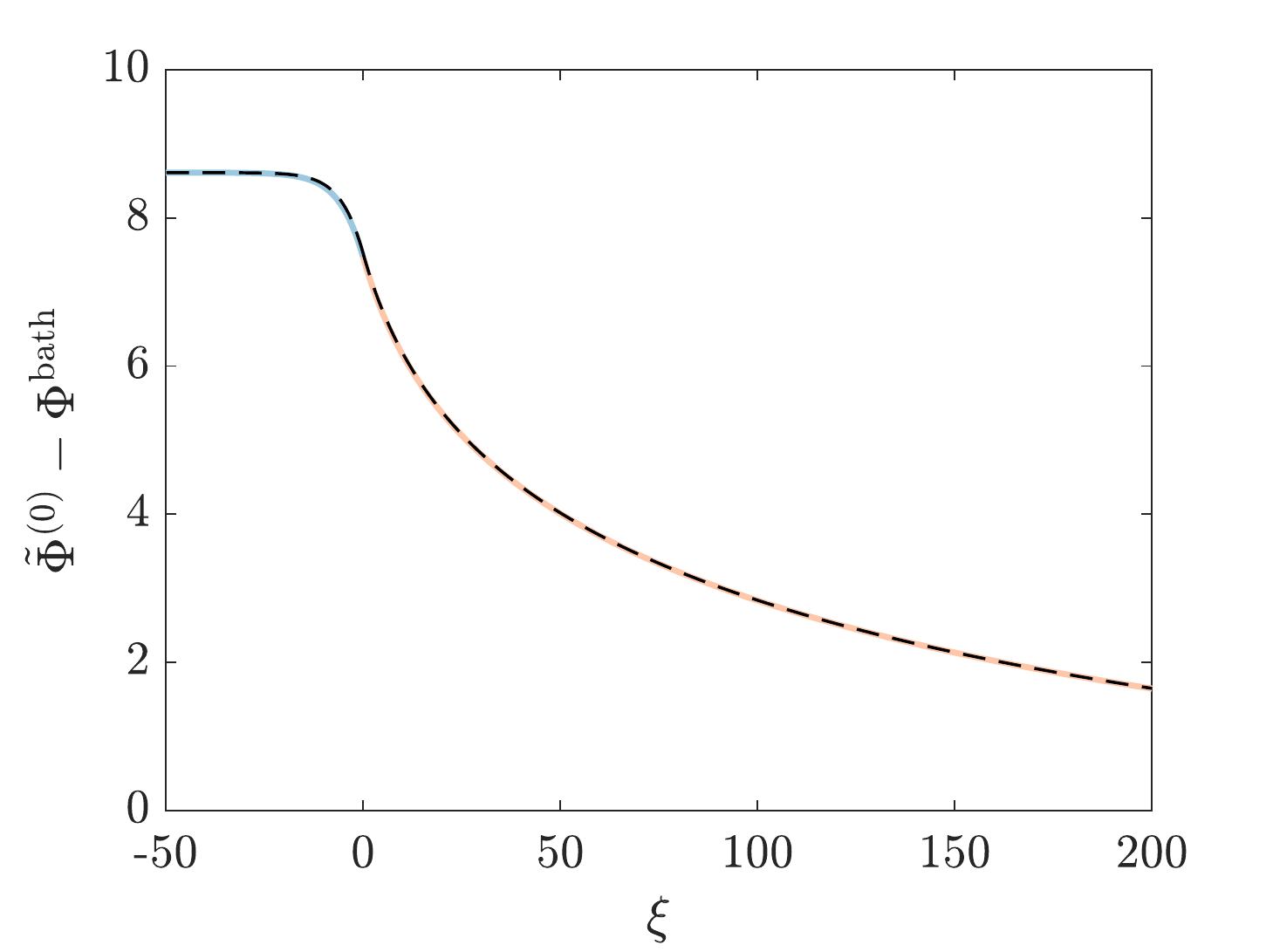}}
  \subfigure[]{\includegraphics[width=0.48\textwidth]{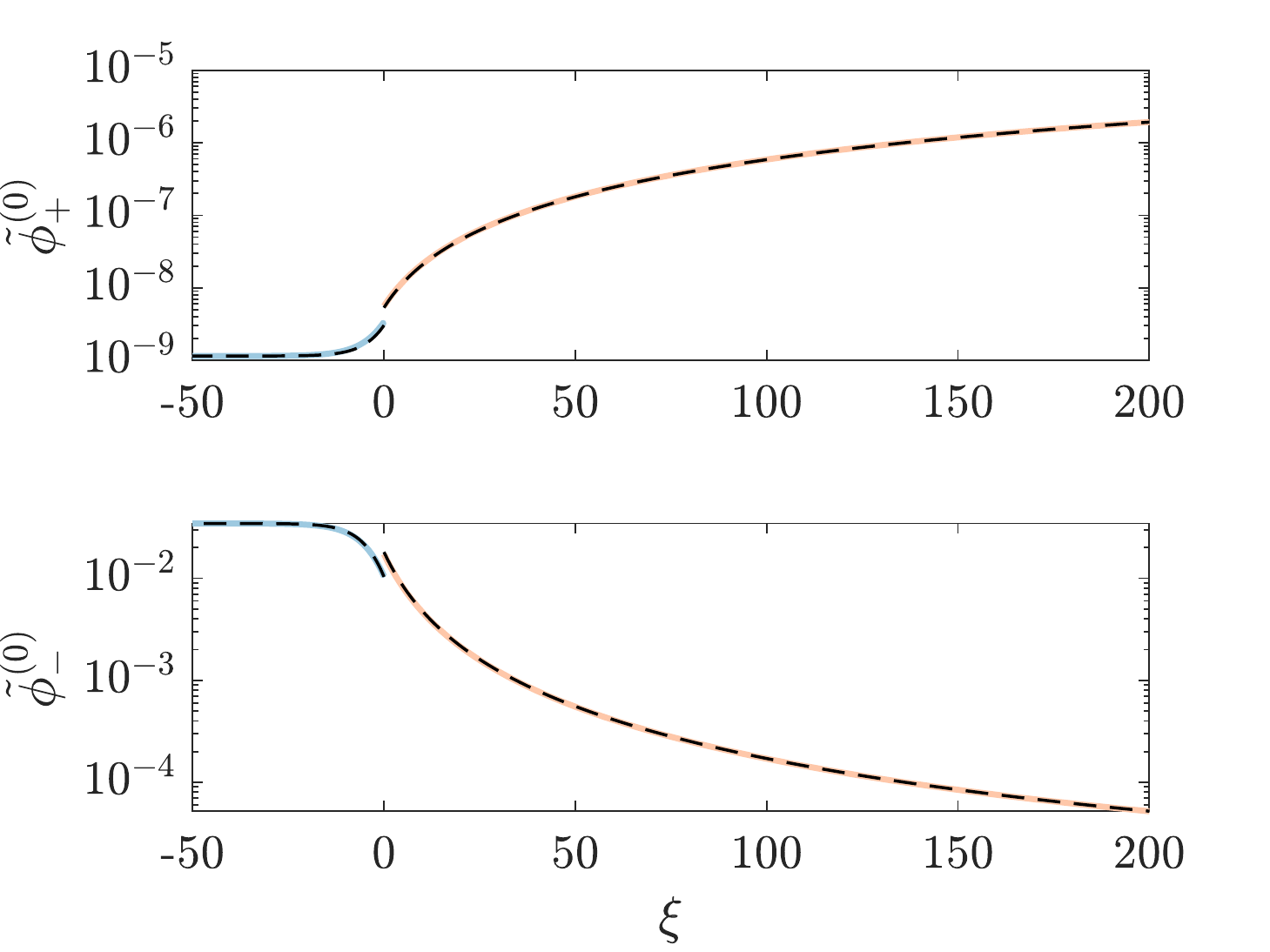}}
  \\
  \subfigure[]{\includegraphics[width=0.48\textwidth]{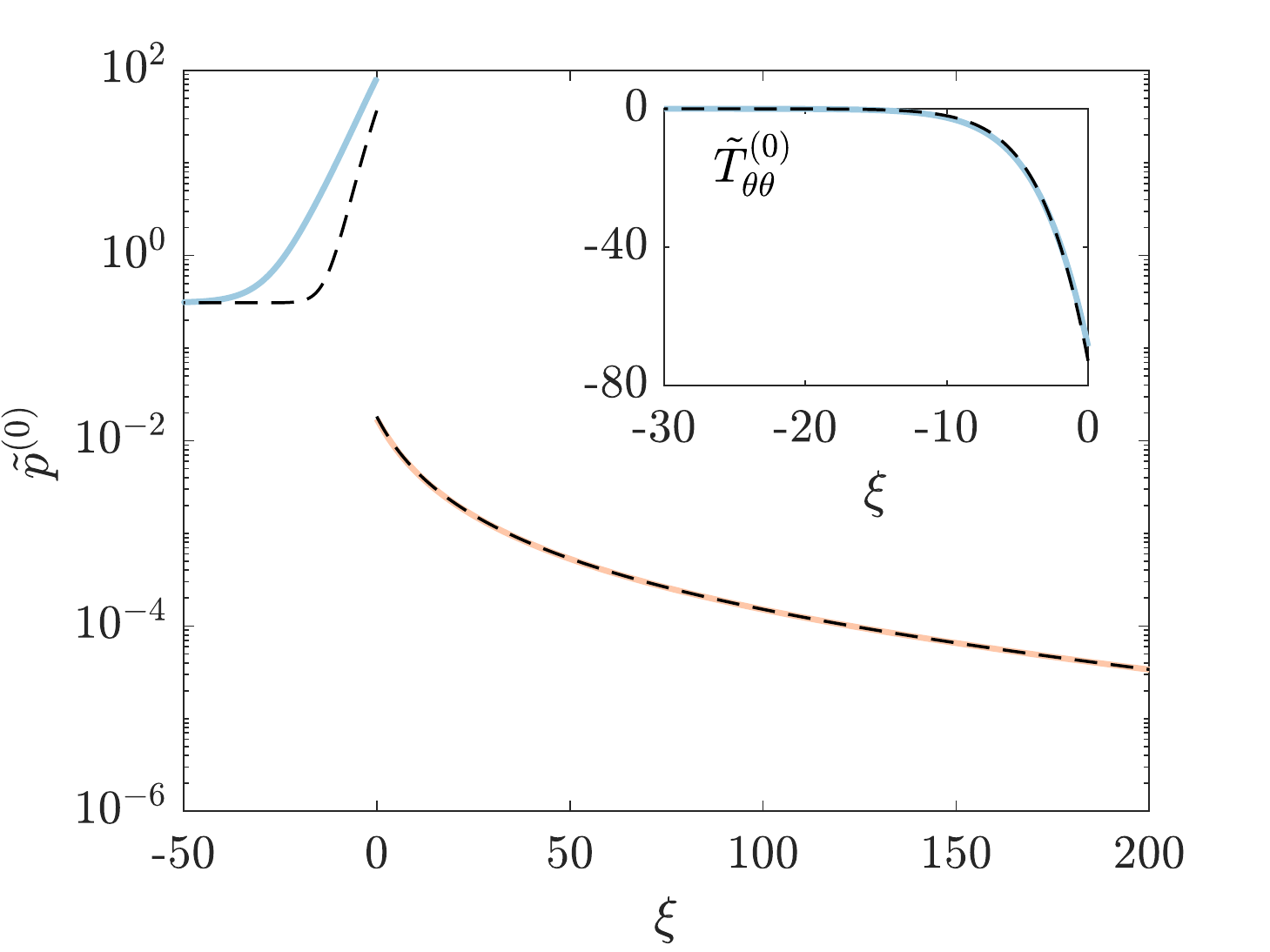}\label{fig:cyl_c}}
  \subfigure[]{\includegraphics[width=0.48\textwidth]{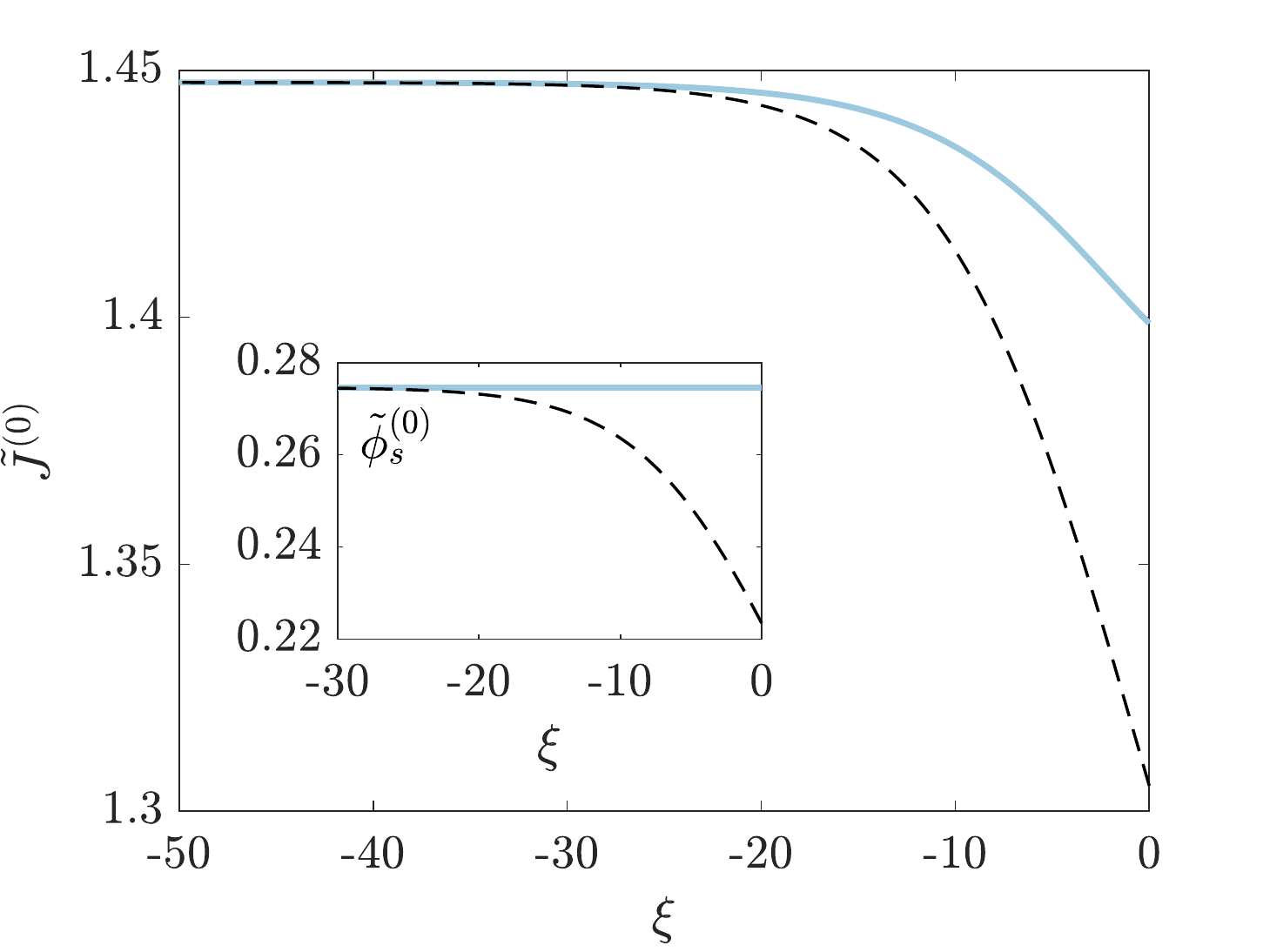}\label{fig:cyl_d}}
  \caption{Numerical solution of the inner problem with far-field conditions
    corresponding to the collapsed state. 
    Solid and dashed lines represent solutions
    to models with $\omega \gg \beta$ and $\omega = 0$, respectively. 
    Parameters: $\chi = 1.2$, $\G = 0.0005$, $\alphaf = 0.05$, $\bath{\phi}_+ = 10^{-5}$, $\lambda_z = 1$, $\epsilon_r = 1$, $z_{\pm} = \pm 1$, and $z_f = 1$.}
  \label{fig:cyl_collapsed}
\end{figure}

In Fig.~\ref{fig:cyl_collapsed}, we plot the inner solutions when the
outer solution corresponds to the collapsed state. 
The solid and dashed lines correspond to the cases
$\omega \gg \beta$ and $\omega = 0$, respectively.
For this parameter set, the value of $\omega$ does not lead
to noticeable changes in the electric potential and ion fractions;
see Fig.~\ref{fig:cyl_collapsed} (a)--(b). However, substantial
differences arise in the gel pressure and the solvent fraction; see
Fig.~\ref{fig:cyl_collapsed} (c)--(d). In the case when $\omega = 0$,
the gel pressure balances a large Maxwell stress.
This large pressure causes a local
decrease in the solvent fraction and a minor collapse of the gel
(Fig.~\ref{fig:full_valid} (c)), which can be rationalised in
terms of \eqref{omega:phi_s}. At equilibrium, the osmotic pressure
$\t{\Pi}_s$ must balance the mechanical
pressure $\t{p}$. To compensate for the increase in mechanical pressure
that arises from the Maxwell stresses, the osmotic pressure must decrease,
which drives solvent out of the gel and causes it to shrink.
When $\omega \gg \beta$, gradients in the
solvent fraction are energetically penalised; thus, the solvent fraction
remains uniform across the EDL. From a mechanical perspective, this
penalisation occurs through the development of a large Korteweg stress,
which counters the opposing effects of the Maxwell stress in order to maintain
a uniform solvent fraction. The mechanical contribution from the Korteweg
stress manifests as an increase in the gel pressure compared to the $\omega = 0$
case, as seen in Fig.~\ref{fig:cyl_collapsed} (c). Although the solvent fraction
is constant across the EDL when $\omega \gg \beta$, the swelling ratio
still decreases relative to the bulk value (Fig.~\ref{fig:cyl_collapsed} (c))
due to the variation in ionic content (Fig.~\ref{fig:cyl_collapsed} (b)).

The inset of Fig.~\ref{fig:cyl_collapsed} (c) shows the total hoop stress
in the gel, which is the same in both models owing to the
strong similarities in the
electric potential. Due to the large Maxwell stresses, the gel experiences a
substantial compressive hoop stress, which leads to the
intriguing possibility of localised mechanical instabilities in the EDL.

\begin{figure}
  \centering
  \subfigure[]{\includegraphics[width=0.48\textwidth]{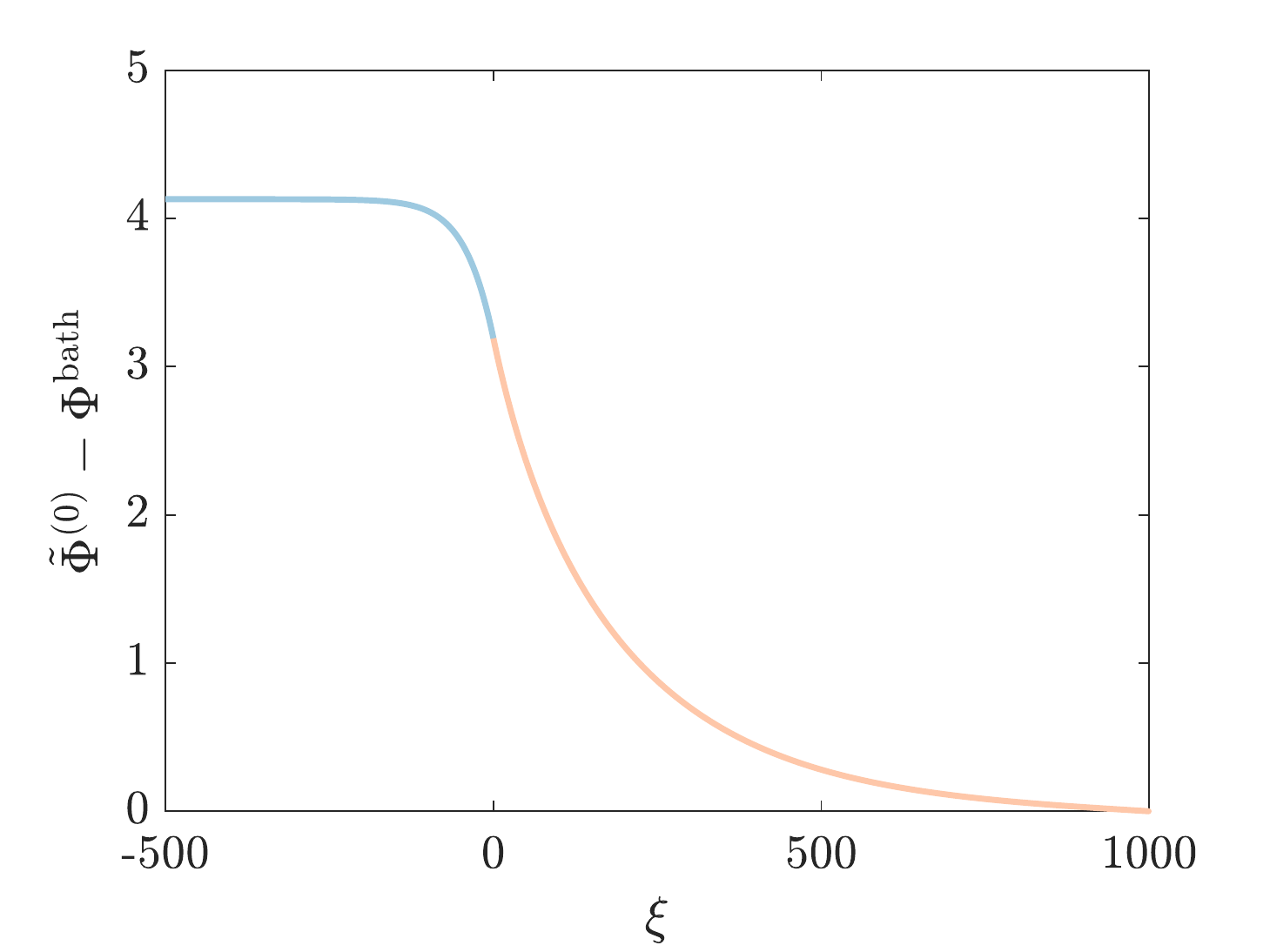}}
  \subfigure[]{\includegraphics[width=0.48\textwidth]{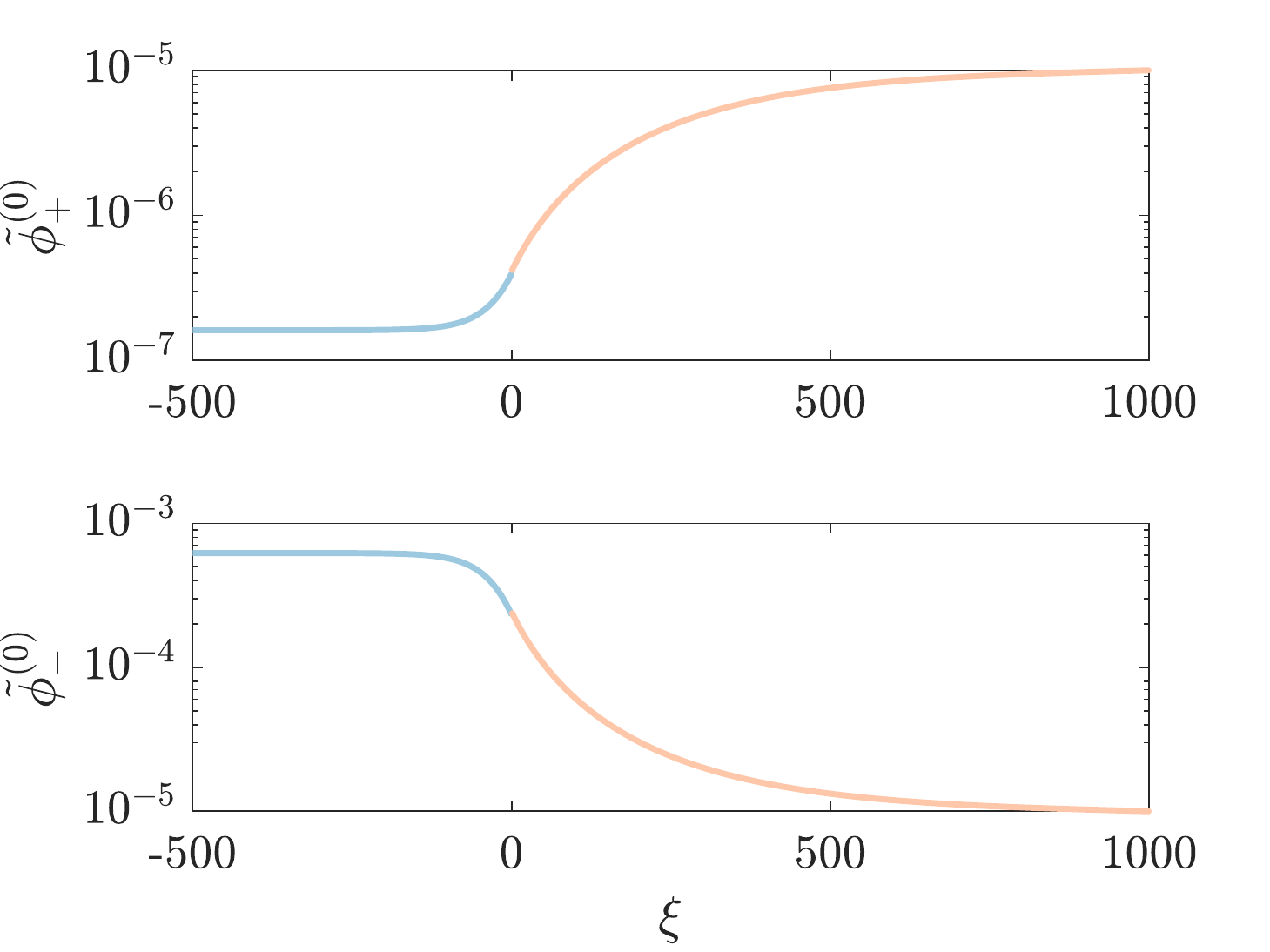}}
  \\
  \subfigure[]{\includegraphics[width=0.48\textwidth]{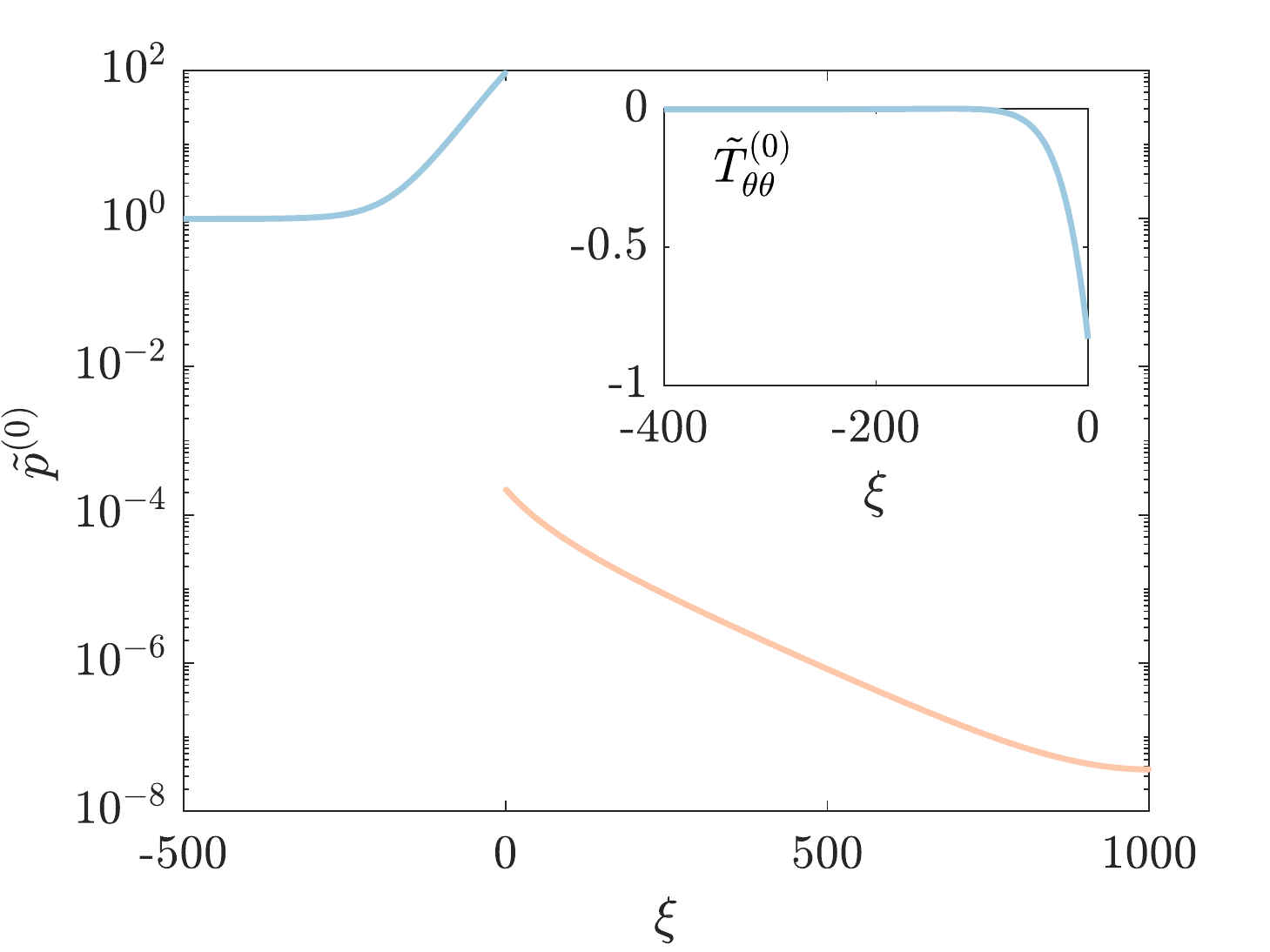}}
  \subfigure[]{\includegraphics[width=0.48\textwidth]{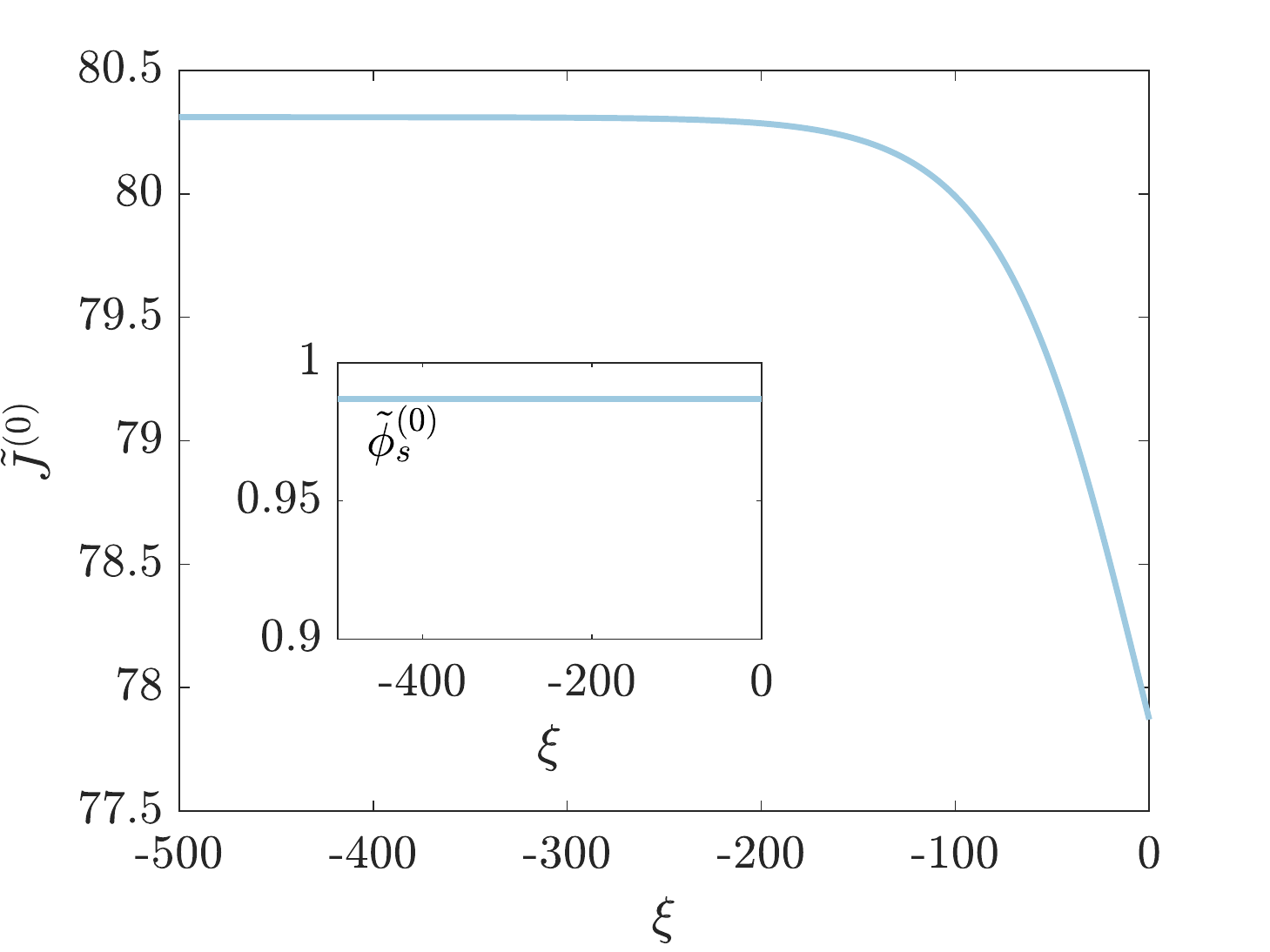}}
  \caption{Numerical solution of the inner problem with far-field conditions
    corresponding to the swollen state when $\omega \gg \beta$.
    Parameter values are the same as in Fig.~\ref{fig:cyl_collapsed}: $\chi = 1.2$, $\G = 0.0005$, $\alphaf = 0.05$, $\bath{\phi}_+ = 10^{-5}$, $\lambda_z = 1$, $\epsilon_r = 1$, $z_{\pm} = \pm 1$, and $z_f = 1$.}
  \label{fig:cyl_swollen}
\end{figure}

In Fig.~\ref{fig:cyl_swollen}, we show the numerical solution of the inner
problem with $\omega \gg \beta$ when the outer solution corresponds
to the highly swollen state.  The qualitative features of the solution
are similar to those shown in Fig.~\ref{fig:cyl_collapsed}, where the outer
solution corresponds to the collapsed state. However, an important
difference is that the concentration of anions has decreased by more
than a factor of ten due to the reduction in the volume fraction of fixed
charges when the gel is highly swollen. 
Consequently, the EDL in the gel has increased in thickness
by a roughly factor of ten to approximately $250 \beta$ (or 25~nm). The gradient
in the electric potential in the gel is therefore ten times weaker,
resulting in a 100-fold reduction in the Maxwell stresses and
the total hoop stress. Despite these decreases, the pressure in the gel
remains large because of the Korteweg stresses. Due to convergence issues,
it was not possible to compute
the corresponding inner solution when $\omega = 0$.

\begin{figure}
  \centering
  \subfigure[$\bath{\phi}_+ = 10^{-5}$]{\includegraphics[width=0.32\textwidth]{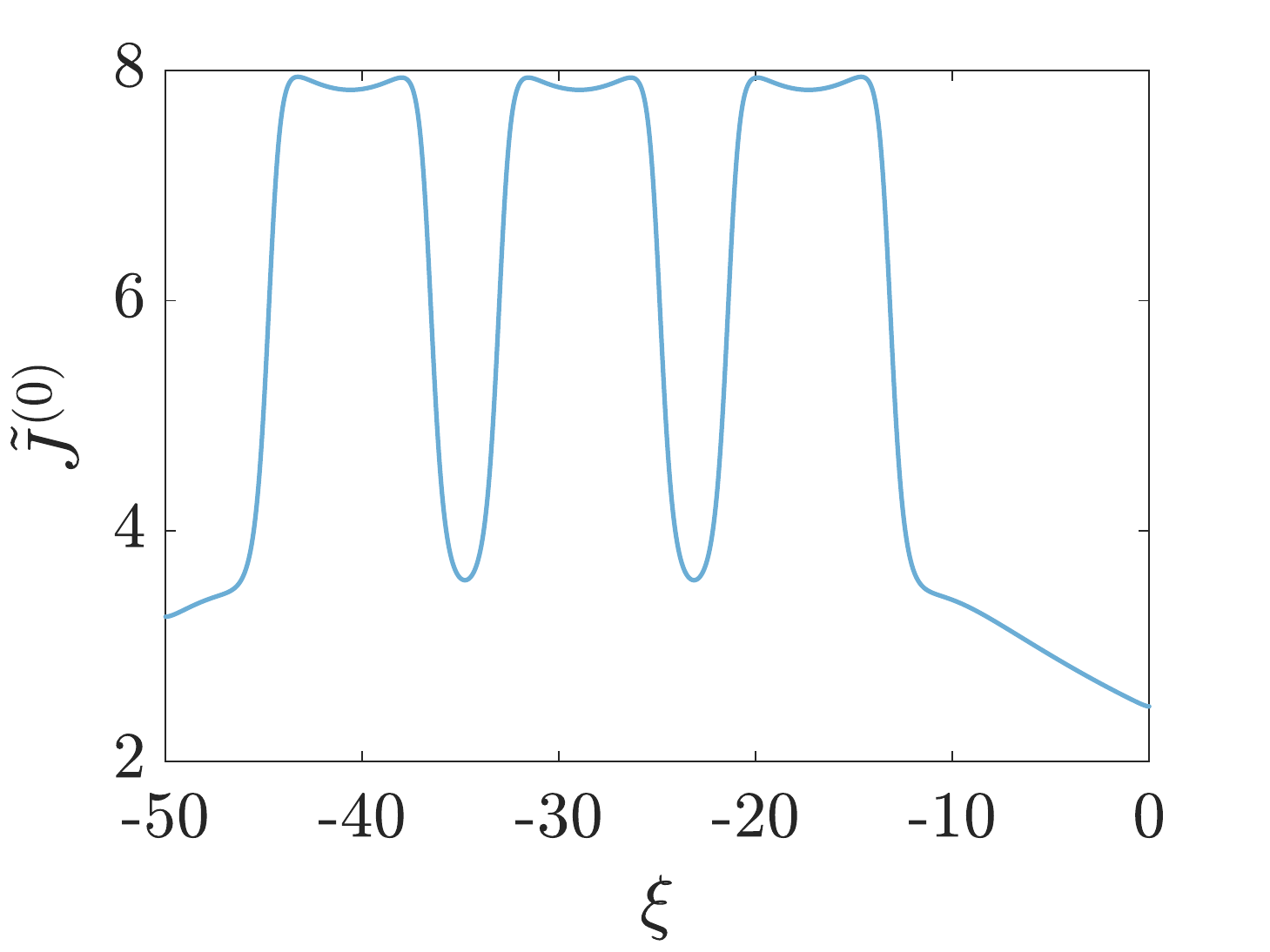}}
  \subfigure[$\bath{\phi}_+ = 6.6\cdot 10^{-4}$]{\includegraphics[width=0.32\textwidth]{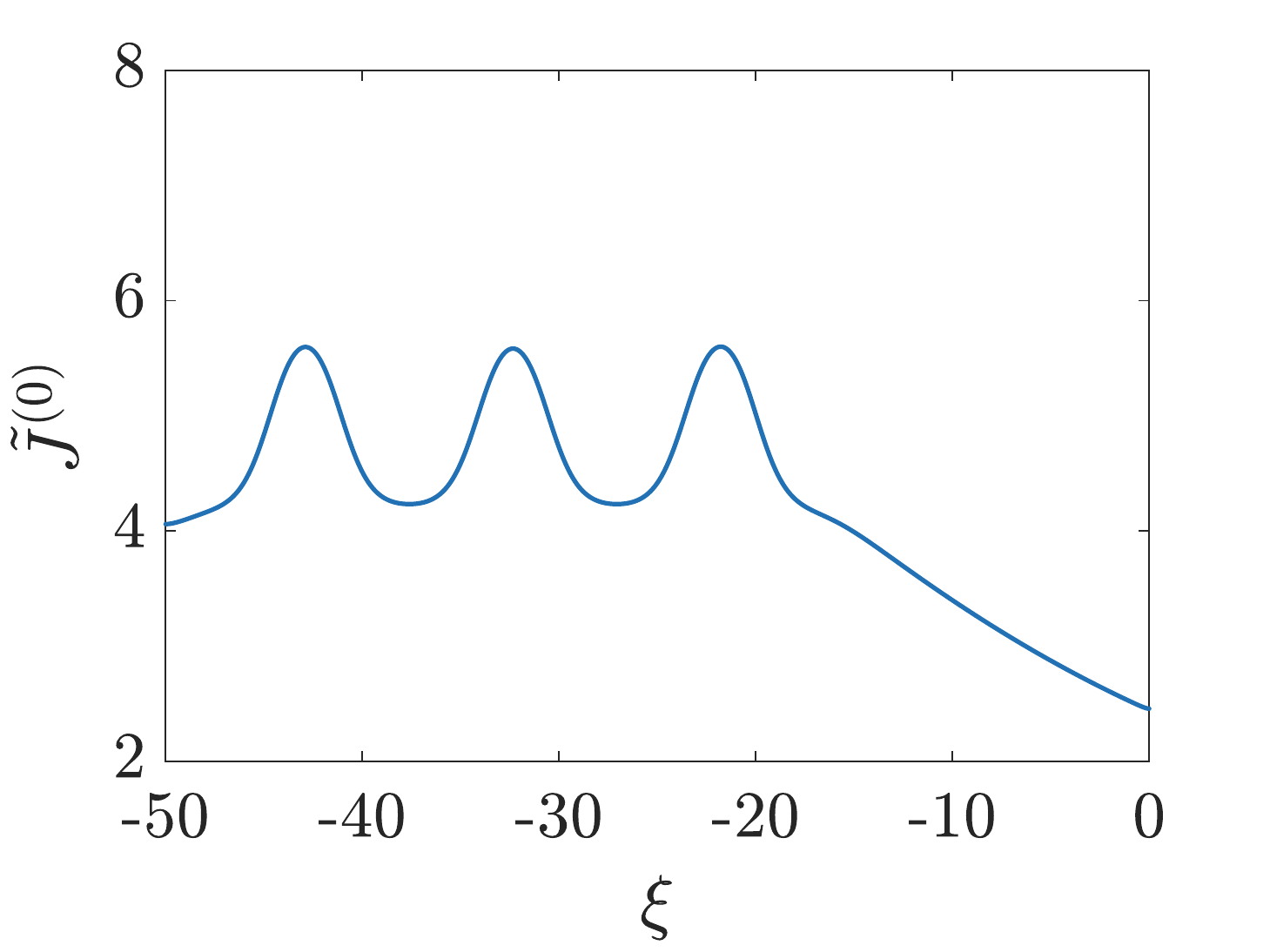}}
  \subfigure[$\bath{\phi}_+ = 10^{-3}$]{\includegraphics[width=0.32\textwidth]{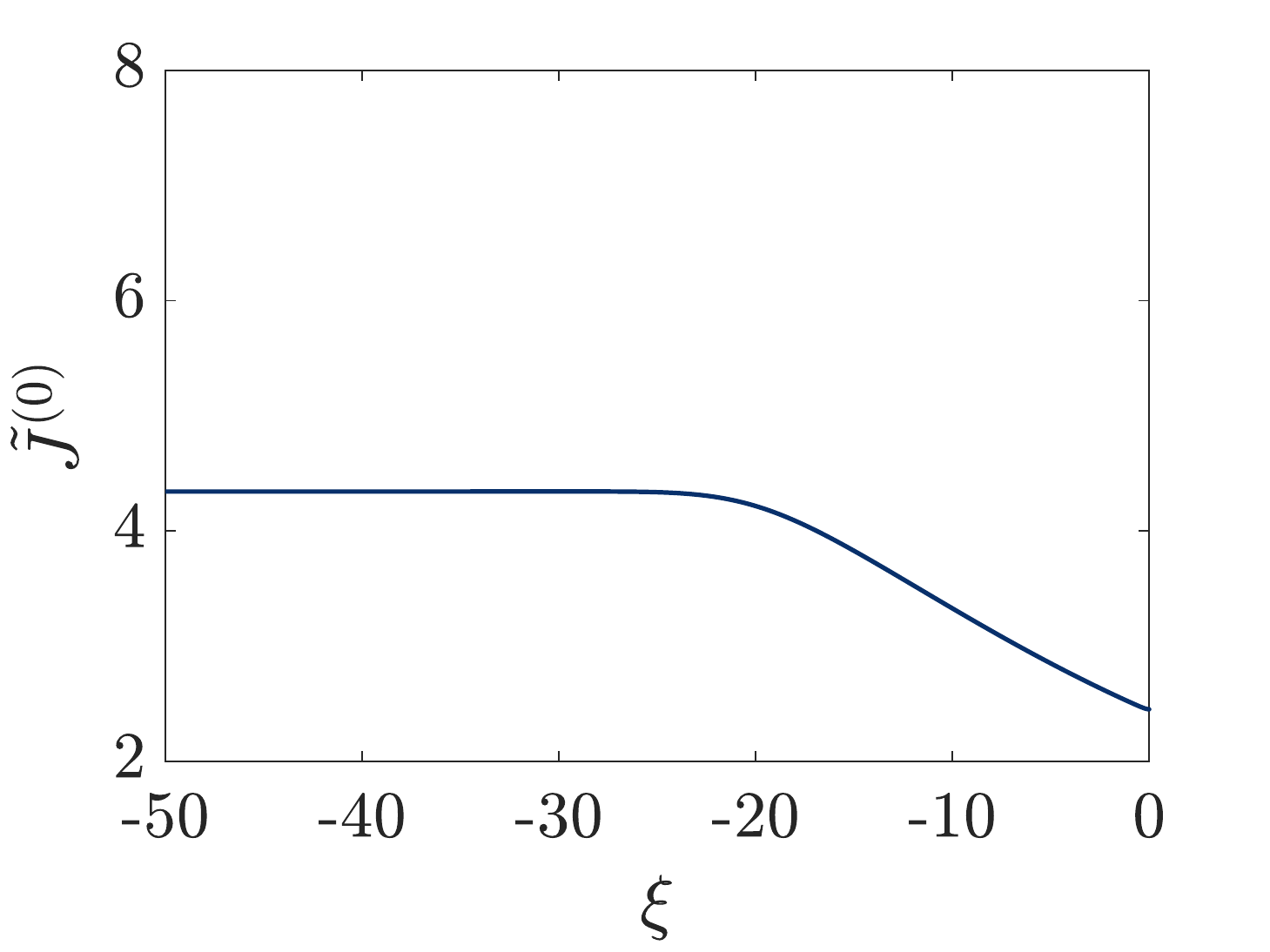}}
  \subfigure[$\bath{\phi}_+ = 10^{-5}$]{\includegraphics[width=0.32\textwidth]{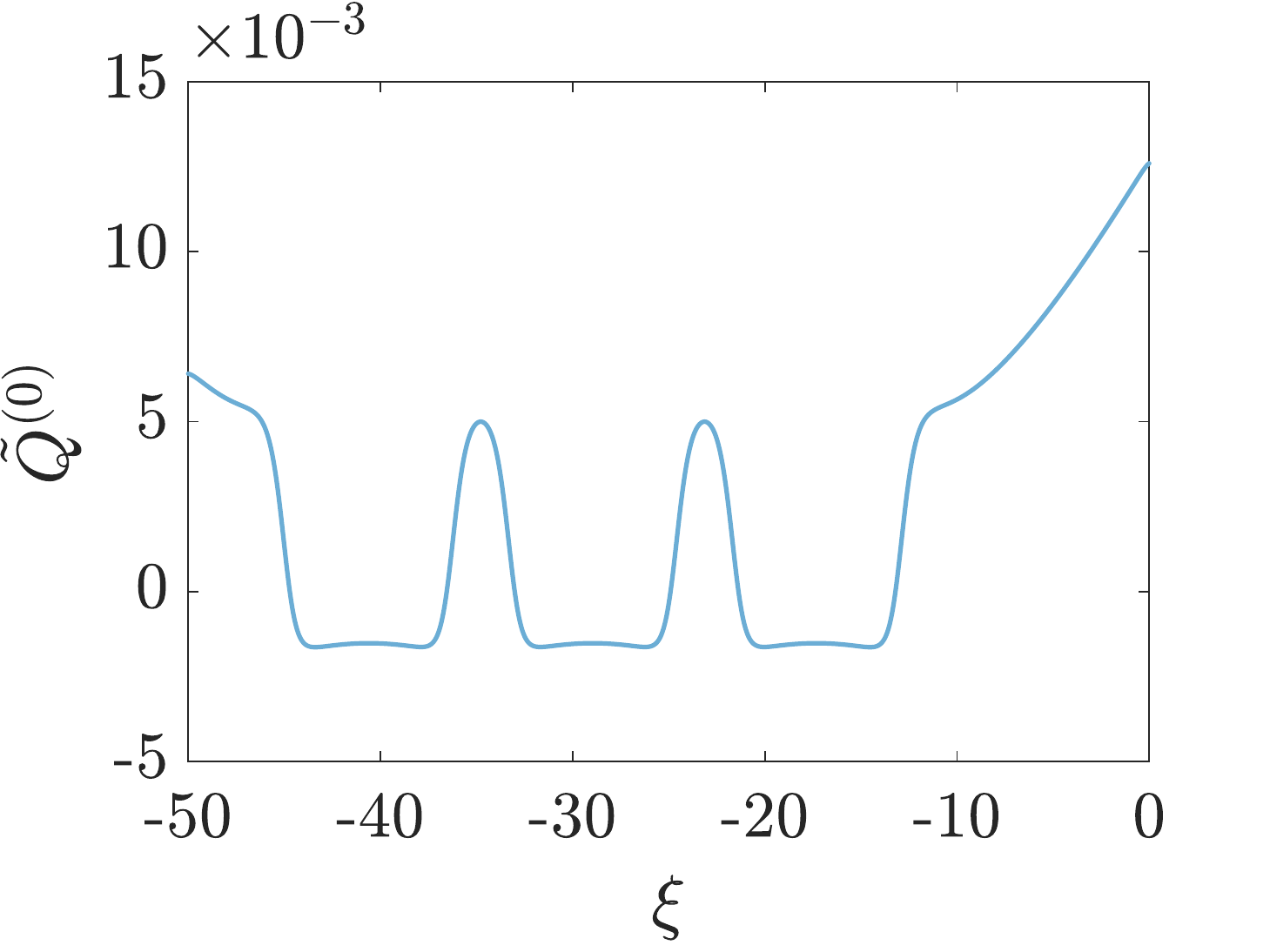}}
  \subfigure[$\bath{\phi}_+ = 6.6\cdot 10^{-4}$]{\includegraphics[width=0.32\textwidth]{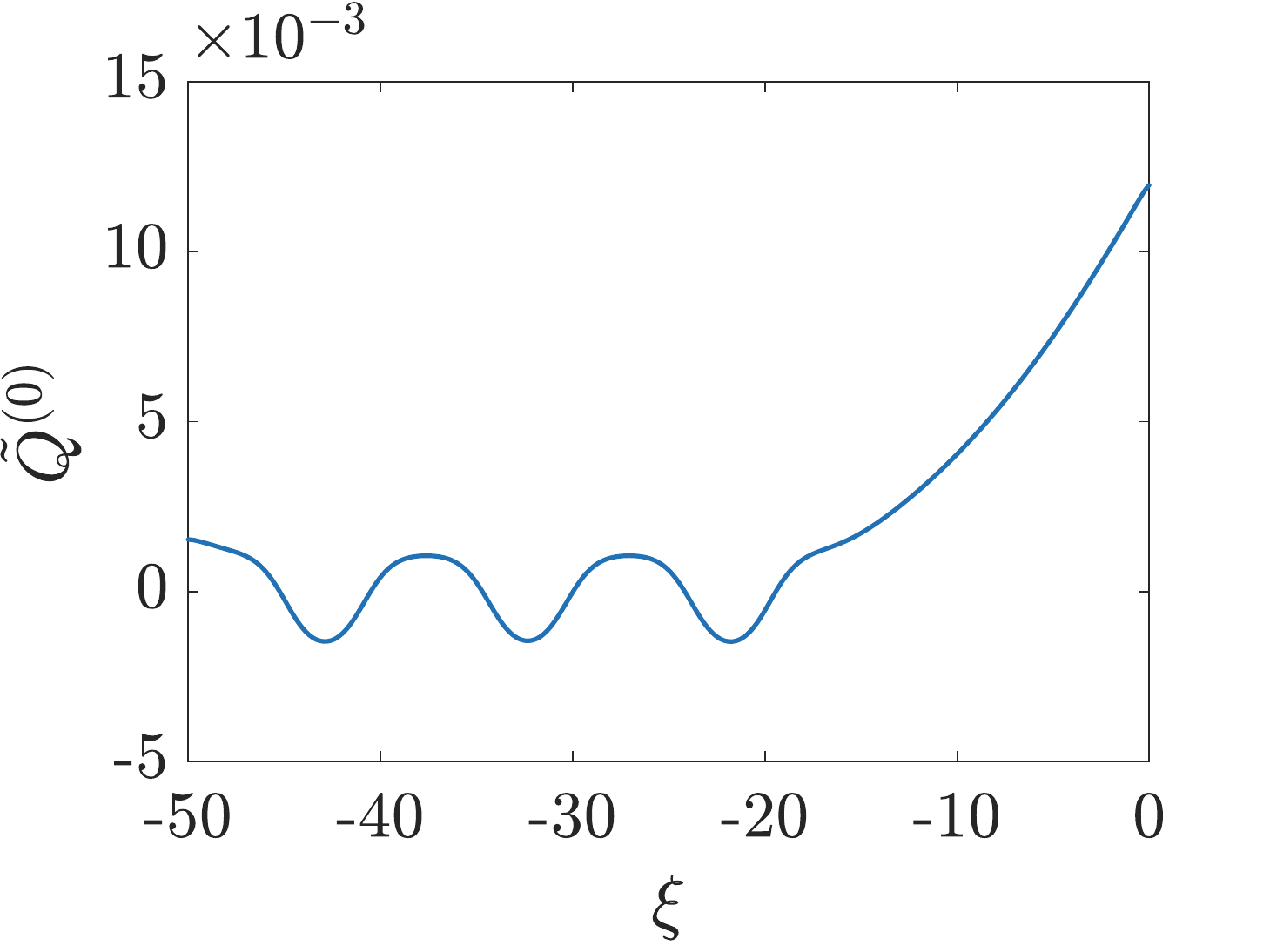}}
  \subfigure[$\bath{\phi}_+ = 10^{-3}$]{\includegraphics[width=0.32\textwidth]{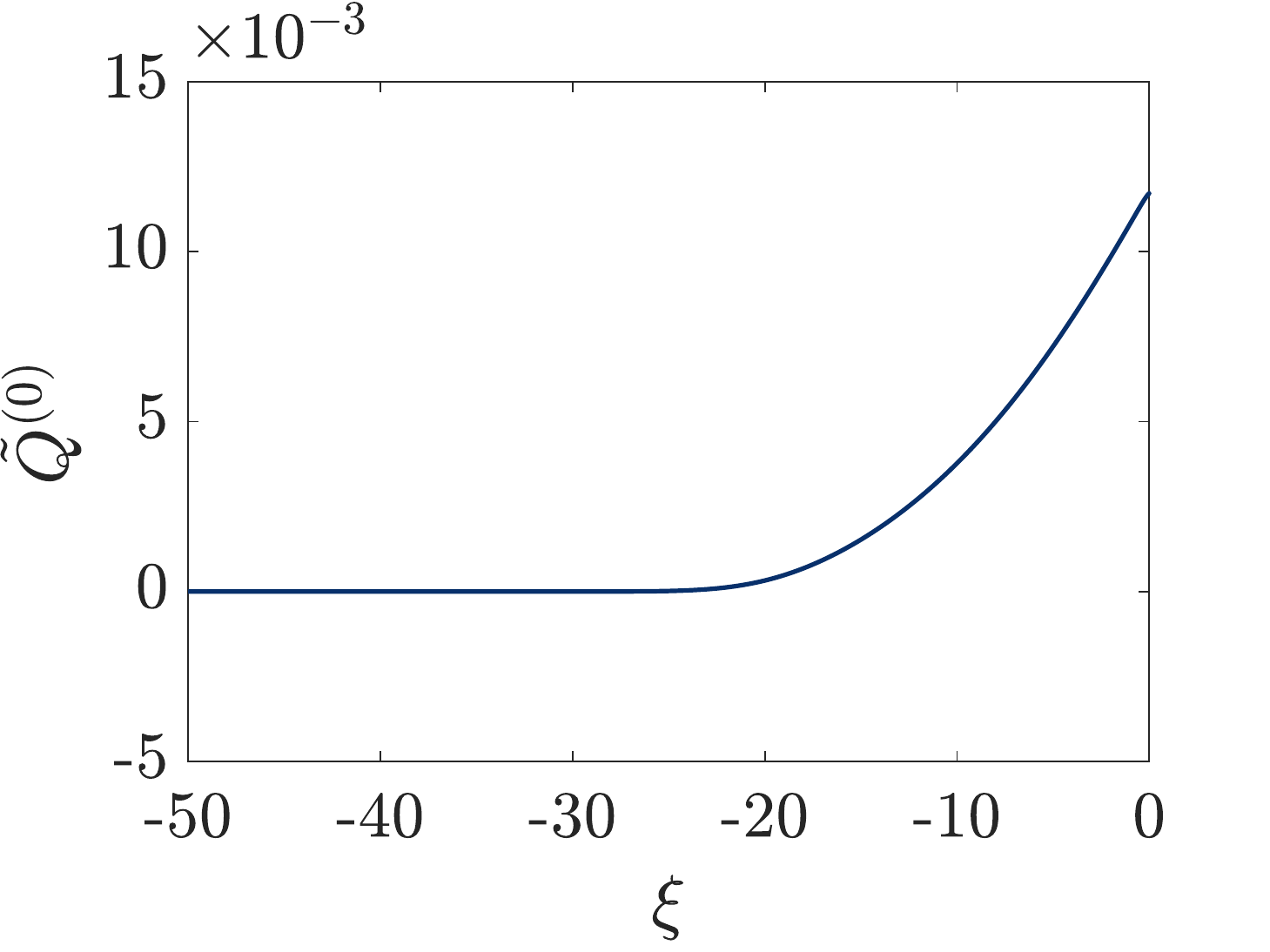}}
  \caption{Phase separation in the inner region. (a)--(c)
    The swelling ratio and (d)--(f) the total
    electric charge computed using the intermediate asymptotic model when the
    Debye length is comparable to the Kuhn length. We have
    taken $\omega = \Omega \beta$ with $\Omega = 10^{-1}$.
    The remaining parameters
    are $\chi = 0.7$, $\G = 4\cdot 10^{-3}$, $\alphaf = 0.04$, 
    $z_{\pm} = \pm 1$, $z_f = 1$, $\epsilon_r = 1$, and $\lambda_z = 1$.}
  \label{fig:cyl_ps}
\end{figure}

To understand the origin of these numerical difficulities,
we consider an intermediate asymptotic limit
where $\omega = \Omega \beta$, with $\Omega = O(1)$ as $\beta \to 0$.
Full details of
this limit are beyond the scope of this work; however, for the purpose of
this discussion it suffices to say that the inner problem in the gel
amounts to changing \eqref{inner:gel:mu_s} or \eqref{omega:phi_s} to
\subeq{
\begin{align}
  \t{\Pi}_s^{(0)} + \G \t{p}^{(0)} + \Omega^2 \pdd{\t{\phi}_s^{(0)}}{\xi} =
  \bath{\mu}_s,
\end{align}
where we have used the equality of the equilibrium chemical potentials $\gel{\mu}_s = \bath{\mu}_s$. The pressure \eqref{inner:gel:p_0} can be evaluated using
a Korteweg stress given by
\begin{align}
  \vec{n} \cdot \t{\tens{T}}_K^{(0)}\cdot \vec{n} = \G^{-1} \Omega^2
  \left[\t{\phi}_s^{(0)}\pdd{\t{\phi}_s^{(0)}}{\xi} - \frac{1}{2}\left(\pd{\t{\phi}_s^{(0)}}{\xi}\right)^2\right].
\end{align}}
The intermediate asymptotic model was solved using 
a second parameter set that reduces the degree of swelling
that occurs in the gel and forces the 
the outer problem to have only a single branch of solutions. Thus,  
the gel monotonically and continuously decreases in volume as the
salt fraction in the bath $\bath{\phi}_+$ increases. The inner problem was
solving using the intermediate model at three specific values of 
$\bath{\phi}$ using a value of $\Omega = 0.1$.
The swelling ratio $\t{J}^{(0)}$ and total charge $\t{Q}^{(0)} = \t{\phi}^{(0)}_{+} - \t{\phi}_{-}^{(0)} + z_f \t{\phi}_f^{(0)}$ are computed and plotted as
functions of space in Fig.~\ref{fig:cyl_ps}.
In this case, decreasing the salt fraction in the bath from $\bath{\phi}_+ = 10^{-3}$ triggers the onset of phase separation, which gives rise to
an array of electrically charged structures that spans the entire domain
of the inner problem.
Charge neutrality is not recovered in the far field, even if the
domain used to numerically solve the inner problem is increased,
meaning that the inner solution cannot be matched with
the homogeneous outer solutions computed from \eqref{cyl:eq}.
We therefore posit that homogeneous outer solutions do not always exist
in the limit $\beta \to 0$ with $\omega = O(\beta)$ or $\omega = 0$.
The lack of a homogeneous outer solution could explain the difficulties
in numerically solving the inner problem using the same parameters as in
Fig.~\ref{fig:cyl_swollen} when $\omega = 0$.

\begin{figure}
  \centering
  \subfigure[]{\includegraphics[width=0.45\textwidth]{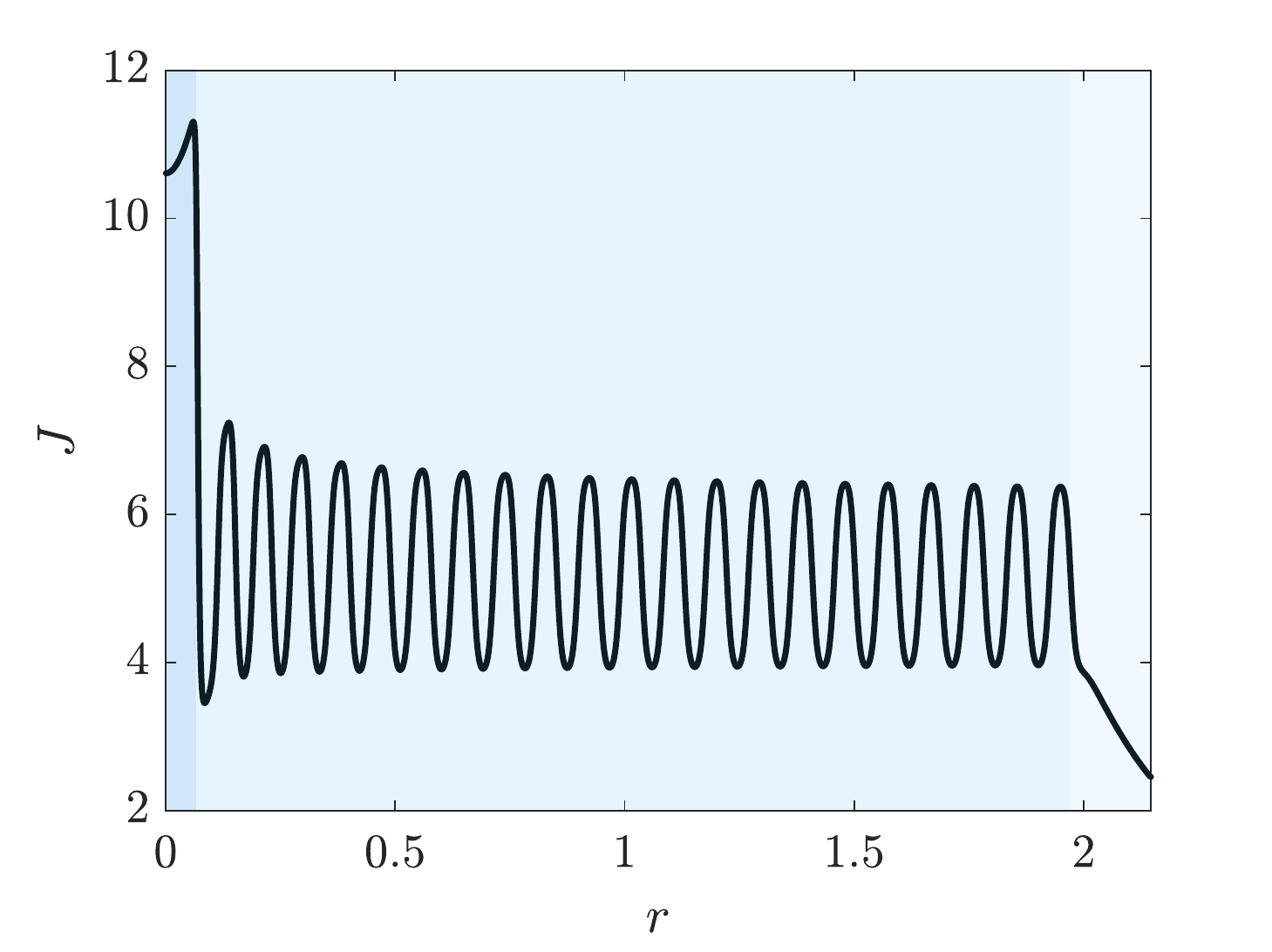}}
  \subfigure[]{\includegraphics[width=0.45\textwidth]{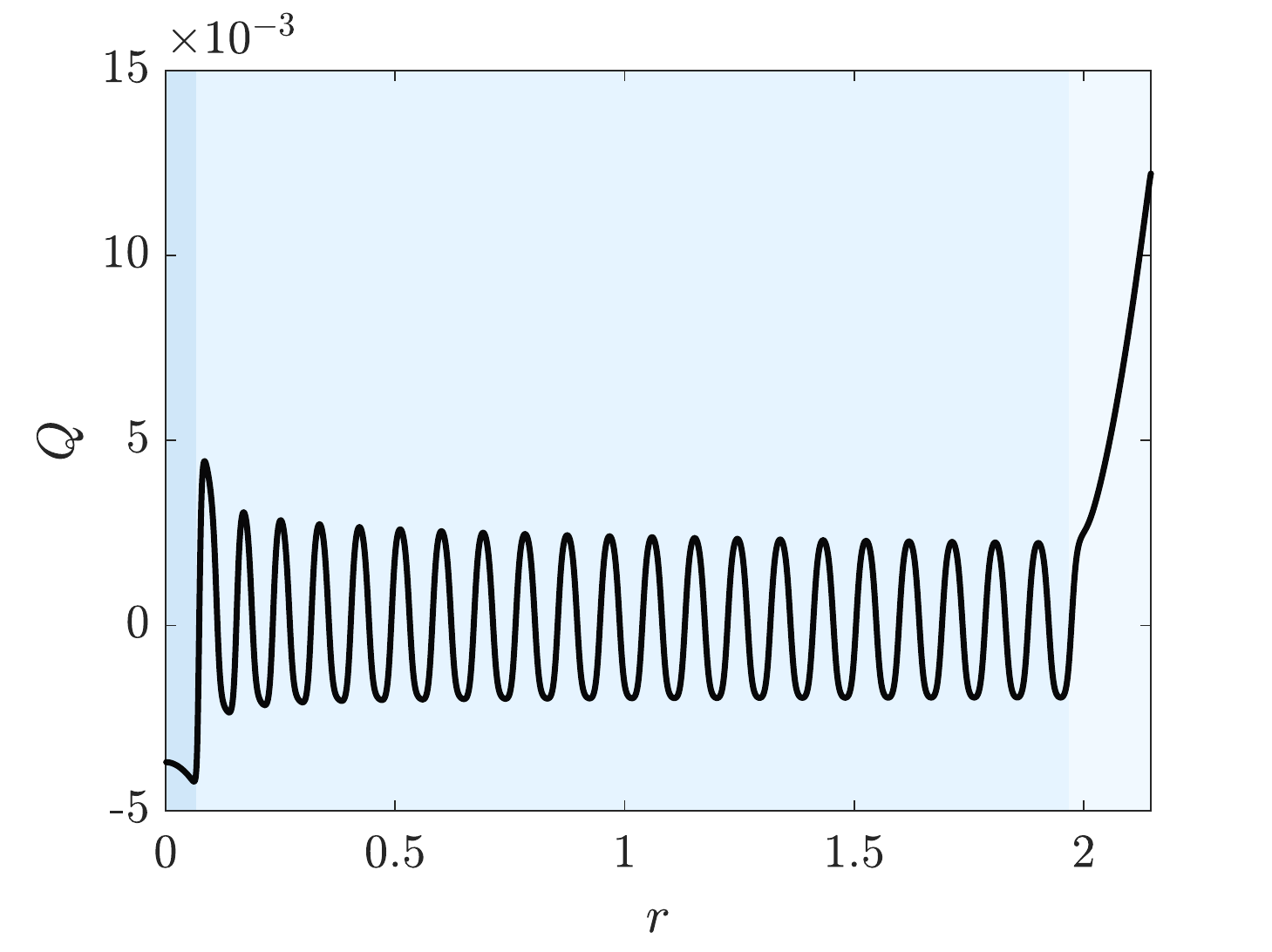}}
  \caption{Phase separation drives the breakdown of charge neutrality
    in the gel when
    the Debye length is comparable to the Kuhn length. (a) The swelling ratio and (b) the total electric charge
    computed from the full steady problem in cylindrical coordinates.
    The parameter values are $\beta = 10^{-2}$, $\omega = 10^{-3}$,
    $\bath{\phi}_+ = 6.6 \cdot 10^{-4}$, $\chi = 0.7$, $\G = 4\cdot 10^{-3}$, $\alphaf = 0.04$, 
    $z_{\pm} = \pm 1$, $z_f = 1$, $\epsilon_r = 1$, and $\lambda_z = 1$.}
  \label{fig:full_ps}
\end{figure}

To explore the hypothesis that the bulk of the gel may not be homogeneous
and electrically neutral at equilibrium, 
we solved the full steady problem with $\beta = 10^{-2}$ and $\omega = 10^{-3}$.
The salt fraction in the bath was set to $\bath{\phi}_+ = 6.6\cdot 10^{-4}$,
corresponding to the parameters in Fig.~\ref{fig:cyl_ps} (b) and (e).
The swelling ratio $J$ and the total charge $Q$, which are shown in
Fig.~\ref{fig:full_ps}, reveal that phase separation occurs throughout
the entire gel and gives rise to a periodic arrangement of electrically
charged domains. Using numerical integration, we find that the total amount of
electric charge contained within a pair of adjacent domains is on the order
of $10^{-7}$.
Thus, the gel effectively separates
into three distinct regions consisting of an electrically negative,
highly swollen core ($0 < r < 0.73$); a moderately swollen interior that is electrically neutral on average ($0.073 < r < 2.0$); and a positively charged, collapsed shell ($2.0 < r < 2.1$). Overall, the gel carries
a net positive charge which exactly balances the
net negative charge in the bath to ensure that charge neutrality holds on a
global scale. 
The pointwise breakdown of charge neutrality across the gel indicates that it
is not always appropriate to decompose the problem into inner and outer
regions that are characterised by the local charge density of the gel.

%% file: conclusions.tex
\section{Discussion and conclusion}
\label{sec:conclusions}

Asymptotic and numerical methods are used to
study the EDL that forms at the interface between a salt
bath and a polyelectrolyte gel. The gel is described using a phase-field
model, which introduces an additional length scale, the Kuhn length,
into the problem.  The Kuhn length measures the thickness of diffuse internal
interfaces that can form due to phase separation within the gel. 
The ratio of the non-dimensional Kuhn and Debye lengths, $\omega$ and
$\beta$, has
a profound influence on the structure of equilibrium solutions that has
not been reported before.

When $\omega \gg \beta$, there is a high energy cost associated with
gradients in the solvent concentration. Therefore, the
leading-order solvent volume fraction is uniform across the EDL.
Importantly, the complex
interplay between mechanics, electrostatics, and thermodymamics,
which can result in phase separation, is suppressed. This interplay is captured
in the contributions to the solvent chemical potential from the osmotic
and mechanical pressures, which do not enter at leading order. 
When applying
the asymptotic framework to a cylindrical gel, it is possible to match the
inner solutions to electrically neutral, homogeneous outer solutions in all
of the considered cases.
In contrast, when $\omega = 0$, the leading-order solvent fraction in the EDL
is set by the between the osmotic and mechanical pressures. In this
case, it is not always possible to compute a numerical solution to the
inner problem.

Our preliminary investigation of the intermediate asymptotic limit where
$\beta \to 0$ with
$\omega = O(\beta)$  reveals that phase separation can result in highly
heterogeneous gels consisting of repeating pairs of positively
and negatively charged domains. The breakdown of charge neutrality means that
the inner region effectively spans the entire gel.
The difficulties in numerically computing inner solutions with $\omega = 0$
are therefore attributed to the gel undergoing phase separation and the loss of
homogeneous outer solutions.

In Celora \etal\cite{Celora_modelling_2021}, we used continuation methods
to track numerical solutions of the full steady problem as the salt
fraction in the bath is varied in the regime when $\omega$ and $\beta$
are comparable.
We found that the breakdown of charge neutrality in the gel occurs via a
cascade of saddle-node bifurcations associated with spatially localised
modes of phase separation. A more in-depth analysis of the asymptotic limit
$\beta \to 0$ with $\omega = O(\beta)$ and the bifurcation structure will
be an interesting and insightful area of future work.

Physically,
the breakdown of electroneutrality due to phase separation when the
Kuhn and Debye lengths can be rationalised as follows. Phase separation
leads to the formation of diffuse interfaces that separate domains
with distinct compositions and electric potentials. The gradient in the
electric potential across the diffuse interface generates an electric field.
When the Kuhn and Debye lengths are commensurate, the electric field near
the diffuse interface will be of sufficient magnitude to trigger the formation
of an internal EDL. If the Kuhn length greatly exceeds the Debye
length, then the electric field is too weak to generate an EDL and hence
the gel remains electrically neutral. 

Typical models of polyelectrolyte gels do not account for
phase separation and thus
implicitly set $\omega = 0$. Homogeneous and hence electrically neutral
solutions that neglect the EDL are often sought and compared
against experimental data.
However, our results show that these homogeneous `solutions' may be
asymptotically inconsistent because there is no inner
solution in the EDL that can be matched to them.
In fact, when $\omega = 0$, the bulk behaviour of the
gel can be strongly coupled to the behaviour in the EDL and thus the latter
must be considered when constructing model solutions. 
The extensive use of
homogeneous, electroneutral solutions to characterise the response of highly
swollen polyelectrolyte gels is more
consistent with the assumption that $\omega \gg \beta$, as this limit enables
the successful matching of inner and outer solutions and prohibits the
breakdown of electroneutrality in the bulk of the gel.

A key outcome of this work is the systematic derivation of an
electroneutral model for a polyelectrolyte gel with consistent jump
conditions across the gel-bath interface that can capture phase separation.
This model was derived in the limit $\beta \to 0$ with $\beta \ll \omega$.
In Celora \etal\cite{Celora_SIAP_2021},
we use our electroneutral model 
to study the rich variety of dynamics
that can occur when a polyelectrolyte gel in contact with a salt bath
undergoes phase separation.
Given importance of electroneutral models in the applied
literature, the results presented in this paper
will increase our understanding of how asymptotic methods
can be used to derive consistent jump conditions across dynamic EDLs that form at the free interfaces of complex materials,
including those which undergo large elastic deformations.

%% file: appendix.tex
\begin{appendix}

\section{Summary of the governing equations in dimensional form} 
\label{app:dimensional}

\subsection{Bulk equations for the gel}
Conservation of solvent and ions is given by
\begin{align}
  \pd{c_m}{t} + \nabla \cdot (c_m \vec{v}_m) = 0
\end{align}
for $\allm$, where
$c_m$ is the (current) concentration (number of molecules per unit current volume). The velocity $\vec{v}_m$ is related to the network velocity $\vec{v}_n$ and the diffusive flux $\vec{j}_m$ according to
\begin{align}
  c_m(\vec{v}_m - \vec{v}_n) = \vec{j}_m.
\end{align}
Due to incompressibility, the determinant of the deformation tensor is
\begin{align}
  J = 1 + \sum_{\allm} \nu C_m = \left(1 - \sum_{\allm} \nu c_m\right)^{-1},
  \label{eqn:dim_ic}
\end{align}
where $C_m = J c_m$ is the nominal concentration of each mobile species and $\nu$ is the molecular volume, i.e. the volume of an individual molecule. For simplicity, we assume that all of the molecules are roughly the same size.
The diffusive fluxes of the solvent and ions are given by
\subeq{
\begin{align}
  \vec{j}_s &= -\frac{D_s(J)}{k_B T} \sum_{\allm} c_m \nabla \mu_{m}, \\
  \vec{j}_\pm &= -\frac{D_{\pm} c_{\pm}}{k_B T} \nabla \mu_{\pm} + \frac{c_{\pm}}{c_s}\vec{j}_s,
\end{align}
}
where $T$ is temperature, $k_B$ is Boltzmann's constant, $D_s$ is the diffusivity of solvent in a polymer network, and $D_{\pm}$ are the diffusivity of ions in a pure solvent bath. A common functional form of the solvent diffusivity is
$D_s = D_s^0 J^{a}$ with $a = 1.5$; see Bertrand \etal\cite{Bertrand2016}.
The chemical potential of solvent can be written as
\begin{align}
  \mu_s = \mu_s^0 + \nu(p + \Pi_s) - \gamma \nabla^2 c
\end{align}
where $p$ is the mechanical pressure, $\Pi_s$ is the osmotic pressure of the solvent,
\begin{align}
  \Pi_s = \frac{k_B T}{\nu}\left[\log(\nu c_s) + \frac{\chi(1 - \nu c_s)}{J} + \frac{1}{J}\right],
\end{align}
and $\chi$ is the Flory interaction parameter. 
The chemical potential of ions is given by
\begin{align}
  \mu_{\pm} = \mu_{\pm}^0 + \nu(\Pi_{\pm} + p) \pm e \Phi,
\end{align}
where $\Phi$ is the electric potential, $e$ is the elementary charge, and $\Pi_{\pm}$
is the osmotic pressure
\begin{align}
  \Pi_\pm = \frac{k_B T}{\nu}\left[\log(\nu c_\pm) + \frac{1}{J}(1 - \chi \nu c_s)\right].
\end{align}
The quantities $\mu_m^0$ are reference values of the chemical potential.
The electric potential satisfies
\begin{align}
  -\epsg\nabla^2 \Phi = e(c_+ - c_- + z_f c_f)
\end{align}
where $\epsg$ is the electrical permittivity of the gel and $c_f$ is the current concentration of fixed charges. Mechanical equilibrium leads to
\begin{align}
  \nabla \cdot \tens{T} = 0,
\end{align}
where the Cauchy stress tensor $\tens{T}$ can be decomposed into four contributions
\begin{align}
  \tens{T} = \tens{T}_e + \tens{T}_K + \tens{T}_M - p \tens{I},
\end{align}
associated with the elastic stress $\tens{T}_e$, the Korteweg stress $\tens{T}_K$,
the Maxwell stress $\tens{T}_M$, and the isotropic fluid pressure. These three
stress tensors are given by
\subeq{
  \begin{align}
    \tens{T}_e &= G J^{-1}(\tens{B} - \tens{I}), \\
  \tens{T}_K &= \gamma \left[\left(\frac{1}{2}|\nabla c_s|^2 + c_s \nabla^2 c_s\right)\tens{I} - \nabla c_s \otimes \nabla c_s\right], \\
  \tens{T}_M &= \epsg\left(\nabla \Phi \otimes \nabla \Phi - \frac{1}{2}|\nabla \Phi|^2 \tens{I}\right),
\end{align}
}
where $G$ and $\gamma$ play the role of a shear modulus and surface energy, respectively. The left Cauchy--Green tensor is defined as $\tens{B} = \tens{F} \tens{F}^T$. In Eulerian coordinates, the deformation gradient tensor satisfies
$\tens{F}^{-1} = \nabla \vec{X}$. The velocity of the network can be
determined from
\begin{align}
  \vec{v}_n = -\tens{F}\,\pd{\vec{X}}{t}.
\end{align}

\subsection{Governing equations for the bath}

Conservation of solvent and ions is given by
\begin{align}
  \pd{c_m}{t} + \nabla \cdot (c_m \vec{v}_m) = 0,
\end{align}
for $\allm$. The mixture velocity is defined as
\begin{align}
  \vec{v} = \sum_m \nu c_m \vec{v}_m.
\end{align}
Note that we also have
\begin{align}
  \sum_m \nu c_m = 1, \quad 
  \nabla \cdot \vec{v} = 0.
\end{align}
The velocity of each species can be linked to the diffusive flux via
\begin{align}
  c_m(\vec{v}_m - \vec{v}) = \jb_m, \label{bath:v_j}
\end{align}
which implies that
\begin{align}
  \sum_{\allm} \jb_m = 0.
\end{align}
The diffusive fluxes are defined by
\subeq{
\begin{align}
  \vec{j}_\pm &= -\frac{D_{\pm} c_{\pm}}{k_B T} \left(\nabla \mu_{\pm} -
                \nu \sum_{\allm} c_m \nabla \mu_m\right) + 
                \frac{c_{\pm}}{c_s}\vec{j}_s, \\
  \vec{j}_s &= -\vec{j}_{+} - \vec{j}_{-}.
\end{align}
}
The chemical potentials are given by
\subeq{
\begin{align}
  \mu_s &= \mu_s^0 + \nu (\Pi_s + p), \\
  \mu_{\pm} &= \mu_\pm^0 + \nu(\Pi_{\pm} + p) \pm e \Phi,
\end{align}
}
where
\begin{align}
  \Pi_m = \frac{k_B T}{\nu} \log(\nu c_m).
\end{align}
The electric potential satisfies
\begin{align}
  -\epsb \nabla^2 \Phi = e(c_+ - c_-).
\end{align}
The stress balance in the bath is given by
\begin{align}
  \nabla \cdot \tens{T} = 0,
\end{align}
where $\tens{T} = \tens{T}_v + \tens{T}_M - p \tens{I}$ where
\subeq{
\begin{align}
  \tens{T}_v &= \eta(\nabla \vec{v} + \nabla \vec{v}^T), \\
  \tens{T}_M &= \epsb\left(\nabla \Phi \otimes \nabla \Phi - \frac{1}{2}|\nabla \Phi|^2 \tens{I}\right).
\end{align}}

\subsection{Boundary conditions at the gel-bath interface}
The boundary conditions are discussed in detail in the text. 
Conservation of solvent and ions across the gel-bath interface are given by
\begin{align}
  \left[c_m(\vec{v}_m \cdot \vec{n} - V_n)\right]^{\vec{x}=\vec{r}^+}_{\vec{x} = \vec{r}^{-}} = 0,
\end{align}
where $V_n$ is the normal velocity of the interface.
The kinematic boundary condition for the velocity of the polymer network
is
\begin{align}
  \left[\vec{v}_n \cdot \vec{n} - V_n\right]_{\vec{x} = \vec{r}^{-}} = 0.
\end{align}
Continuity of chemical potential implies that
\begin{align}
  \left[\mu_m\right]^{\vec{x} = \vec{r}^{+}}_{\vec{x} = \vec{r}^{-}} = 0.
\end{align}
The variational condition for the solvent concentration is
\begin{align}
  \left[\nabla c_s \cdot \vec{n}\right]_{\vec{x} = \vec{r}^{-}} = 0.
\end{align}
Conservation of normal and tangential momentum gives
\begin{align}
  \left[\tens{T}\cdot \vec{n}\right]^{\vec{x} = \vec{r}^{+}}_{\vec{x} = \vec{r}^{+}} = 0.
\end{align}
The slip condition reads as
\begin{align}
  \left[\vec{v}\cdot \vec{t}_i\right]^{\vec{x} = \vec{r}^{+}}_{\vec{x} = \vec{r}^{-}} = 0.
\end{align}
We impose continuity of electrical potential and electric displacement
\subeq{
\begin{align}
  [\Phi]^{\vec{x} = \vec{r}^+}_{\vec{x} = \vec{r}^-} = 0, \\
  \left[-\epsilon\nabla \Phi \cdot \vec{n}\right]^{\vec{x} = \vec{r}^+}_{\vec{x} = \vec{r}^-} = 0.
\end{align}
}
and therefore do not account for surface charges on the gel.


\section{Conventions and identities}
\label{app:conventions}
A vector $\vec{v}$ is written in component form as
$\vec{v} = v_i \vec{e}_i$. Similarly, a tensor $\tens{T}$ is written in
  component form as
  $\tens{T} = {\sf T}_{ij} \vec{e}_i \otimes \vec{e}_j$. The gradient of the
  vector $\vec{v}$ is defined in Cartesian coordinates as
  \begin{align}
    \nabla \vec{v} = \pd{}{x_j}\left( v_i \vec{e}_i\right)\otimes \vec{e}_j.
  \end{align}
  Similarly, the tensor divergence is defined as
  \begin{align}
    \nabla \cdot \tens{T} = \pd{}{x_i}\left({\sf T}_{jk} \vec{e}_j \otimes \vec{e}_k\right)\vec{e}_i,
  \end{align}
  which can be evaluated using the property of the dyadic product
  $(\vec{a}\otimes\vec{b})\vec{c} = (\vec{b}\cdot\vec{c})\vec{a}$. Given two vectors
  $\vec{a} = a_i \vec{e}_i$ and $\vec{b} = b_j \vec{e}_j$ and a tensor
  $\tens{T} = T_{kl} \vec{e}_k \otimes \vec{e}_l$, we write
  \begin{align}
    \vec{a} \cdot \tens{T} \cdot \vec{b} = (a_i {\sf T}_{kl} b_j)(\vec{e}_l \cdot \vec{e}_j)(\vec{e}_i \cdot \vec{e}_k),
  \end{align}
  which collapses to $a_i T_{ij} b_j$ if the basis vectors are orthonormal.



  \input{inner_3d.tex}

  \input{plane_strain.tex}

  \input{dilute.tex}

\input{full_cylinder.tex}

\end{appendix}

%% file: inner_3d.tex
\section{Transformation of the derivatives in the inner region}
\label{app:inner}
We derive the asymptotic expressions in \eqref{eqn:inner_diff}
for the time derivative, gradient, and Laplacian in the inner region.
We will use the convention of summing over repeated indices. Greek
indices range from 1 to 2. 

In the inner problem we write
\subeq{
\begin{align}
  \vec{x} &= \vec{r}(s_1, s_2, t) + \beta \xi \vec{n}(s_1, s_2, t), \\
  t &= t',
\end{align}
}
where $\vec{r}$ denotes the location of the gel-bath interface and
$\vec{n}$ is the unit normal vector pointing from the gel into the bath.
The tangent and normal vectors are defined as
\begin{align}
  \vec{t}_\alpha = \pd{\vec{r}}{s_\alpha}, \quad \vec{n} = \frac{\vec{t}_1 \times \vec{t}_2}{\|\vec{t}_1 \times \vec{t}_2\|}. \label{app:inner:tn}
\end{align}
The normal velocity of the interface is defined as
$V_n = \vec{n} \cdot \p_{t'}\vec{r}$. 

Before proceeding with the transformation, it is helpful to summarise some
key definitions and results from differential geometry. 
The components of the metric tensor are defined as $g_{\alpha \nu} = \vec{t}_\alpha \cdot
\vec{t}_\nu$.
We let $g^{\alpha \nu}$ denote the components of the inverse of the
metric tensor. The curvature tensor has components
\begin{align}
  K_{\alpha \nu} = -\vec{n} \cdot \pd{\vec{t}_\alpha}{s_\nu} = \pd{\vec{n}}{s_\nu}\cdot
  \vec{t}_\alpha.
  \label{app:inner:K}
\end{align}
The metric tensor, its inverse, and the  curvature tensor are
all symmetric. The shape operator is defined as
$S^{\gamma}_{\nu} = g^{\gamma \alpha} K_{\alpha\nu}$.
The eigenvalues of the shape operator, $\kappa_{1}$ and $\kappa_{2}$, define
the principal curvatures of the surface. Similarly, the trace of the shape
operator is related to the mean curvature of the surface, $\kappa = (\kappa_1 + \kappa_2)/2$, through the relation $S^{\alpha}_{\alpha} = 2\kappa$.
By ensuring that
the normal vector $\vec{n}$ computed from \eqref{app:inner:tn} points into
the bath, the principal curvatures of a spherical gel will be positive. 

A straightforward application of the chain rule shows that
\begin{align}
  \pd{}{s_\alpha} &= \left(\vec{t}_\alpha + \beta \xi \pd{\vec{n}}{s_\alpha}\right)\cdot
  \nabla, \\
  \pd{}{\xi} &= \beta \vec{n} \cdot \nabla, \\
  \pd{}{t'} &= \pd{}{t} + \left(\pd{\vec{r}}{t'} + \beta \xi \pd{\vec{n}}{t'}\right)\cdot \nabla.
\end{align}
We now exploit the fact that $\beta \ll 1$ and write the differential operators
$\nabla$ and $\p_t$ as asymptotic series of the form
$\nabla = \beta^{-1} \nabla^{(-1)} + \nabla^{(0)} + \beta \nabla^{(1)} + O(\beta^2)$ and $\p_t = \beta^{-1} \p_t^{(-1)} + \p_{t}^{(0)} + O(\beta)$.

The $O(\beta^{-1})$ problem for the del operator is
\subeq{
\begin{align}
  0 = \vec{t}_\alpha \cdot \nabla^{(-1)}, \\
  \pd{}{\xi} = \vec{n} \cdot \nabla^{(-1)},
\end{align}}
which has the solution
\begin{align}
  \nabla^{(-1)} = \vec{n} \pd{}{\xi}.
\end{align}
The corresponding problem for the time derivative is trivial to solve and
has solution
\begin{align}
  \p_t^{(-1)} = -V_n \pd{}{\xi}.
\end{align}

The $O(1)$ problem for the del operator is given by
\subeq{
\begin{align}
  \pd{}{s_\alpha} &= \vec{t}_\alpha \cdot \nabla^{(0)} + \xi \vec{n}\cdot\pd{\vec{n}}{s_\alpha}
  \pd{}{\xi}, \label{app:inner:1t} \\
  0 &= \vec{n} \cdot \nabla^{(0)}. \label{app:inner:1n}
\end{align}
}
Since $\vec{n}$ is a unit vector, we have that
$\vec{n}\cdot \p_{s_\alpha} \vec{n} = (1/2) \p_{s_\alpha}(\vec{n}\cdot\vec{n}) = 0$,
implying the final term in \eqref{app:inner:1t} vanishes.
Equation \eqref{app:inner:1n} implies that $\nabla^{(0)}$ lies in the tangent
plane and thus has the form $\nabla^{(0)} = a_\alpha \vec{t}_\alpha$.
Inserting this
solution in \eqref{app:inner:1t} and solving gives
\begin{align}
  \nabla^{(0)} =  g^{\alpha \nu} \vec{t}_\alpha\, \pd{}{s_\nu} \equiv \nabla_s,
  \label{app:inner:n0}
\end{align}
where $\nabla_s$ is the surface gradient. The $O(1)$ contribution to the
time derivative can be calculated as
\begin{align}
  \p_{t}^{(0)} = \pd{}{t'} - \pd{\vec{r}}{t'}\cdot \nabla_s.
\end{align}

The $O(\beta)$ problem for the del operator, after minor simplification,
is given by
\subeq{
\begin{align}
  \vec{t}_\alpha \cdot \nabla^{(1)} &= -\xi \pd{\vec{n}}{s_\alpha} \cdot \nabla^{(0)}, \\
  \vec{n} \cdot \nabla^{(1)} &= 0.
\end{align}}
By following the same strategy as the $O(1)$ problem, substituting
the solution in \eqref{app:inner:n0}, and using \eqref{app:inner:K} and the
definition of the shape operator, we find that
\begin{align}
  \nabla^{(1)} = -\xi S^{\alpha}_{\gamma} g^{\gamma\nu} \vec{t}_\alpha \pd{}{s_\nu}.
\end{align}

Using these asymptotic expansions, we can construct the Laplacian
$\nabla^2 = \nabla \cdot \nabla$. In doing so, we will use the fact that
the tangent and normal vectors $\vec{t}_\alpha$ and $\vec{n}$ are independent of the coordinate $\xi$. As a result, $\nabla^{(-1)}\cdot \nabla^{(-1)} = \p_{\xi\xi}$,
$\nabla^{(-1)} \cdot \nabla^{(0)} = 0$, and
$\nabla^{(-1)} \cdot \nabla^{(1)} = 0$. Moreover,
\subeq{
\begin{align}
  \nabla^{(0)}\cdot \nabla^{(-1)}
  &=
    g^{\alpha\nu} \vec{t}_\alpha \cdot \pd{\vec{n}}{s_\nu}\pd{}{\xi} =
    S^{\alpha}_{\alpha} \pd{}{\xi} = 2 \kappa \pd{}{\xi}, \\
  \nabla^{(0)}\cdot \nabla^{(0)}
  &=
    g^{\alpha\nu} \vec{t}_\alpha \cdot \pd{}{s_\nu}\left(
    g^{\gamma\delta}\vec{t}_\gamma\pd{}{s_\delta}\right)
    = g^{\alpha\nu} g^{\gamma\delta}\, \vec{t}_\alpha \cdot \pd{\vec{t}_\gamma}{s_\nu}\pd{}{s_\delta} + \pd{}{s_\gamma}\left(g^{\gamma\delta}\pd{}{s_\delta}\right).
    \label{app:inner:n0n0}
\end{align}}
In order to simplify \eqref{app:inner:n0n0}, we express the
derivatives of the tangent vectors as
\begin{align}
  \pd{\vec{t}_\gamma}{s_\nu} = \Gamma^{\epsilon}_{\gamma\nu} \vec{t}_\epsilon - K_{\gamma\nu} \vec{n},
\end{align}
where $\Gamma^{\epsilon}_{\gamma\nu}$ is the Christoffel symbol. In addition, we invoke
the identity
\begin{align}
  \Gamma^{\alpha}_{\gamma\alpha} = \frac{1}{\sqrt{g}}\pd{}{s_\gamma}\left(\sqrt{g}\right),
\end{align}
where $g = g_{11}g_{22} - g_{12}^2$ is the determinant of
the metric tensor. Thus, we find that $\nabla^{(0)} \cdot \nabla^{(0)} = \nabla_s^2$, where
\begin{align}
  \nabla^2_s = \frac{1}{\sqrt{g}}\pd{}{s_\gamma}\left(\sqrt{g} g^{\gamma\delta} \pd{}{s_\delta}\right)
  \label{app:inner:lap_s}
\end{align}
is the surface Laplacian (or Laplace--Beltrami operator). Finally, we
have that
\begin{align}
  \nabla^{(1)}\cdot \nabla^{(-1)} = -\xi S^{\alpha}_{\gamma} g^{\gamma\nu}\,\vec{t}_\alpha \cdot
  \pd{\vec{n}}{s_\nu}\pd{}{\xi} = -\xi S^{\alpha}_{\gamma} g^{\gamma\nu} K_{\alpha\nu} \pd{}{\xi}
  = -\xi S^{\alpha}_{\gamma} S^{\gamma}_{\alpha} \pd{}{\xi} = -\xi (\kappa_\alpha \kappa_\alpha) \pd{}{\xi}.
\end{align}
The last equality is obtained by noticing that $S^{\alpha}_{\gamma} S^{\gamma}_{\alpha}$ is the
trace of the square of the shape operator and thus $S^{\alpha}_{\gamma} S^{\gamma}_{\alpha} =
\kappa_\alpha \kappa_\alpha$.

%% file: plane_strain.tex
\section{Specialisation to plane-strain problems}
\label{sec:ps}

\begin{figure}
  \centering
  \includegraphics[width=0.5\textwidth]{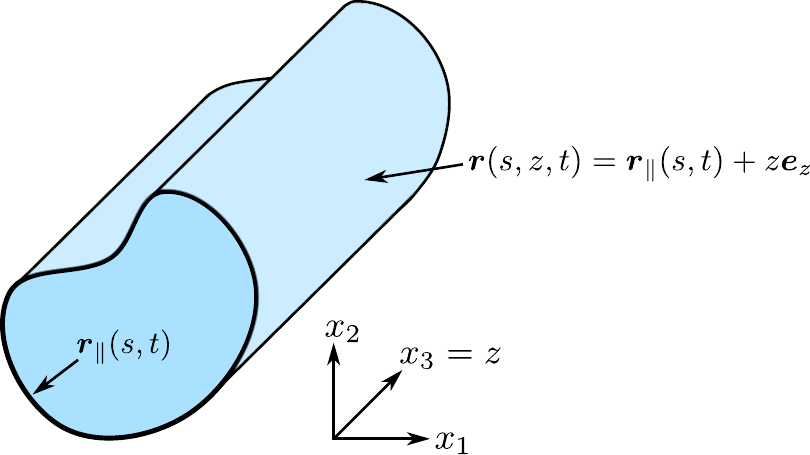}
  \caption{The geometry of a cylindrical hydrogel with arbitrary cross section.
    The quantity
    $\pl{\vec{r}}$ represents the one-dimensional gel-bath
    interface formed at each cross section and it is parametrised by its
    arclength $s$. The full two-dimensional gel-bath interface $\vec{r}$
    is parametrised in terms of $s$ and the axial coordinate $z$.
    The coordinates $x_1$ and $x_2$ lie in the cross-sectional plane.}    
  \label{fig:plane_strain}
\end{figure}

The governing equations can be simplified in the case of
plane-strain problems. We thus consider cylindrical geometries with
arbitrary cross sections as shown in Fig.~\ref{fig:plane_strain}. We assume
that the unit vectors $\vec{e}_1$ and $\vec{e}_2$ span the cross-sectional
plane
and that $\vec{e}_3$ is aligned with the axial direction. For clarity,
we write $\vec{e}_3 \equiv \vec{e}_z$ and let $z \equiv x_3$ denote the axial
coordinate. Using this construction, any vector $\vec{u}$
can be decomposed into components $\pl{\vec{u}}$ and $u_z$
that lie in the cross-sectional plane and in the axial direction according to
$\vec{u} = \pl{\vec{u}} + u_z \vec{e}_z$, with
$\pl{\vec{u}} = u_\alpha \vec{e}_\alpha$. 

In plane-strain probems, the deformation gradient tensor can be written as
$\tens{F}(\vec{x},t) = \tens{F}_{\parallel}(\pl{\vec{x}},t) + \lambda_z \vec{e}_z \otimes \vec{E}_Z$, where $\pl{\tens{F}} = (\pl{\nabla} \pl{\vec{X}})^{-1}$ is the in-plane deformation gradient tensor,
$\pl{\vec{X}} = X_{\alpha}(\pl{\vec{x}},t) \vec{E}_{\alpha}$, and
$\lambda_z = (\pdf{X_z}{z})^{-1}$ corresponds to a constant stretch or
compression that is imposed in the axial direction.
The Jacobian can be decomposed
as $J = \pl{J} \lambda_z$ where $\pl{J} = \det \pl{\tens{F}}$.  We assume that
all variables, except for $X_z$, are independent of the axial coordinate $z$.

The outer problem is trivial to formulate and will not be discussed in detail.
Instead, we focus on the inner problem for the gel mechanics building upon the results from Sec.~\ref{sec:inner}.
The gel-bath interface can be parametrised in terms
of $s_1 = s$ and $s_2 = z$ as $\vec{r}(s_1,s_2,t) = \pl{\vec{r}}(s,t) + z \vec{e}_z$.
As illustrated in Fig.~\ref{fig:plane_strain}, the quantity $\pl{\vec{r}}$ represents the one-dimensional gel-bath interface
formed at each cross section, which is parametrised in terms of its
arclength $s$. The corresponding unit tangent vectors are
$\vec{t}_1 = \pdf{\pl{\vec{r}}}{s} \equiv \vec{t}$ and $\vec{t}_2 = \vec{e}_z$
and
satisfy $\vec{t}\cdot\vec{e}_z = 0$. The triad $\{\vec{t}, \vec{e}_z, \vec{n}\}$ thus forms an orthonormal basis.
From the calculations in Appendix \ref{app:inner},
it follows that the principal curvatures of the surface are
given by $\kappa_1 = -\vec{n}\cdot \p_s \vec{t}$ and $\kappa_2 = 0$ and the derivatives in the inner region
transform according to \eqref{eqn:inner_diff} with
$\nabla_s = \vec{t}\,\p_s + \vec{e}_z \p_{z}$ and
$\nabla^2_s = \p_{ss} + \p_{zz}$.

To calculate the in-plane deformation gradient tensor in the inner layer, we again
introduce the Lagrangian analogues of the gel-bath interface $\pl{\vec{r}}$,
its arclength
$s$, and the unit normal and tangent vectors $\vec{n}$ and $\vec{t}$; these are
denoted by $\pl{\vec{R}}$, $S$, $\vec{N}$, and $\vec{T} = \p_S \pl{\vec{R}}$,
respectively. By repeating the calculations in Sec.~\ref{sec:inner}, we find
that the deformation gradient tensor is diagonal and given by
\begin{align}
  \pl{\t{\tens{F}}}^{(0)} =
  \left(\pd{\t{\Xi}^{(0)}}{\xi}\right)^{-1} \vec{n} \otimes \vec{N}
  + \gel{\lambda_s} \vec{t} \otimes \vec{T},
  \label{ps:F}
\end{align} 
where $\gel{\lambda}_s = (\p_s \t{S}^{(0)})^{-1}$ is analogous to the
surface deformation gradient and quantifies stretching of material elements
in the tangential direction. To calculate $\gel{\lambda}_s$, we match
\eqref{ps:F} to the outer solution and use the fact that $\vec{t}$ and
$\vec{T}$ are unit vectors to obtain
\begin{align}
  \gel{\lambda}_s = \vec{t} \cdot \gel{\tens{F}} \cdot \vec{T}.
  \label{ps:lambda_s}
\end{align}
The in-plane determinant is readily given by
\begin{align}
  \pl{\t{J}}^{(0)} = \left(\pd{\t{\Xi}^{(0)}}{\xi}\right)^{-1} \gel{\lambda}_s.
\end{align}
while the in-plane components of the elastic stress tensor are 
\begin{align}
  \t{\tens{T}}_{e,\parallel}^{(0)} &= \frac{1}{\t{J}^{(0)}}\left(\pl{\t{\tens{B}}}^{(0)}-\vec{n}\otimes\vec{n} - \vec{t}\otimes\vec{t}\right),
	\label{ps:T}
\end{align}
where $\pl{\t{\tens{B}}} = \pl{\t{\tens{F}}}^{(0)}(\pl{\t{\tens{F}}}^{(0)})^T$.


%% file: dilute.tex
\section{Simplification of the equilibria for cylindrical gels}
\label{app:dilute}

The nonlinear system for the outer solution \eqref{cyl:eq} can be
greatly simplified in the limit of a dilute salt, $\bath{\phi}_{+} \ll \phi_f$.
Balancing terms in the electroneutrality condition \eqref{cyl:eq_en} gives
\begin{align}
  \gel{\Phi} - \bath{\Phi} \sim \log\left(\frac{z_f \phi_f}{\bath{\phi}_+}\right)
  + \G\left(\frac{1}{\lambda_z}-\frac{1}{\gel{J}}\right)
  + \frac{1}{\gel{J}}(1-\chi \gel{\phi}_s),
\end{align}
where we have assumed that $\G / \gel{J}$ at most $O(1)$ in size.
The ion fractions in the gel are approximately given by
\begin{align}
  \gel{\phi}_+ \sim \frac{(\bath{\phi}_+)^2}{z_f \phi_f}
  \exp\left[-2\G\left(\frac{1}{\lambda_z} - \frac{1}{\gel{J}}\right) -\frac{2}{\gel{J}}(1-\chi \gel{\phi}_s)\right],
  \qquad
  \gel{\phi}_- \sim z_f \phi_f,
\end{align}
showing that the anions, to leading order in $\bath{\phi}_+$, balance the
fixed charges on the polymer chains. 
Since the cation fraction $\gel{\phi}_+$ will be extremely small relative to the
anion fraction $\gel{\phi}_{-}$, the Jacobian determinant then reduces to
\begin{align}
  \gel{J} \sim \frac{1 + z_f \alphaf}{1 - \gel{\phi}_s},\label{cy1:eq_J}
\end{align}
where we have used $\phi_f = \alphaf / \gel{J}$.
The solvent fraction can then be obtained by solving
\begin{align}
  \label{cyl:eq_red}
  \log \gel{\phi}_s + \frac{1 - \gel{\phi}_s}{1 + z_f \alphaf} +
  \frac{\chi (1 - \gel{\phi}_s)^2}{1 + z_f \alphaf} + \G\left(\frac{1}{\lambda_z}
  - \frac{1 - \gel{\phi}_s}{1 + z_f \alphaf}\right) = -2 \bath{\phi}_+,
\end{align}
and used to evaluate the Jacobian determinant, ion fractions, and jump in
electric potential. The black dashed lines in Fig.~\ref{fig:cylindrical_eq}
represent solutions of \eqref{cy1:eq_J}-\eqref{cyl:eq_red}, which are in very good agreement
with the full nonlinear system \eqref{cyl:eq}. 

%% file: full_cylinder.tex
\section{The steady problem in cylindrical coordinates}
\label{app:full_cylinder}

In this section the full system of equations are specialised to a stationary
axisymmetric situation in cylindrical coordinates. In this case, all
of the fluxes and velocities are equal to zero and the chemical potentials
are spatially uniform. As in Sec.~\ref{sec:cylinder}, we consider a monovalent
salt with $z_{\pm} = \pm 1$. The cylindrical hydrogel is assumed to be
constrained in the axial direction such that the axial stretch is fixed
to $\lambda_z = 1$. 

\subsection{The bath problem}

In the far field ($r \to \infty$) we set $\Phi = 0$, $p = 0$, $\phi_{+} = \phi_{-} = \bath{\phi}_+$. Since the chemical potentials are uniform, matching to the far field gives $\mu_{\pm} = \log \phi_{\pm} + \epsilon_r \beta^2 p \pm \Phi = \log\ \bath{\phi}_{+}$. The ionic volume fractions can therefore be expressed as 
\begin{align}
  \phi_{\pm} = \bath{\phi}_{+}\exp(-\epsilon_r \beta^2 p \mp \Phi).
  \label{fc:bath:ions}
\end{align}
Substituting \eqref{fc:bath:ions} into the Poisson--Boltzmann equation for
the potential \eqref{nd:bath:Phi} leads to
\begin{align}
  -\frac{\epsilon_r \beta^2}{r}\td{}{r}\left(r \td{\Phi}{r}\right) =
  -2 \bath{\phi}_{+} \sinh(\Phi) \exp(-\epsilon_r \beta^2 p).
\end{align}
The radial component of the stress balance \eqref{nd:bath:div_T_nc}
simplifies to
\begin{align}
  \frac{1}{r}\td{\Phi}{r}\td{}{r}\left(r \td{\Phi}{r}\right) = \td{p}{r}.
\end{align}
These equations can be combined to determine the pressure:
\begin{align}
  p = \epsilon_r^{-1}\beta^{-2} \log\left(1 - 2 \bath{\phi}_{+}(1 - \cosh \Phi)\right).
\end{align}
The electric potential therefore satifies the equation
\begin{align}
  \frac{\epsilon_r \beta^2}{r}\td{}{r}\left(r \td{\Phi}{r}\right) =
  \frac{2 \bath{\phi}_{+} \sinh(\Phi)}{1 - 2 \bath{\phi}_{+}(1 - \cosh \Phi)}.
\end{align}

\subsection{The gel problem}
The chemical potentials in the gel can be written as
\subeq{
\begin{align}
  \log(1 - 2 \bath{\phi}_+) &= \Pi_s + \G p - \frac{\omega^2}{r}\td{}{r}\left(r\td{\phi_s}{r}\right), \label{fc:gel:mu_s}
  \\
  \log \bath{\phi}_{+}  &= \Pi_{\pm} + \G p \pm \Phi,
\end{align}
}
where we have used the continuity of chemical potentials across the
gel-bath interface \eqref{nd:bc:mu_m}. The osmotic pressures
$\Pi_m$ are defined in \eqref{nd:gel:Pi}. The electric potential satisfies
\begin{align}
  -\frac{\beta^2}{r}\td{}{r}\left(r \td{\Phi}{r}\right)
  = \phi_{+} - \phi_{-} + z_f \phi_f,
  \label{fc:gel:Phi}
\end{align}
with $\phi_f = C_f / J$. The deformation gradient tensor is written as
\begin{align}
  \tens{F} = \left(\td{R}{r}\right)^{-1} \vec{e}_r \otimes \vec{e}_r +
  \frac{r}{R} \vec{e}_\theta \otimes \vec{e}_\theta
  + \vec{e}_z \otimes \vec{e}_z.
\end{align}
The incompressibility condition simplifies to
\begin{align}
  R \td{R}{r} = \frac{r}{J} = r (1 - \phi_s - \phi_{+} - \phi_{-}).
\end{align}
The radial and orthoradial
elastic stresses are denoted as
$T_{e,rr} = \vec{e}_r \cdot \tens{T}_e \cdot \vec{e}_r$ and
$T_{e,\theta\theta} = \vec{e}_\theta \cdot \tens{T}_e \cdot \vec{e}_\theta$
and can be expressed as
\begin{align}
  T_{e,rr} = \frac{R}{r}\left(\left(\td{R}{r}\right)^{-1} - \td{R}{r}\right),
  \quad 
  T_{e,\theta\theta} = \td{R}{r}\left(\frac{r}{R} - \frac{R}{r}\right).
\end{align}
The radial component of the stress balance \eqref{nd:gel:div_T}
can be written as
\begin{align}
  \td{T_{e,rr}}{r} + \frac{T_{e,rr} - T_{e,\theta\theta}}{r}
  + \omega^2 \G^{-1} \phi_s\td{}{r}\left[\frac{1}{r}\td{}{r}\left(r \td{\phi_s}{r}\right)\right] +
  \frac{\beta^2 \G^{-1}}{r}\td{\Phi}{r}\td{}{r}\left(r \td{\Phi}{r}\right) = \td{p}{r}
\end{align}
and can be simplified through the use of \eqref{fc:gel:mu_s} and
\eqref{fc:gel:Phi} to 
\begin{align}
  \td{T_{e,rr}}{r} + \frac{T_{e,rr} - T_{e,\theta\theta}}{r}
  + \G^{-1} \phi_s\td{\Pi_s}{r}
  - 
  \G^{-1}(\phi_{+} - \phi_{-} + \alpha_f J^{-1})\td{\Phi}{r} = (1 - \phi_s) \td{p}{r}.
\end{align}

\subsection{Boundary conditions}

At the origin of the hydrogel ($r = 0$) we impose
\begin{align}
  R = 0, \quad \pd{\phi_s}{r} = 0, \quad \pd{\Phi}{r} = 0.
\end{align}
The first of these ensures that the Lagrangian origin is mapped to the
Eulerian origin. 

Due to the formulation of the model in terms of Eulerian coordinates, the deformed radius of the gel, $a$, is an unknown. Hence the steady problem is, in fact, a free boundary problem. The gel radius is implicitly defined by the equation
$R(r=a) = 1$, where we have scaled the undeformed radius of the
gel to one through a suitable non-dimensionalisation. At the gel-bath
interface, continuity of electric potential and electric displacement leads to
\begin{align}
  \Phi|_{r=a^{-}} = \Phi|_{r = a^+}, \quad 
  \left.\pd{\Phi}{r}\right|_{r=a^{-}} =  \epsilon_r \left.\pd{\Phi}{r}\right|_{r=a^{+}}.
\end{align}
In addition, the variational condition \eqref{nd:bc:grad_phi} becomes
\begin{align}
  \left.\pd{\phi_s}{r}\right|_{r = a^-} = 0. \label{fc:gel:variational}
\end{align}
Continuity of stress at the free boundary implies that
\begin{align}
  \left[\G T_{e,rr} + \frac{\beta^2}{2}\left(\pd{\Phi}{r}\right)^2 + \phi_s (\Pi_s - \bath{\mu}_s) - \G (1 - \phi_s) p\right]_{r=a^{-}} =
  \epsilon_r \beta^2\left[\frac{1}{2}\left(\pd{\Phi}{r}\right)^2 - p\right]_{r = a^+},
\end{align}
where $\bath{\mu}_s = \log(1 - \bath{\phi}_+)$ and \eqref{fc:gel:mu_s} along
with \eqref{fc:gel:variational} have been used to simplify the
Korteweg stress.

\subsection{Numerical treatment}

To numerically solve this problem, we use a Landau transformation and write $\hat{r} = r / a$. In addition, we rescale the Lagrangian radial coordinate as $\hat{R} = R / a$. The deformation gradient tensor is invariant under this transformation. However, the position of the free boundary is now explicitly determined by $\hat{R}(\hat{r}=1) = 1 / a$. The equations are discretised using finite differences and simultaneously solved using Newton's method with damping.